\documentclass[aps,prd,preprintnumbers,groupedaddress,nofootinbib,amssymb,notitlepage,eqsecnum]{revtex4-2}
\usepackage{here}
\usepackage[dvipdfmx]{graphicx}
\usepackage{amsmath,amsthm,amssymb}
\usepackage{bm}
\usepackage{color}

\usepackage{amsfonts}
\usepackage{dcolumn}
\usepackage{hyperref}
\allowdisplaybreaks[1]
\usepackage{stackengine}

\newcommand{\be}{\begin{equation}}  
\newcommand{\ee}{\end{equation}}
\newcommand{\ba}{\begin{eqnarray}}
\newcommand{\ea}{\end{eqnarray}}

\newcommand{\rd}{{\rm d}}

\newcommand{\cL}{{\cal L}}
\newcommand{\bem}{\begin{bmatrix}}
\newcommand{\eem}{\end{bmatrix}}
\newcommand{\Mpl}{M_{\rm Pl}}

\allowdisplaybreaks

\begin{document}

\preprint{RIKEN-iTHEMS-Report-25, YITP-25-152, WUCG-25-11}

\title{Vector Horndeski black holes in nonlinear electrodynamics}

\author{Che-Yu Chen$^{1}$\footnote{{\tt 
b97202056@gmail.com}}}
\author{Antonio De Felice$^{2}$\footnote{{\tt antonio.defelice@yukawa.kyoto-u.ac.jp}}}
\author{Shinji Tsujikawa$^{3}$\footnote{{\tt tsujikawa@waseda.jp}}}
\author{Taishi Sano$^{3}$\footnote{{\tt t.sano@ruri.waseda.jp}}}

\affiliation{$^1$RIKEN iTHEMS, Wako, Saitama 351-0198, Japan\\
$^2$Center for Gravitational Physics and Quantum Information, 
Yukawa Institute for Theoretical Physics, Kyoto University, 
606-8502, Kyoto, Japan\\
$^3$Department of Physics, Waseda University, 
3-4-1 Okubo, Shinjuku, Tokyo 169-8555, Japan}

\begin{abstract}

On a spherically symmetric and static background, we study the existence of linearly stable black hole (BH) solutions in nonlinear electrodynamics (NED) with a Horndeski vector-tensor (HVT) coupling, with and without curvature singularities at the center ($r=0$). Incorporating the electric charge $q_E$ and the magnetic charge $q_M$, we first show that nonsingular BHs can exist only if $q_M = 0$. We then study the stability of purely electric BHs by analyzing the behavior of perturbations in the metric and the vector field. Nonsingular electric BHs are unstable due to a Laplacian instability in the vector perturbation near the regular center. In the absence of the HVT coupling ($\beta=0$), singular BHs in power-law NED theories can be consistent with all linear stability conditions, while Born-Infeld BHs encounter strong coupling due to a vanishing propagation speed as $r \to 0$. In power-law NED and Born-Infeld theories with $\beta \neq 0$, the electric fields for singular BHs are regular near $r=0$, while the metric functions behave as $\propto r^{-1}$.
Nevertheless, we show that Laplacian instabilities occur for regions inside the outer horizon $r_h$, unless the HVT coupling constant $\beta$ is significantly smaller than $r_h^2$. For $\beta \neq 0$, 
we also reconstruct the NED Lagrangian so that one of the metric functions takes the Reissner-Nordstr\"om form. 
In this case, there exists a branch where all squared propagation speeds are positive, 
but the ghost and strong coupling problems are present 
around the BH center.
Thus, the dominance of the HVT coupling generally leads to BH instability in the high-curvature regime.

\end{abstract}

\date{\today}

%\pacs{04.50.Kd, 95.36.+x, 98.80.-k}

\maketitle

%%%%%%%%%%%%%%%%%%%%%%%%%%%%%%%%%
\section{Introduction}
\label{introsec}
%%%%%%%%%%%%%%%%%%%%%%%%%%%%%%%%%

The existence of black holes (BHs) is a fundamental prediction of Einstein's General Relativity (GR) \cite{Schwarzschild:1916uq}. 
In the absence of rotation, the BH solution in Einstein-Maxwell theory is described by the 
Reissner-Nordstr\"om (RN) spacetime \cite{Reissner:1916cle}, characterized by a mass $M$ and electric and/or magnetic charges $q$.
The stability of RN solutions can be analyzed by considering linear perturbations of both the metric and the electromagnetic field \cite{Moncrief:1974gw,Moncrief:1974ng,Zerilli:1974ai,Moncrief:1975sb}. In total, there are four dynamical perturbations: two tensor modes from the gravitational sector and two vector modes from the electromagnetic sector. All of these dynamical degrees of freedom (DOFs) propagate at the speed of light and are free from ghost instabilities in both 
timelike and spacelike regions.
Consequently, the RN BH solution is stable against linear perturbations on a static and spherically
symmetric (SSS) background.

In Einstein-Maxwell theory, the Lagrangian is given by the 
Einstein-Hilbert term, 
$\Mpl^2 R/2$, together with the electromagnetic field term, 
$F = -F_{\mu\nu} F^{\mu\nu}/4$, 
where $\Mpl$ is the reduced Planck mass,  
$R$ is the Ricci scalar, and  $F_{\mu\nu} = \partial_\mu A_\nu - \partial_\nu A_\mu$ denotes the field strength of a vector field $A_\mu$.
One can consider BH solutions in a more general framework where 
the Maxwell Lagrangian, $F$, is extended to a nonlinear function, 
${\cal L}(F)$ \cite{Ayon-Beato:1998hmi,Ayon-Beato:1999kuh,Ayon-Beato:2000mjt,Dymnikova:2004zc,Ansoldi:2008jw,Balart:2014cga,Balart:2014jia,Fan:2016hvf,Rodrigues:2018bdc,Maeda:2021jdc,Yajima:2000kw,Fernando:2003tz,Cai:2004eh,Dey:2004yt,Kruglov:2017mpj}.
Such a scheme, which is known as 
nonlinear electrodynamics (NED), 
encompasses Euler-Heisenberg 
theory \cite{Heisenberg:1936nmg}
and Born-Infeld theory \cite{Born:1934gh}. 
In Einstein-NED theory, the behavior of the electromagnetic field on the SSS background is modified compared to that in Einstein-Maxwell theory, leading to corresponding changes in the background metric.
This allows one to distinguish between the two theories through observations of gravitational waves \cite{LIGOScientific:2016aoc} and BH shadows \cite{EventHorizonTelescope:2019dse}.

In Einstein-NED theory, there have been attempts to construct nonsingular BHs for which the curvature scalars remain finite at the center ($r=0$) \cite{Ayon-Beato:1998hmi,Ayon-Beato:1999kuh,Ayon-Beato:2000mjt,Dymnikova:2004zc,Ansoldi:2008jw,Balart:2014cga,Balart:2014jia,Fan:2016hvf,Rodrigues:2018bdc,Maeda:2021jdc}. With suitable choices of the NED Lagrangian, 
${\cal L}(F)$, it is possible to obtain regular BH solutions carrying an electric or magnetic charge.
However, an analysis of BH perturbations near the regular center reveals that vector-field perturbations invariably develop a Laplacian instability in such nonsingular BHs \cite{DeFelice:2024seu,DeFelice:2024ops}.
Since this instability triggers the rapid growth of metric perturbations, nonsingular BHs in Einstein-NED theory cannot remain stable.

In Einstein-NED theory, the vector field $A_\mu$ is not directly coupled to gravity. 
To overcome the difficulties in constructing regular BHs without instabilities, one may introduce couplings between the vector field and the curvature tensors. 
It is desirable to construct an interacting Lagrangian whose Euler-Lagrange equations remain second order, thereby avoiding Ostrogradsky instabilities \cite{Ostrogradsky:1850fid}.
Moreover, maintaining the $U(1)$ gauge invariance of the vector field prevents the propagation of an additional longitudinal mode.\footnote{For example, in generalized Proca theories with broken $U(1)$ symmetry \cite{Heisenberg:2014rta,Tasinato:2014eka,BeltranJimenez:2016rff,Allys:2016jaq}, BH solutions with cubic vector Galileons \cite{Heisenberg:2017xda,Heisenberg:2017hwb} exhibit pathological behavior in the longitudinal scalar perturbation \cite{DeFelice:2024bdq}.}
Indeed, Horndeski derived a unique vector-gravity interaction that satisfies both of these properties \cite{Horndeski:1976gi}. 
The corresponding Lagrangian, known as the Horndeski 
vector-tensor (HVT) term, is
${\cal L}_{\rm HVT} = 
\beta L^{\mu \nu \rho \sigma} F_{\mu \nu}F_{\rho \sigma}$,
where $\beta$ is a coupling constant and $L^{\mu \nu \rho \sigma}$ denotes the double dual Riemann tensor.

In Einstein-HVT theory, BH solutions on the SSS background have been studied in 
Refs.~\cite{Horndeski:1978ca,Mueller-Hoissen:1988cpx,Balakin:2007am,Verbin:2020fzk,Chen:2024hkm} (see Refs.~\cite{Esposito-Farese:2009wbc,Barrow:2012ay,BeltranJimenez:2013btb} for cosmological applications). 
The HVT coupling does not eliminate the curvature singularity at $r=0$, but it modifies the behavior of the metric functions near the center compared to the 
RN solution. In Ref.~\cite{Chen:2024hkm}, the stability of BHs in Einstein-HVT theories was analyzed outside the outer event horizon. This analysis constrains the coupling $\beta$ and the electric and magnetic charges by requiring the absence of ghosts and Laplacian instabilities.
However, it remains to be seen whether the stability of BHs is also ensured inside the outer horizon under the bounds derived in Ref.~\cite{Chen:2024hkm}.

In this paper, we study the BH solutions and their stabilities against linear perturbations in 
Einstein-NED-HVT theory given by 
the total Lagrangian 
${\cal L}_{\rm T}=\Mpl^2 R/2+{\cal L}(F)+\beta L^{\mu \nu \rho \sigma}
F_{\mu \nu}F_{\rho \sigma}$.
First, we aim to clarify whether linearly stable, nonsingular BHs can be realized by incorporating both the NED and HVT couplings.
For this purpose, we consider the magnetic charge $q_M$ in addition to the electric 
charge $q_E$. 
We will show that the realization of nonsingular BHs at the background level requires $q_M = 0$. 
Next, we investigate the linear stability of purely electrically charged BHs that possess regular centers. We find that the squared propagation speed of vector-field perturbations in the even-parity sector, $c_{\Omega 4}^2$, is always negative for such nonsingular electric BHs. 
Thus, in Einstein-NED-HVT theory, there are no regular charged BH solutions that satisfy all theoretically consistent conditions. 

Although nonsingular BHs without theoretical pathologies do not exist in Einstein-NED-HVT theory, it may still allow the presence of linearly stable singular BHs with curvature singularities at $r=0$. To address this issue, we focus on purely electrically charged BHs and study their stability against linear perturbations. 
For this purpose, we formulate a BH perturbation theory valid in both timelike and spacelike regions. In the presence of a singularity at $r=0$, the linear perturbation theory is expected to break down below the effective-field-theory (EFT) length scale 
$r_{\rm EFT}$. 
Since nonlinear perturbations are significant only close to $r=0$, the EFT scale should lie well below the outer event horizon radius, $r_h$. The linear perturbation theory is valid not only for distances $r > r_h$ but also in the regime $r_{\rm EFT} < r < r_h$, where $0<r_{\rm EFT} \ll r_h$.

We will first revisit the stability of BHs in Einstein-Maxwell-HVT theory by analyzing the behavior of perturbations in the spacelike region.
We find that there exists a characteristic distance $r_g$, dependent on $\beta$, below which both ghost and Laplacian instabilities appear. 
To avoid these instabilities, we require that 
$r_g < r_{\rm EFT}$.
This puts a tight constraint on the coupling $\beta$, typically $|\beta| \lesssim r_{\rm EFT}^3/r_h$, so that $|\beta| \ll r_h^2$ for $r_{\rm EFT} \ll r_h$. 
In such cases, the observational signatures of the HVT coupling for $r > r_h$, such as in BH quasinormal modes, are likely to be difficult to detect.

In Einstein-NED theory without the HVT coupling ($\beta=0$), we also investigate the stability of BHs in power-law NED, with ${\cal L}(F)=F+a_p F^p$, and in Born-Infeld theory, with ${\cal L}(F)=(2/b) \left(1-\sqrt{1-bF}\right)$, 
where $a_p$, $p$, and $b$ are 
constants. In power-law NED with an integer $p \ge 2$, we show that all linear stability conditions are satisfied for $a_p > 0$ at any distance $r$, in both timelike and spacelike regions. In Born-Infeld theory, 
$c_{\Omega 4}^2$ approaches 0 as $r \to 0$, giving rise to  
a strong coupling problem. To avoid this issue, the Born-Infeld coupling is constrained to be small, 
i.e., $b<2r_{\rm EFT}^4/q_E^2$. 
When the HVT coupling is present ($\beta \neq 0$), we find that $c_{\Omega 4}^2 < 0$ near $r=0$ for both power-law NED and Born-Infeld theories. Therefore, unless the coupling $\beta$ is sufficiently small, as mentioned above, the presence of the HVT coupling renders the BH solutions unstable. For $\beta \neq 0$, we also reconstruct the NED Lagrangian ${\cal L}(F)$ so that one of the metric functions takes the same form as the RN BH. In this case, Laplacian instabilities are absent for all $r>0$, but ghosts appear, along with strong coupling near $r=0$. 
Therefore, when the HVT coupling dominates at high curvature, it typically induces ghosts, Laplacian instabilities, or strong coupling.

This paper is organized as follows.
In Sec.~\ref{backsec}, we derive 
the field equations of motion on 
the SSS background in 
Einstein-NED-HVT theory. 
In Sec.~\ref{NSBHsec}, we investigate the possibility of realizing nonsingular BHs and show that the existence of consistent background solutions requires $q_M=0$. 
In Sec.~\ref{NSBHsec0}, we then discuss electrically charged BH solutions with curvature singularities at $r=0$ for several subclasses of Einstein-NED-HVT theory. 
In Sec.~\ref{BHpersec}, we obtain
the linear stability conditions for 
electrically charged BHs by analyzing the behavior of perturbations in both 
timelike and spacelike regions.
In Sec.~\ref{elesinstasec}, we show that nonsingular electric BHs in Einstein-NED-HVT theory exhibit Laplacian instabilities due to negative values of $c_{\Omega 4}^2$. In Sec.~\ref{elesstasec}, we examine the linear stability of BHs 
singular at $r=0$ for the subclasses of Einstein-NED-HVT theory introduced in Sec.~\ref{NSBHsec0}.
Sec.~\ref{consec} is devoted to conclusions.

%%%%%%%%%%%%%%%%%%%%%%%%%%%%%%%%%%%%%%%%%%%%%%%%%%%%%%%%%%%%%%%
\section{Background equations in 
Einstein-NED-HVT theory}
\label{backsec}
%%%%%%%%%%%%%%%%%%%%%%%%%%%%%%%%%%%%%%%%%%%%%%%%%%%%%%%%%%%%%%%

We study theories described by the action 
\be
{\cal S}=\int {\rm d}^4 x \sqrt{-g} \left[ 
\frac{\Mpl^2}{2}R+{\cal L}(F)
+\beta L^{\mu \nu \rho \sigma} 
F_{\mu \nu}F_{\rho \sigma}  \right]\,,
\label{action}
\ee
where $g$ is the determinant of the metric tensor $g_{\mu \nu}$, 
$\Mpl$ is the reduced Planck mass, 
and $R$ is the Ricci scalar.
The Lagrangian ${\cal L}$ is a function 
of the electromagnetic field strength 
$F=-F_{\mu \nu}F^{\mu \nu}/4$, 
where $F_{\mu \nu}=\partial_{\mu}A_{\nu}
-\partial_{\nu}A_{\mu}$ and $A_{\mu}$ is 
a gauge field.
The last term in Eq.~(\ref{action}) is the 
HVT interaction \cite{Horndeski:1978ca}, 
where $L^{\mu \nu \rho \sigma}$ is the double dual Riemann tensor defined by
\be
L^{\mu\nu\rho\sigma}=\frac{1}{4}
\mathcal{E}^{\mu\nu\alpha\beta}
\mathcal{E}^{\rho\sigma\gamma\delta} 
R_{\alpha\beta\gamma\delta}\,.
\ee
Here, $R_{\alpha\beta\gamma\delta}$ and $\mathcal{E}^{\mu\nu\alpha\beta}$  
are the Riemann tensor and the 
anti-symmetric Levi-Civita tensor, 
respectively, with $\mathcal{E}^{0123}=-1/\sqrt{-g}$ and $\mathcal{E}_{0123}=\sqrt{-g}$.

NED corresponds to a Lagrangian ${\cal L}$ 
that depends nonlinearly on $F$. 
In Einstein gravity with the 
Euler-Heisenberg Lagrangian  
${\cal L} = F + \alpha_2 F^2$ \cite{Heisenberg:1936nmg}, 
where the term $\alpha_2 F^2$ represents 
a correction to the Maxwell term $F$, it is known that hairy BH solutions exist 
\cite{Yajima:2000kw}, in which 
the RN geometry is modified by this correction.
Hairy BHs are also present for Einstein gravity with the Born-Infeld Lagrangian 
${\cal L}=(2/b) (1-\sqrt{1-b F})$ \cite{Born:1934gh}, as 
discussed in Refs.~\cite{Fernando:2003tz,Cai:2004eh,Dey:2004yt,Kruglov:2017mpj}.
These hairy BH solutions have a curvature singularity at the center.
By choosing specific forms of the function ${\cal L}(F)$, Einstein-NED theory admits nonsingular BHs that are regular at their centers \cite{Ayon-Beato:1998hmi,Ayon-Beato:1999kuh,Ayon-Beato:2000mjt,Bronnikov:2000vy,
Dymnikova:2004zc}. However, these solutions suffer from Laplacian instabilities in the angular direction, so the nonsingular metric cannot remain 
stable \cite{DeFelice:2024seu,DeFelice:2024ops}.

In this paper, we incorporate the HVT coupling $\beta L^{\mu \nu \rho \sigma} 
F_{\mu \nu}F_{\rho \sigma}$ into Einstein-NED theories and investigate the existence of 
hairy BH solutions that are free from linear instabilities.
This HVT coupling represents a distinctive interaction between the vector field and gravity, 
preserving the field equations up to second order while maintaining $U(1)$ gauge invariance 
\cite{Horndeski:1976gi}.
We note that hairy BH solutions free from both ghosts and Laplacian instabilities 
outside the outer event horizon were found in Ref.~\cite{Chen:2024hkm}, 
although they possess curvature 
singularities at their centers.
Our first aim in this paper is to investigate whether nonsingular BHs 
free from instabilities can exist in theories described by the action (\ref{action}). 
If no such solutions are found, we will then explore the parameter space in which singular BHs satisfy the linear stability conditions in both timelike and spacelike regions.

We consider a SSS background described 
by the line element 
\be
\rd s^2=-f(r) \rd t^{2} +h^{-1}(r) \rd r^{2}+ 
r^{2} \left( \rd \theta^{2}+\sin^{2}\theta\,\rd\varphi^{2} 
\right)\,,
\label{metric_bg}
\ee
where $f$ and $h$ depend on the radial coordinate $r$. 
On this background, we choose the 
following vector-field configuration
\be
A_{\mu}=\left[ A_0 (r), 0, 0, -q_M \cos \theta \right]\,,
\ee
where $A_0$ is a function of $r$, and $q_M$ is a constant 
representing the magnetic charge. 
In theories described by the action (\ref{action}), 
the $U(1)$ gauge symmetry permits us to set the radial 
component of the vector field $A_1(r)$, to 0.
The electromagnetic field strength 
is expressed as\footnote{The magnetic charge must be constant to ensure compatibility with spherical symmetry. 
Indeed, if $q_M$ is a function of $r$, the electromagnetic field strength takes the form 
$F = h A_0'^2/(2f) - q_M^2/(2r^4) - h q_M'^2/(2r^2 \tan^2\theta)$. Since the assumption of spherical symmetry requires $F$ to be independent of $\theta$, 
it follows that $q_M' = 0$, i.e., 
$q_M = \mathrm{constant}$.}
\be
F=\frac{hA_0'^2}{2f}-\frac{q_M^2}{2r^4}\,,
\label{Fre}
\ee
where a prime denotes differentiation 
with respect to $r$.

Since the determinant of the metric is 
$g = -(f/h)\, r^4 \sin^2 \theta$, 
the requirement of a positive $-g$ 
in the action (\ref{action}) imposes the condition
\be
\frac{f}{h}>0\,.
\label{fhcon}
\ee
This inequality implies that $f$ and $h$ must change sign 
simultaneously at the horizons (characterized by $f=0=h$). 
Therefore, there are only two possible regions: 
(i) a time-like region with $f>0$ and $h>0$, and 
(ii) a space-like region with $f<0$ and $h<0$. 
Since we are interested in BH physics in both time-like 
and space-like regions, we will not fix the sign of $f$ (or $h$) 
in the following discussion. 

Varying Eq.~(\ref{action}) with respect to $f$, $h$, 
and $A_0$, the resulting background equations 
of motion are
\ba
& &
h'=\frac{r^4(1-h)(\Mpl^2 f-4\beta h A_0'^2)
+r^6(f\cL-h A_0'^2 \cL_{,F})
+24 \beta q_M^2 fh}
{rf(\Mpl^2 r^4+4\beta q_M^2)}\,,
\label{back1}\\
& &
f'=\frac{r^3 [f(1-h)\Mpl^2
+r^2 (f\cL-h A_0'^2 \cL_{,F})
+4\beta h (3h-1)A_0'^2]}
{h(\Mpl^2 r^4+4\beta q_M^2)}\,,
\label{back2}\\
& & 
\left( \sqrt{\frac{h}{f}}
\left[ r^2\cL_{,F}
-8 \beta (h-1) \right]A_0' \right)'=0\,.
\label{back3}
\ea
Here, we use the notation $\cL_{,F} \equiv \mathrm{d}\cL/\mathrm{d}F$.
Equation (\ref{back3}) can be integrated to give 
\be
A_0'=\frac{q_E}
{r^2\cL_{,F}-8 \beta (h-1)}\sqrt{\frac{f}{h}}\,,
\label{back4}
\ee
where $q_E$ is an integration constant 
characterizing the electric charge.
For the electrically charged BH, the non-vanishing 
electric field $A_0'(r)$ affects the metric 
components $f$ and $h$ through the right-hand 
sides of Eqs.~(\ref{back1}) and (\ref{back2}). 
For the purely magnetically charged BH, we have $A_0' = 0$, 
so that the terms proportional to $A_0'^2$ 
in Eqs.~(\ref{back1}) and (\ref{back2}) vanish. 
Combining Eq.~(\ref{back1}) with 
Eq.~(\ref{back2}), we find 
\be
\frac{f'}{f}-\frac{h'}{h}
=\frac{8\beta (r^4 h A_0'^2-3q_M^2 f)}
{rf(\Mpl^2 r^4+4\beta q_M^2)}\,.
\label{fh}
\ee
In the absence of the HVT coupling ($\beta=0$), 
it follows that $f'/f=h'/h$. 
Imposing asymptotically flat boundary conditions, 
$f \to 1$ and $h \to 1$ as 
$r \to \infty$, we then obtain $f=h$. 
However, the presence of the HVT coupling 
generally leads to a difference 
between $f$ and $h$.

For later convenience, we derive an equation 
involving $A_0'$, $f$, $h$, and their derivatives 
with respect to $r$. Using Eq.~(\ref{back4}), 
$\cL_{,F}$ can be expressed in terms of $A_0'$. 
Substituting this relation into Eq.~(\ref{back1}) 
and solving for $\cL$, we obtain
\be
\cL=\frac{q_E r^4 h\sqrt{f/h}\,A_0'
+\Mpl^2 r^5 f h'
+r^4 (h-1)(\Mpl^2 f+4\beta h A_0'^2)
+4\beta q_M^2 f (rh'-6h)}
{r^6 f}\,.
\label{cL}
\ee
By differentiating Eqs.~(\ref{Fre}) and 
(\ref{cL}) with respect to $r$, 
we can compute $\cL_{,F} = \cL'(r)/F'(r)$ as a function of $r$. 
Substituting this expression for $\cL_{,F}$ 
into Eq.~(\ref{back4}) allows us to eliminate $\cL_{,F}$, yielding
\ba
& &
\Mpl^2 r^4 A_0' \left( 2h-2-r^2 h'' \right) 
fh \sqrt{f/h}+2q_E f \left( r^4 h A_0'^2
+f q_M^2 \right) \nonumber \\
& &
-4 \beta h \sqrt{f/h}\, A_0' \left[ 
r^5 hh'A_0'^2 -2r^4 h(h - 1)A_0'^2 
+q_M^2 r^2 fh'' 
- 11q_M^2 rfh' + 4 q_M^2f (8h + 1)
\right]=0\,.
\label{A0eq}
\ea
For the purely electric BH characterized 
by $q_E \neq 0$ and $q_M=0$, and with $A_0'$ nonvanishing, 
Eq.~(\ref{A0eq}) possesses two branches of solutions, 
given by 
\be
A_{0 \pm}'=-\frac{q_E}
{4\beta (2h-2-rh')} \sqrt{\frac{f}{h}}
\left( 1 \pm \sqrt{1-\xi} \right)\,,
\label{A0so}
\ee
where 
\be
\xi \equiv \frac{4\Mpl^2 
\beta (2h-2-rh')(2h-2-r^2 h'')}
{q_E^2}\,.
\ee
The double signs in the subscript of $A_0'$ 
correspond to the same order of signs as those 
on the right-hand side of Eq.~(\ref{A0so}).
Taking the limit $\beta \to 0$ in the 
plus branch of Eq.~(\ref{A0so}), we obtain
\be
A_{0+}'=-\frac{q_E}
{2 (2h-2-rh')}\sqrt{\frac{f}{h}}\beta^{-1}
+{\cal O}(\beta^0)\,,
\ee
which diverges as $\beta \to 0$. 
Thus, a finite electric field in NED 
is not recovered in the continuous limit $\beta \to 0$. 
On the other hand, for the minus branch 
of Eq.~(\ref{A0so}), we have
\be
A_{0-}'=-\frac{\Mpl^2 
(2h-2-r^2h'')}{2q_E}\sqrt{\frac{f}{h}}
+{\cal O}(\beta)\,,
\ee
which smoothly approaches the NED case 
as $\beta \to 0$.

For the purely magnetic BH ($q_E=0$ and 
$q_M \neq 0$), Eq.~(\ref{back4}) gives 
\be
A_0'=0\,,
\ee
at any distance $r$. 
This solution is indeed consistent with Eq.~(\ref{A0eq}).

%%%%%%%%%%%%%%%%%%%%%%%%%%%%%%%%%%%%%%
\section{Nonsingular black holes}
\label{NSBHsec}
%%%%%%%%%%%%%%%%%%%%%%%%%%%%%%%%%%%%%%

In this section, we investigate the possibility of 
realizing nonsingular BHs in  
Einstein-NED-HVT theory at the background level. 
To avoid divergences in the Ricci scalar $R$, 
the squared Ricci tensor $R_{\mu \nu} R^{\mu \nu}$, 
and the Kretschmann scalar 
$R_{\mu \nu \rho \sigma} R^{\mu \nu \rho \sigma}$, 
the metric components in Eq.~(\ref{metric_bg}) 
should be expanded around $r = 0$, as \cite{Frolov:2016pav,DeFelice:2024ops} \footnote{We will focus on nonsingular geometries that evade 
Penrose's singularity theorem 
by breaking global hyperbolicity. 
Nonsingular BHs that preserve 
global hyperbolicity are also possible. 
See Ref.~\cite{Carballo-Rubio:2019fnb} for 
the classification of these geometries.} 
\be
f(r)=f_0+\sum_{n=2}^{\infty} f_n r^n\,,\qquad 
h(r)=1+\sum_{n=2}^{\infty} h_n r^n\,,
\label{fhexpansion}
\ee
where $f_0$, $f_n$, and $h_n$ are constants. 
We require $h(0) = 1$ to avoid a conical singularity at $r = 0$. 
For $f_0 = 0$, the curvature scalars diverge, 
so we impose the condition $f_0 \neq 0$. 
Moreover, the inequality (\ref{fhcon}) 
must hold at $r = 0$, which implies
\be
f_0 > 0\,.
\ee
In the following, we first show that the existence 
of nonsingular BHs requires $q_M = 0$. 
We then examine the behavior of purely 
electrically charged BHs near $r = 0$.

\subsection{Proof of $q_M=0$}

Solving Eq.~\eqref{fh} for $A_0'^2$, it follows that  
\be
A_0'^2=\frac{r(hf'-fh')(\Mpl^2 r^4+4\beta q_M^2)+24 \beta q_M^2 fh}{8\beta r^4 h^2}\,,\label{eq:A0_prime_sq}
\ee
which is valid for $\beta\neq0$. 
By substituting Eq.~\eqref{eq:A0_prime_sq} into the 
$A_0'^2$ terms of Eq.~\eqref{A0eq}, the equation 
simplifies to a linear form in $A_0'$.
We then obtain
\begin{align}
& A_{0}' \,\{2 [2 (52 \beta  q_{M}^{2}-\Mpl^{2} r^{4}) f 
-r {f'} (\Mpl^{2} r^{4}+4 \beta  q_{M}^{2})] h^{2}
+\{2 [(r^{5} \Mpl^{2}-28 \beta  r q_{M}^{2}) {h'}+\left( \Mpl^{2}r^{6}+4 \beta q_{M}^{2} \,r^{2} \right) {h''}\nonumber\\
&+2 \Mpl^{2} r^{4}+40 \beta  q_{M}^{2}] f +r {f'} (r {h'}+2) (\Mpl^{2} r^{4}+4 
\beta q_{M}^{2})\} h -r f {h'} (r {h'}+2) \left(\Mpl^{2} r^{4}+4 \beta  q_{M}^{2}\right)\} \nonumber\\
&+\frac{q_E \{[r {h'} (\Mpl^{2} r^{4}+4 \beta q_{M}^{2})-32 q_{M}^{2} \beta h] f -r h {f'} (\Mpl^{2} r^{4}+4 \beta  q_{M}^{2})\}}{2\beta 
\sqrt{h/f} }=0\,.
\label{A0fh}
\end{align}
This relation uniquely determines $A_0'$ as a function of $f$ and $h$. 
By solving Eq.~\eqref{A0fh} for $A_0'$, squaring the result, 
and equating it with Eq.~\eqref{eq:A0_prime_sq}, 
we obtain an equation that involves only $f$, $h$, and 
their derivatives with respect to $r$. 
For nonsingular BH solutions, this equation can be used 
to fix the series coefficients of $f$ in Eq.~\eqref{fhexpansion} 
in terms of those of $h$ (or vice versa). 
Following this procedure, we obtain
\be
0=-3q_M^2f_0 r^{-4} 
+ \mathcal{O}(r^{-2})\,.
\ee
Since $f_0>0$, this implies that 
\be
q_M=0\,,
\ee
independently of the value of $q_E$. 
This argument excludes both purely magnetic ($q_E = 0$, $q_M \neq 0$) 
and dyonic ($q_E \neq 0$, $q_M \neq 0$) BHs.
The absence of nonsingular BHs for $q_M \neq 0$ arises 
from the nonvanishing right-hand side of Eq.~\eqref{fh} 
when $\beta \neq 0$. For the purely magnetic BH with 
$A_0'=0$, we have  
\be
\frac{f'}{f}-\frac{h'}{h}
=-\frac{24\beta q_M^2}{r(\Mpl^2 r^4+4\beta q_M^2)}\,.
\label{fhred}
\ee
Integrating this equation with respect to $r$, we obtain 
\be
\frac{f}{h}={\cal C} \left( 1+\frac{4\beta q_M^2}
{\Mpl^2 r^4} \right)^{3/2}\,,
\label{fhre}
\ee
where ${\cal C}$ is an integration constant. 
We require ${\cal C} \neq 0$ to prevent $f(r)$ from 
vanishing when $h(r) \neq 0$.
Using the expansion of $h(r)$ in Eq.~\eqref{fhexpansion}, 
Eq.~\eqref{fhre} yields the following expansion 
for $f(r)$ around the BH center:
\be
f(r)={\cal C}\left( \frac{4\beta q_M^2}
{\Mpl^2} \right)^{3/2}r^{-6}+{\cal O}(r^{-4})\,.
\label{frma}
\ee
This behavior is incompatible with the requirement 
for the metric component $f(r)$ in Eq.~\eqref{fhexpansion}. 
As seen from Eq.~\eqref{fhre}, this arises from the nonvanishing HVT coupling $\beta$. 
In the absence of the HVT coupling ($\beta = 0$), nonsingular magnetic BHs can exist.

For the dyon BH, the electric 
field $A_0'$ contributes to Eq.~\eqref{fh}. 
As long as $A_0'(r)$ behaves as $A_0'(r) \propto r^p$ with $p<-2$ 
around $r=0$, the behavior on the right-hand side of Eq.~(\ref{fh}) 
differs from that in Eq.~(\ref{fhred}). 
In this case, near $r=0$, Eq.~\eqref{fh} 
can be approximated as

\be
\frac{f'}{f}-\frac{h'}{h}
\simeq \frac{2r^3h A_0'^2}{fq_M^2} 
\simeq \frac{2{\cal C}_0^2 r^{2p+3}}
{f_0 q_M^2}\,,
\ee
where we have substituted $A_0' = {\cal C}_0 r^p$, 
with ${\cal C}_0$ being a constant. 
The integrated solution around $r=0$ is 
\be
\frac{f}{h}={\cal C}_1 \exp \left[ 
\frac{{\cal C}_0^2}{f_0 q_M^2(p+2) 
r^{-2(p+2)}} \right]\,,
\label{fhre3}
\ee
where ${\cal C}_1$ is an integration constant. 
For $p<-2$, the right-hand side of Eq.~\eqref{fhre3} 
exponentially vanishes as $r\to0$, so a solution with 
$A_0'(r)\propto r^p$ is incompatible with the regular 
metric ansatz in Eq.~\eqref{fhexpansion}.
We note that Eq.~\eqref{fhre3} applies to dyonic BHs 
with $q_M \neq 0$ but not to purely electric BHs with $q_M = 0$, 
which will be discussed separately in Sec.~\ref{pureelesec}.

\subsection{Purely electric BHs}
\label{pureelesec}

For purely electric BHs, we use Eq.~\eqref{A0so} to 
estimate the behavior of the electric field near $r = 0$. 
By substituting Eq.~\eqref{fhexpansion} with $h_3 \neq 0$ 
into Eq.~\eqref{A0so} and expanding near $r=0$, 
the $\pm$ branches of Eq.~(\ref{A0so}) yield
\ba
A_{0+}'(r) &=& \frac{q_E \sqrt{f_0}}
{2\beta h_3}r^{-3} 
-\frac{q_E \sqrt{f_0} h_4}
{\beta h_3^2}r^{-2}+{\cal O}(r^{-1})\,,
\label{A0p} \\ 
A_{0-}'(r) &=& \frac{2\Mpl^2 \sqrt{f_0}h_3}
{q_E}r^3+\frac{5\Mpl^2 \sqrt{f_0}h_4}{q_E} 
r^4+{\cal O}(r^5)\,, 
\label{A0m}
\ea
respectively. 
While $A_{0+}'(r)$ diverges at $r=0$, $A_{0-}'(r)$ 
approaches 0 as $r \to 0$. 
The latter property is similar to that of 
nonsingular BHs in NED with $\beta = 0$ \cite{DeFelice:2024seu}.
From Eq.~(\ref{fh}), we obtain
\be
\frac{f'}{f}-\frac{h'}{h}
=\frac{8\beta h A_0'^2}{\Mpl^2 rf}
=\frac{q_E^2 (1 \pm \sqrt{1-\xi})^2}
{2\Mpl^2 \beta r (2h-2-rh')^2}\,.
\label{fE}
\ee
Using the expansions in Eqs.~\eqref{A0p} and \eqref{A0m} 
around $r = 0$ and integrating Eq.~\eqref{fE} with respect to $r$, 
the leading-order contribution to $f/h$ can be estimated as
\ba
\frac{f}{h} &=& {\cal C} \exp \left( 
-\frac{q_E^2}{3 \Mpl^2 \beta h_3^2 r^6}
\right)\,,\qquad {\rm for} \quad 
A_0'=A_{0+}'\,,\label{fh1} \\
\frac{f}{h} &=& {\cal C}  \exp \left( 
\frac{16 \Mpl^2 \beta h_3^2 r^6}{3q_E^2}
\right)\,,\qquad~ {\rm for} \quad 
A_0'=A_{0-}'\,,\label{fh2}
\ea
where ${\cal C}$ is an integration constant.
In the limit $r \to 0$, the right-hand 
side of Eq.~(\ref{fh1}) approaches 0 for $\beta>0$, 
while it diverges to $\pm \infty$ for $\beta < 0$, 
depending on the sign of ${\cal C}$.
This behavior is incompatible with Eq.~\eqref{fhexpansion}, 
which requires $f/h \to f_0 > 0$ as $r \to 0$ with finite $f_0$. 
Thus, the branch $A_0' = A_{0+}'$ does not lead to nonsingular BHs. 
Conversely, Eq.~\eqref{fh2} can match the expansions of the metric 
components in Eq.~\eqref{fhexpansion} at $r=0$ by choosing ${\cal C} = f_0$.
Accordingly, it is possible to realize regular electric BHs 
in the branch $A_0' = A_{0-}'$.

In the following, we will focus on the branch $A_0'=A_{0-}'$. 
Taking the minus branch of Eq.~\eqref{A0so} and 
substituting it into Eqs.~\eqref{Fre} and \eqref{cL}, 
the field strength $F$ and the Lagrangian $\cL$ are expressed as
\ba
F &=& \frac{q_E^2 (1-\sqrt{1-\xi})^2}
{32 \beta^2 (2h-2-rh')^2}\,,\label{Fgene}\\
\cL &=& \frac{\Mpl^2 (rh'+h-1)}{r^2}
-\frac{q_E^2[rh'(\sqrt{1-\xi}-1)+\xi (h-1)]}
{4\beta r^2 (2h-2-rh')^2}\,,\label{Lgene}
\ea
both of which depend on $h$ but not on $f$. 
Substituting the expanded solution $h(r) = 1 + \sum_{n=2}^{\infty} h_n r^n$ 
around $r = 0$ into the above expressions for $F$ and $\cL$, we obtain
\ba
F &=& \frac{2\Mpl^4 h_3^2}{q_E^2}r^6
+\frac{10\Mpl^4 h_3 h_4}{q_E^2}r^7+{\cal O}(r^8)\,,\\
\cL &=& 3\Mpl^2 h_2+6\Mpl^2 h_3 r+{\cal O}(r^2)\,,
\ea
which are both finite at $r=0$.

For a given nonsingular metric function $h(r)$, the field strength 
$F$ and the Lagrangian $\cL$ are determined as functions of $r$ 
using Eqs.~\eqref{Fgene} and \eqref{Lgene}.
As long as $r$ can be explicitly expressed as a function of $F$, it is possible to reconstruct the Lagrangian $\cL$ as a function of $F$. 
We note that many of the regular metrics proposed in the literature have $h_3 = 0$ \cite{Bardeen:1968,Ayon-Beato:1998hmi,Dymnikova:2004zc,Hayward:2005gi}. 
In this case, the dominant term of $A_{0+}'$ behaves as 
$q_E \sqrt{f_0}/(4 \beta h_4 r^4)$ near $r = 0$, 
which leads to $f/h = \exp\big[-q_E^2/(16 \Mpl^2 \beta h_4^2 r^8)\big]$ 
at leading order. Thus, this branch is not compatible with the 
regular metric functions in Eq.~\eqref{fhexpansion} at $r = 0$.
For $h_3=0$, the other branch $A_{0-}'$ has the following 
behavior around $r=0$:
\ba
A_{0-}' &=& 
\frac{5\Mpl^2 \sqrt{f_0}h_4}
{q_E}r^4+\frac{9\Mpl^2 \sqrt{f_0}h_5}
{q_E}r^5+{\cal O}(r^6)\,,\\
F &=& \frac{25\Mpl^4 h_4^2}
{2q_E^2}r^8+\frac{45 \Mpl^4 h_4 h_5}
{q_E^2}r^{9}+{\cal O}(r^{10})\,,\\
{\cal L} &=& 3\Mpl^2 h_2
+10\Mpl^2 h_4 r^2+{\cal O}(r^3)\,,
\ea
with the leading-order behavior $f/h=f_0 \exp ( 25\Mpl^2 
\beta h_4^2r^8/q_E^2 )$. 
This branch can be consistent with the regular metric ansatz 
in Eq.~\eqref{fhexpansion}. 
Moreover, $A_{0-}'$, $F$, and $\cL$ remain finite at $r = 0$. 
We have thus shown that, for both $h_3 \neq 0$ and $h_3 = 0$, 
the branch $A_{0-}'(r)$ can give rise to nonsingular electric BHs. 
It remains to be seen whether such regular BHs are stable against linear perturbations.

%%%%%%%%%%%%%%%%%%%%%%%%%%%%%%%%%%%%%%
\section{Singular electric BHs}
\label{NSBHsec0}
%%%%%%%%%%%%%%%%%%%%%%%%%%%%%%%%%%%%%%

In this section, we study the properties of background 
BH solutions that have curvature singularities 
at $r = 0$. 
In the following, we focus on the purely 
electric case:
\be
q_E \neq 0\,,\qquad q_M=0\,.
\ee
From Eqs.~\eqref{Fre} and \eqref{back4}, we obtain 
\be
F=\frac{q_E^2}{2\left[r^2\cL_{,F}-8\beta\left(h-1\right)\right]^2}\,.\label{Fele}
\ee
Equations~\eqref{back1} and \eqref{fh} can then be recast as
\be
\Mpl^2\left[r(1-h)\right]'=2F\left[4\beta(1-h)+r^2\cL_{,F}\right]-r^2\cL\,,\label{hele}
\ee
and
\be
\frac{f'}{f}-\frac{h'}{h}=\frac{16\beta F}{\Mpl^2 r}\,,\label{fhele}
\ee
respectively.

In general, for a given $\cL(F)$, the right-hand sides of Eqs.~\eqref{hele} and \eqref{fhele} can be expressed in terms of $F$ and $r$. By combining this with the implicit relation between $F$ and $r$ given by Eq.~\eqref{Fele}, the metric functions $h$ and $f$ can then be determined numerically or perturbatively. Analytic solutions may be possible, particularly in the case $\beta = 0$, where an explicit expression for $F(r)$ can be obtained from Eq.~\eqref{Fele} for certain forms of $\cL(F)$.

\subsection{Maxwell-HVT theory}
\label{MHVTback}

We begin with singular BH solutions in Maxwell-HVT 
theory with the Einstein-Hilbert term, where the 
vector sector is described by the standard Maxwell Lagrangian
\be
{\cal L}=F\,,
\ee
with $\beta \neq 0$. The solutions outside the outer BH horizon 
have already been investigated in Ref.~\cite{Horndeski:1978ca,Mueller-Hoissen:1988cpx,Balakin:2007am,Verbin:2020fzk,Chen:2024hkm}. 
Here, we extend the analysis to include the spacelike regions. 
The background equations for $h$ 
and $f$ obey Eqs.~(\ref{hele}) and (\ref{fhele}), 
with ${\cal L}_{,F}=1$. 
Using Eq.~(\ref{Fele}), 
Eqs.~(\ref{hele}) and (\ref{fhele}) 
can be rewritten, respectively, as
\ba
& &
\Mpl^2\left[r(1-h)\right]'
=\frac{q_E^2}{2[r^2-8\beta(h-1)]}\,,
\label{hele2}\\
& &
\Delta \equiv \frac{f'}{f}-\frac{h'}{h}
=\frac{8 \beta q_E^2}{\Mpl^2 r 
[r^2-8\beta (h-1)]^2}\,.
\label{fhele2}
\ea
To derive the solution to $h(r)$ near $r=0$, we expand it as
\be
h(r)=\sum_{i=-\infty}^{\infty} c_i r^i\,,
\label{hexpanr=0}
\ee
where $c_i$ are constants and $i$ is an integer. 
Substituting Eq.~(\ref{hexpanr=0}) into 
Eq.~(\ref{hele2}) and expanding around $r=0$, 
we can determine the coefficients order by order. 
The resulting solution around the BH center 
is given by
\be
h(r)=-\frac{2m}{r}+1 
-\frac{q_E^2}{64 \beta m \Mpl^2}r
+{\cal O}(r^3)\,,
\label{hr=0}
\ee
where we have set $c_{-1}=-2m$. 
A special case with $m=0$ is possible, 
but in the following 
discussion we focus on $m \neq 0$.
Substituting the solution (\ref{hr=0}) 
into Eq.~(\ref{fhele2}), we find that the 
leading-order term near $r=0$ on the right-hand side of 
Eq.~\eqref{fhele2} is proportional to $r$.
Hence, the difference between $f'/f$ and $h'/h$ 
vanishes as $r \to 0$.  
The solution for $f(r)$, expanded around $r=0$, is given by
\be
f(r)=f_0 \left[ -\frac{2m}{r}+1 
-\frac{3q_E^2}{64 \beta m \Mpl^2}r
+{\cal O}(r^2) 
\right]\,,
\label{fr=0}
\ee
where $f_0$ is an integration constant. 
From Eqs.~(\ref{back4}) and (\ref{Fele}), we obtain 
\be
A_0'(r) \propto r\,,\qquad 
F(r) \propto r^2\,,
\label{A0Ma}
\ee
near $r=0$.  
Thus, the electric field is regularized by the HVT coupling 
in the vicinity of the BH center.

The solutions expanded at spatial infinity, 
which satisfy the boundary condition 
$f(r \to \infty)=1$, are 
given by \cite{Horndeski:1978ca,Chen:2024hkm} 
\ba
h(r) &=& 1-\frac{2M}{r}+\frac{q_E^2}
{2\Mpl^2 r^2}-\frac{2\beta M q_E^2}{\Mpl^2 r^5}+\frac{2\beta q_E^4}
{5\Mpl^4 r^6}+{\cal O}(r^{-8})\,,\label{hlex}\\
f(r) &=& 1-\frac{2M}{r}+\frac{q_E^2}
{2\Mpl^2 r^2}-\frac{2\beta q_E^2}
{\Mpl^2 r^4}+\frac{2\beta M q_E^2}{\Mpl^2 r^5}
-\frac{3\beta q_E^4}
{5\Mpl^4 r^6}+{\cal O}(r^{-7})\,,\label{flex}
\ea
where $M$ is an integration constant. 
The integration constant $f_0$ in Eq.~\eqref{fr=0} is fixed, 
so that the metric satisfies the asymptotic condition $f(r) \to 1$ as $r \to \infty$.
At spatial infinity, the electric field behaves as 
$A_0'(r) \simeq q_E / r^2$.

Near $r=0$, the metric functions behave as 
$h(r) \propto r^{-1}$ and $f(r) \propto r^{-1}$.  
This indicates that the variations of $h(r)$ and $f(r)$ 
near $r=0$ are milder than those in the RN solution, where $h(r) \propto r^{-2}$ and $f(r) \propto r^{-2}$. The absence of the $q_E^2/(2\Mpl^2 r^2)$ term 
in the metric functions around $r=0$ results from the regularization 
of the electric field induced by the HVT coupling. 
Moving from the distant region toward the BH center, the solutions for $h(r)$ and $f(r)$ change from those given in Eqs.~(\ref{hlex}) and (\ref{flex}) to those 
in Eqs.~(\ref{hr=0}) and (\ref{fr=0}).
This occurs at the distance $r_c$ when the magnitude of the $8 \beta (h-1)$ term in 
Eq.~(\ref{hele2}) becomes of the same order as $r^2$.

To avoid divergence of the right-hand sides of Eqs.~(\ref{hele2}) and (\ref{fhele2}), we require that $r^2 - 8\beta(h-1)$ does not vanish at $r=r_c$. 
This condition can be violated when $\beta m < 0$. In fact, for $\beta m < 0$, 
we have not obtained consistent background solutions numerically.
Therefore, in the following discussion, we restrict our attention 
to the case $\beta m > 0$. To quantify the effect of 
the HVT coupling, we define 
\be
R_{\beta} \equiv \frac{|8 \beta (h-1)|}
{r^2}\,,
\ee
where the transition point is characterized by 
the condition $R_{\beta}(r_c)=1$, i.e., 
\be
r_c^2=|8 \beta \left[ h(r_c)-1 \right]|\,.
\ee
In the regime $r \ll r_c$, the solution for $h(r)$ 
is approximated by $h(r) \simeq - 2m/r$.  
By extrapolating this solution up to 
the transition region, one can estimate
\be
r_c \simeq (16 \beta m)^{1/3}\,. 
\label{rc}
\ee
Since there are two possible cases,
(1) $\beta > 0$ and $m > 0$, or
(2) $\beta < 0$ and $m < 0$, we numerically solve 
the background Eqs.~(\ref{hele2}) and (\ref{fhele2}) 
for each case by integrating outward from an initial point 
$r_i$ deep inside the outer horizon 
to a sufficiently large distance.
Note that the sign of $q_E$ is irrelevant, as its contribution 
to the background equations appears only through 
the combination $q_E^2$. 
Therefore, without loss of generality, we take $q_E > 0$.

\subsubsection{$\beta > 0$ {\rm and} $m > 0$}

\begin{figure}[ht]
\begin{center}
\includegraphics[height=2.8in,width=3.4in]{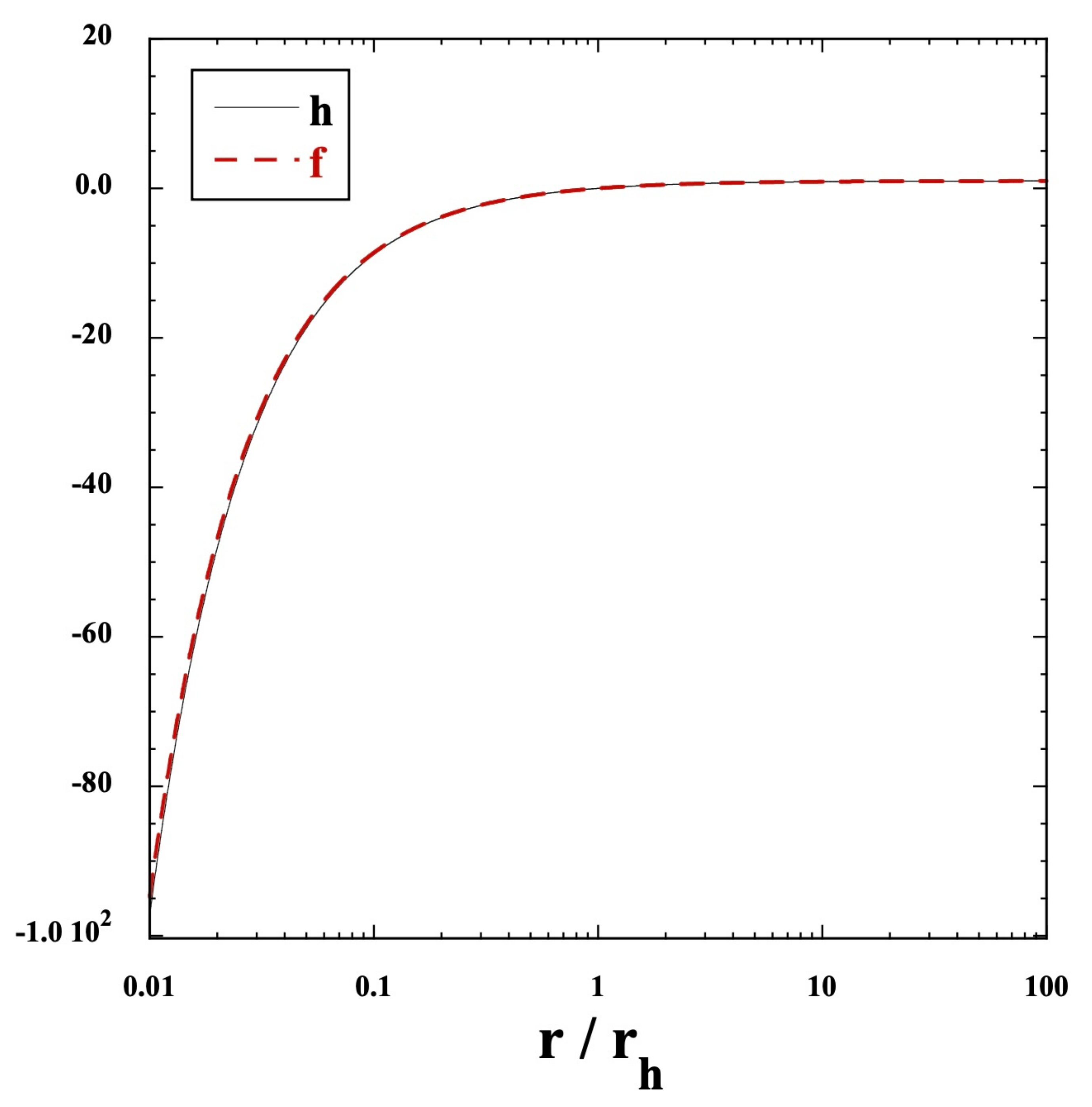}
\includegraphics[height=2.8in,width=3.4in]{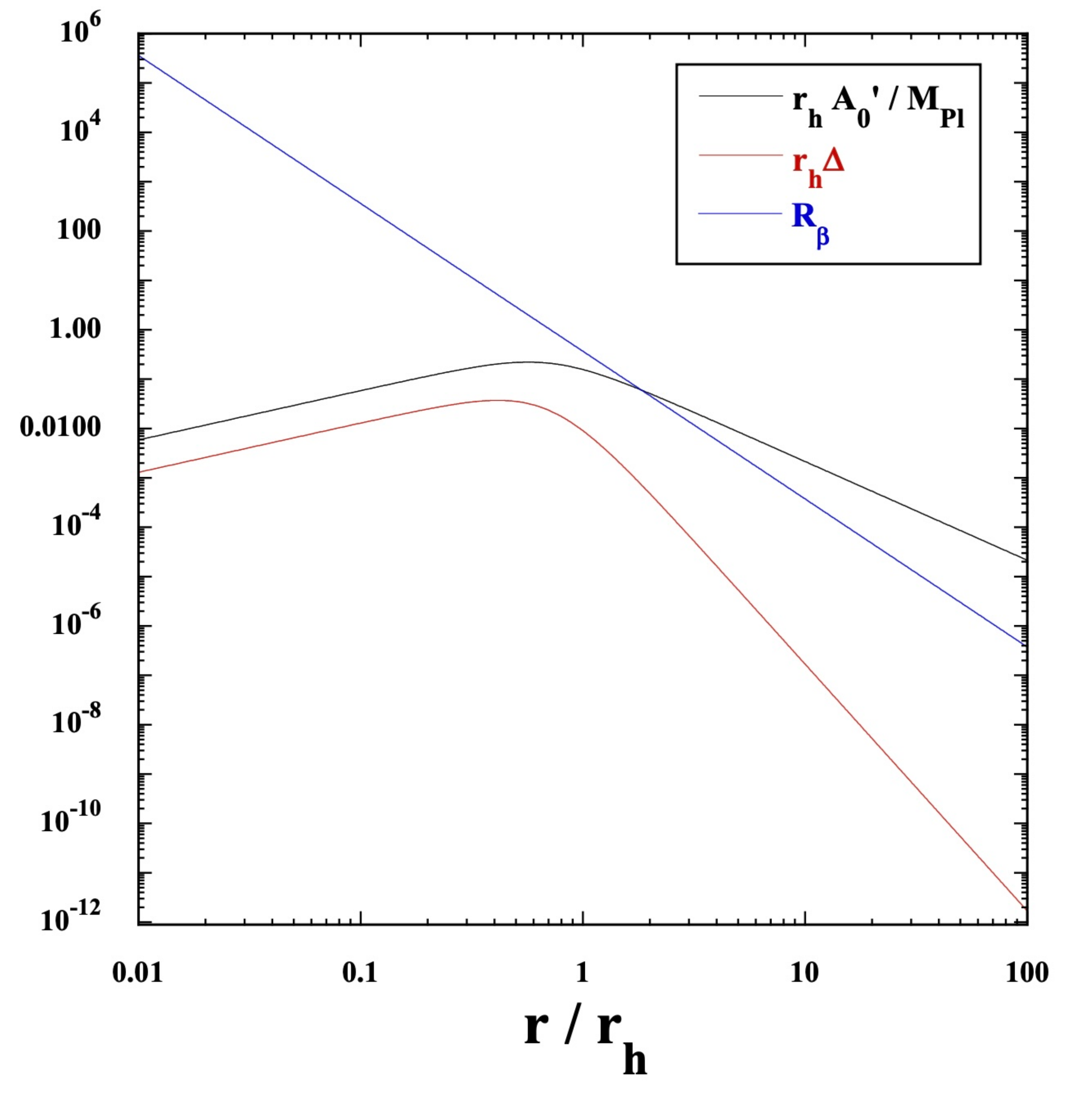}
\end{center}
\caption{\label{fig1} 
(Left panel) Metric components $h$ and $f$ versus $r/r_h$ 
for $\beta = 4.589 \times 10^{-2}r_h^2 $ and $q_E = 0.214 \Mpl r_h$ 
in Maxwell-HVT theory with the 
Einstein-Hilbert term.
The boundary conditions are chosen as $h(r_i) = -99.999$ and $f(r_i) = -97.413$ 
at the distance $r_i = 9.726 \times 10^{-3} r_h$, in which case $m>0$. 
There is a single horizon located at $r = r_h$, where both $h$ and $f$ vanish. 
(Right panel) Plots of $r_h A_0'/\Mpl$, $r_h \Delta$, and $R_{\beta}$ 
versus $r/r_h$ for the same model parameters and boundary conditions 
as in the left panel. The effect of the HVT coupling becomes significant 
when $R_{\beta} > 1$, which corresponds to the region 
$r < 0.71\, r_h$ in the figure.
}
\end{figure}

For $\beta>0$ and $m>0$, the metric function $h$ is largely negative near $r=0$ and increases as a function of $r$. As long as $h$ remains in the region $h<1$ for all distances $r$, the term $r^2 - 8\beta(h-1)$ is always positive, and thus the right-hand sides of Eqs.~(\ref{hele2}) and (\ref{fhele2}) remain finite.
In the left panel of Fig.~\ref{fig1}, we show 
the radial profiles of $h$ and $f$ for 
$\beta = 4.589 \times 10^{-2}r_h^2$ and 
$q_E= 0.214\Mpl r_h$, 
with the boundary conditions specified in the caption. 
We find that both $h$ and $f$ vanish at a horizon located at $r=r_h$. Close to the BH center, $h$ and $f$ are well approximated by Eqs.~(\ref{hr=0}) and (\ref{fr=0}). They smoothly connect to the large-distance solutions (\ref{hlex}) and (\ref{flex}), 
both increasing toward the asymptotic value 1.
For $\beta > 0$ and $m > 0$, there exists only a single horizon, 
in contrast to the RN BH.

In the right panel of Fig.~\ref{fig1}, we plot $r_h A_0'$, $r_h \Delta$, and $R_{\beta}$ as functions of $r/r_h$, using the same model parameters and boundary conditions as in the left panel. The transition distance, defined by the condition $R_{\beta}=1$, is $r_c = 0.71\, r_h$. In the region $r<r_c$, the condition $R_{\beta}>1$ is satisfied, 
so that the dominance of the HVT coupling modifies the background 
solutions to the forms (\ref{hr=0}) and (\ref{fr=0}).
Indeed, the electric field exhibits the behavior $A_0'(r) \propto r$ for 
$r \lesssim r_c$, thereby ensuring regularity in the limit $r \to 0$.
In the left panel of Fig.~\ref{fig1}, 
the difference between $h$ and $f$ is not apparent, but the quantity 
$\Delta = f'/f - h'/h$ is nonvanishing. 
As shown in the right panel, $\Delta$ attains 
its maximum at $r$ close to $r_c$, 
while in the two asymptotic regions 
$r \gg r_c$ and $r \ll r_c$, it 
asymptotically approaches 0.

Since the transition distance is estimated as $r_c \simeq (16 \beta m)^{1/3}$, 
it decreases with smaller values of $\beta$. 
In the numerical simulation presented in Fig.~\ref{fig1}, 
we consider the HVT coupling of order $\beta/r_h^2={\cal O}(10^{-2})$ together with $q_E/(\Mpl r_h)={\cal O}(0.1)$, for which the resulting $r_c$ is comparable to $r_h$.
For smaller values of $\beta$, the transition radius $r_c$ becomes much smaller 
than the horizon radius $r_h$ ($r_c \ll r_h$), and hence the transition 
to the solutions (\ref{hr=0}) and (\ref{fr=0}) takes place deeper inside the horizon.

%%%%%%%%%%%%%%%%%%%%%%%%%%%%%%%%
%%%%%%%%%%%%%%%%%%%%%%%%%%%%%%%%

%
\subsubsection{
$\beta < 0$ {\rm and} $m < 0$}
%

%%%%%%%%%%%%%%%%%%%%%%%%%%%%%%%%
\begin{figure}[ht]
\begin{center}
\includegraphics[height=2.8in,width=3.4in]{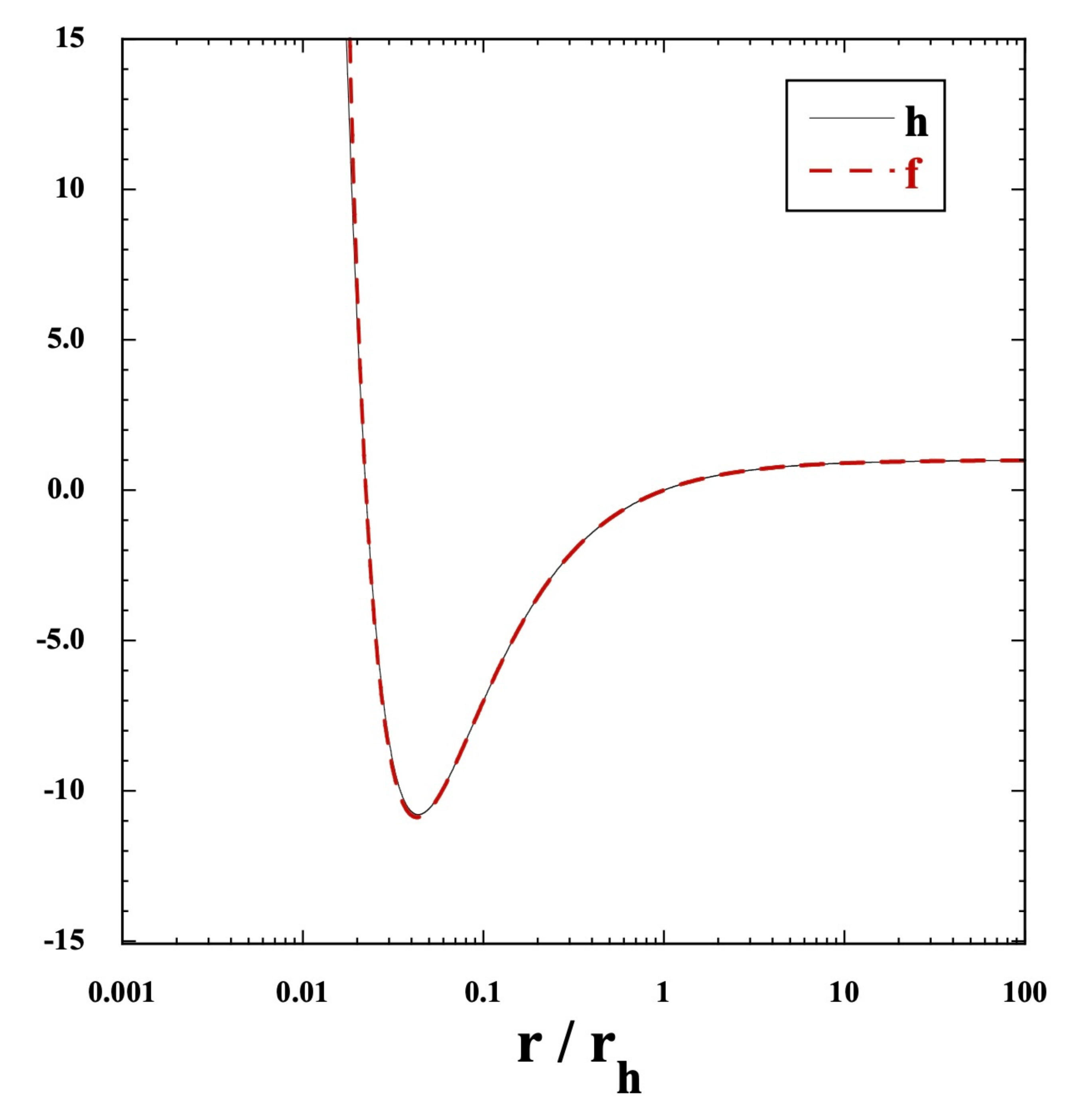}
\includegraphics[height=2.8in,width=3.4in]{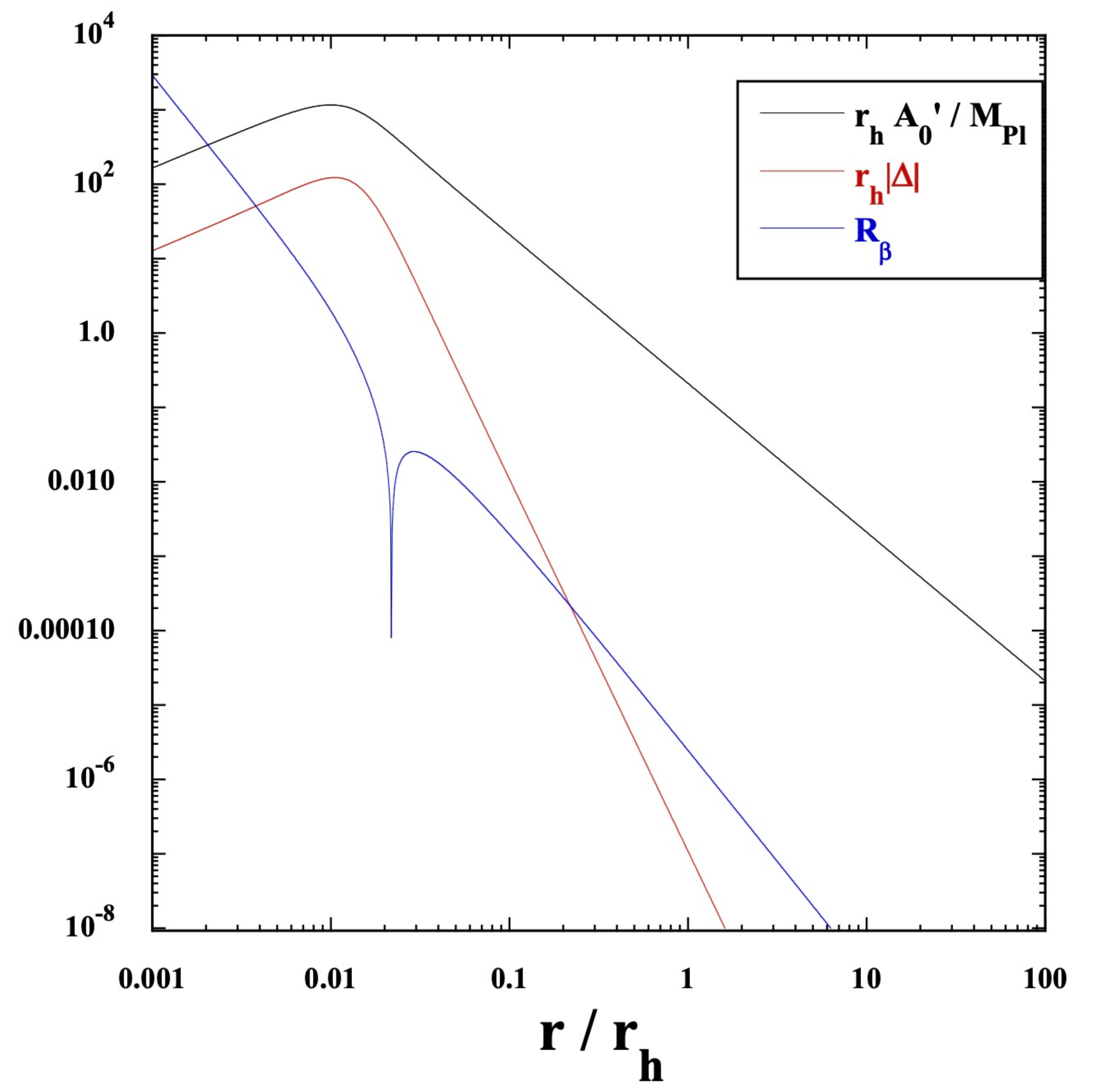}
\end{center}
\caption{\label{fig2}
Plots of the same quantities as in Fig.~\ref{fig1}, 
but with $\beta=-3.062 \times 10^{-7} r_h^2$ and $q_E=0.210 \Mpl r_h$.
The boundary conditions are specified as $h(r_i)=1.500 \times 10^3$ 
and $f(r_i)=7.883 \times 10^3$ at $r_i=7.944 \times 10^{-4}r_h$, 
for which $m<0$.
In this case, two horizons appear, with the outer 
one corresponding to $r_h$.
The quantity $\Delta$ remains negative without crossing 0, 
whereas $R_\beta$ crosses 0 at $h=1$ 
(around $r=2.18 \times 10^{-2} r_h$).
}
\end{figure}
%%%%%%%%%%%%%%%%%%%%%%%%%%%%%%%%

For $\beta<0$ and $m<0$, the metric function $h(r)$ at small distances is positive 
($h(r) \simeq -2m/r$) and decreases as a function of $r$. 
There exist parameter ranges of $\beta$ and $q_E$ in which $h(r)$ decreases monotonically toward the asymptotic value of 1 without crossing 0. This case does not correspond to a BH, so we focus instead on the case where $h(r)$ crosses 0. 
Since $h(r)$ must increase again to approach the asymptotic value of 1, it should possess a minimum, leading to the presence of two horizons. 
In Fig.~\ref{fig2}, we show 
such an example of the BH solution, in which case there 
are two horizons at $r=2.23 \times 10^{-2}r_h$ and $r=r_h$. 
In this case, $R_{\beta}=1$ at $r_c = 1.18 \times 10^{-2} r_h$, so that $h(r) \simeq -2m/r + 1$ for $r < r_c$. When $h(r)$ crosses 1 at $r = 2.18 \times 10^{-2} r_h$, $R_{\beta}$ simultaneously vanishes. 

In the right panel of Fig.~\ref{fig2}, we can confirm 
the behavior $A_0'(r) \propto r$ 
in the region $r \lesssim r_c$.
The quantity $|\Delta|$ attains a maximum near $r = r_c$, where the difference between $f'/f$ and $h'/h$ is largest. In the left panel of Fig.~\ref{fig2}, both $f$ and $h$ cross 0 twice and approach their asymptotic value of 1. 
The absolute value of $\beta$ is smaller than that in Fig.~\ref{fig1}, i.e., $\beta = -{\cal O}(10^{-7}) r_h^2$, allowing the existence of an inner horizon whose radius is much smaller than $r_h$. 
If $|\beta|$ is chosen to be large, the quantity 
$r^2 - 8\beta (h-1)$ can cross 0 
for $\beta < 0$ in the region $h < 1$. 
This leads to divergences in the right-hand sides of 
Eqs.~(\ref{hele2}) and (\ref{fhele2}). 
The plot in Fig.~\ref{fig2} corresponds to the case 
where no such crossing occurs. 
Therefore, the magnitude of negative $\beta$ is constrained 
from above to ensure consistent background solutions, 
which generally gives the bound $|\beta| \ll r_h^2$ 
(see also Fig.~2 in Ref.~\cite{Chen:2024hkm}).

The parameter ranges for 
$\beta$ and $q_E$ that are compatible with linear stability 
in the region $r>r_h$
have been studied in Ref.~\cite{Chen:2024hkm}.
In Sec.~\ref{elesstasec}, we address stability for $r<r_h$ 
to examine further constraints on the HVT coupling $\beta$.

\subsection{Power-law NED theory 
with $\beta=0$}
\label{powersec}

Let us consider the power-law NED theory in 
which the Lagrangian 
${\cal L}$ is given by 
\be
{\cal L}(F)=F+a_p F^p\,,
\label{powerNED}
\ee
where $a_p$ and $p$ are constants. 
We assume that $p$ is an integer in the range $p \geq 2$.
In this section, we focus on the case without the HVT coupling ($\beta=0$). 
The effect of the nonvanishing HVT coupling on the background BH solution will be discussed in Sec.~\ref{NHVTsec}.

Since we are considering the case $\beta=0$, Eq.~(\ref{fhele}) 
admits the integrated solution 
$f = C h$, where $C$ is a constant. 
Imposing the boundary conditions 
$f(r \to \infty) = h(r \to \infty) = 1$ fixes $C=1$, 
and hence $f(r)=h(r)$. 
We now consider the case $p=2$.
From Eq.~(\ref{back4}), we obtain 
\be
A_0'=\frac{q_E}{r^2(a_2 A_0'^2+1)}\,.
\ee
This is an algebraic equation that allows us 
to express $A_0'$ as a function of $r$. 
The metric component $h$ obeys 
\be
h'=\frac{1-h}{r}-\frac{A_0'^2}
{2\Mpl^2}r-\frac{3a_2 A_0'^4}
{4\Mpl^2}r\,.
\ee
In the following, we assume $q_E>0$ 
without loss of generality. 
We will also consider the case $a_2>0$, 
which is required for the linear stability 
of BHs (see Sec.~\ref{powerstasec}).
The solutions expanded around $r=0$ 
are given by 
\ba
A_0'(r) &=& 
\left( \frac{q_E}{a_2} \right)^{1/3}
\frac{1}{r^{2/3}}-\frac{r^{2/3}}
{3 (q_E a_2^2)^{1/3}}+{\cal O}(r^{10/3})\,,\\
h(r) &=& f(r)=-\frac{2m}{r}
-\frac{9}{4\Mpl^2} 
\left( \frac{q_E^4}{a_2} \right)^{1/3}
\frac{1}{r^{2/3}}
+1+{\cal O}(r^{2/3})\,,
\label{hfr=0}
\ea
where $m$ is an integration constant. 
While Einstein-Maxwell theory yields
$A_0'(r) \propto r^{-2}$, 
the presence of the NED term $a_2 F^2$ 
alters the radial dependence to $A_0'(r) \propto r^{-2/3}$ 
near $r=0$. This also modifies the behavior of the metric functions, such that $h(r)=f(r)\propto r^{-1}$ near $r=0$, 
in contrast to $h(r)=f(r)\propto r^{-2}$ for the RN BH.

At spatial infinity, the solutions can be expanded as
\ba
A_0'(r) &=& \frac{q_E}{r^2}
-\frac{a_2 q_E^3}{r^6}+{\cal O}(r^{-10})\,,\\
h(r) &=& f(r)= 1-\frac{2M}{r}+\frac{q_E^2}{2\Mpl^2 r^2}
-\frac{a_2 q_E^4}{20 \Mpl^2 r^6}
+{\cal O}(r^{-10})\,.
\label{fsoELEA2} 
\ea
In this regime, the coupling $a_2$ acts 
as corrections to the electric field and metric
functions of the RN solution.

%%%%%%%%%%%%%%%%%%%%%%%%%%%%%%%%
\begin{figure}[ht]
\begin{center}
\includegraphics[height=3.0in,width=3.4in]{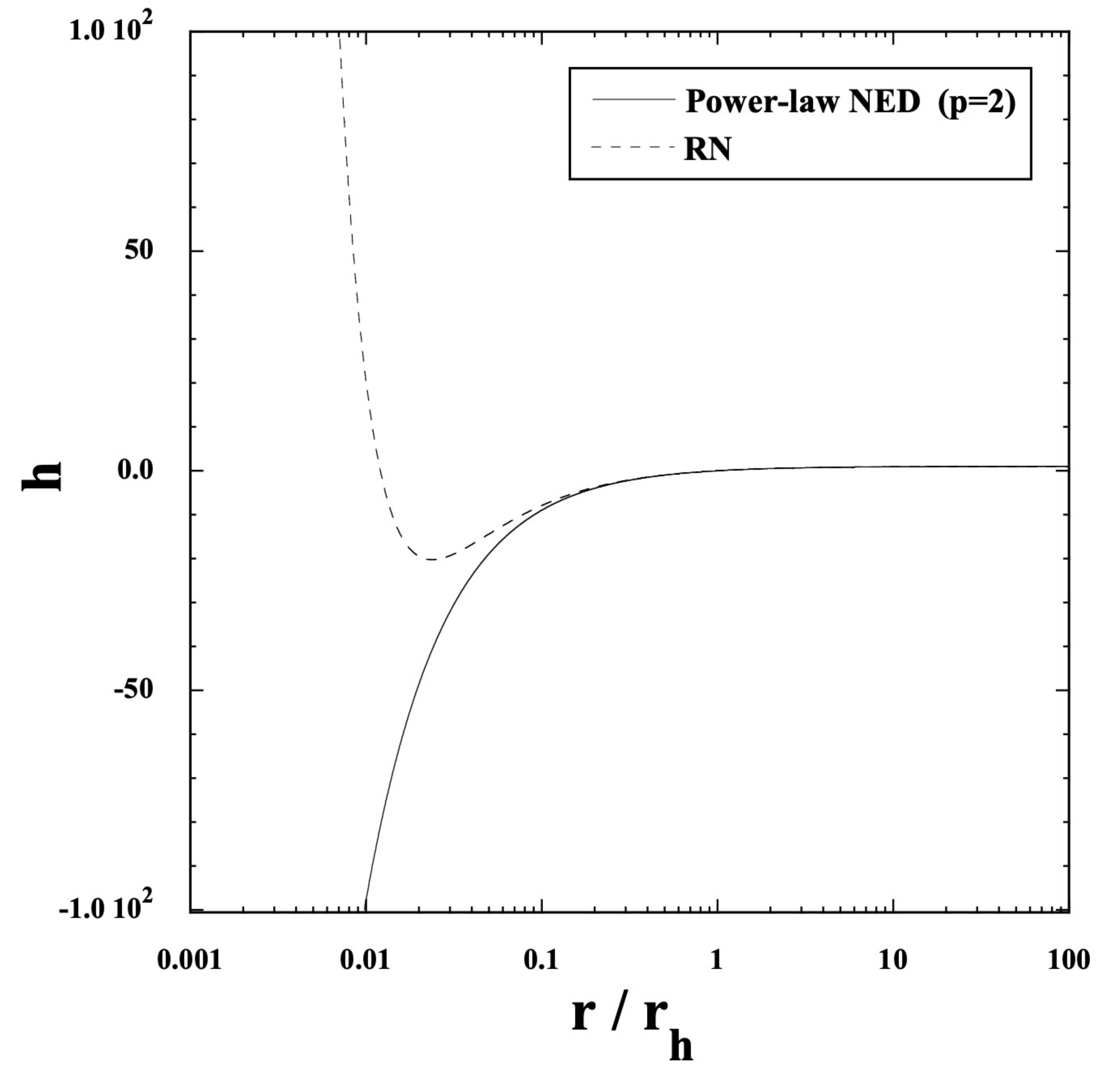}
\includegraphics[height=3.0in,width=3.4in]{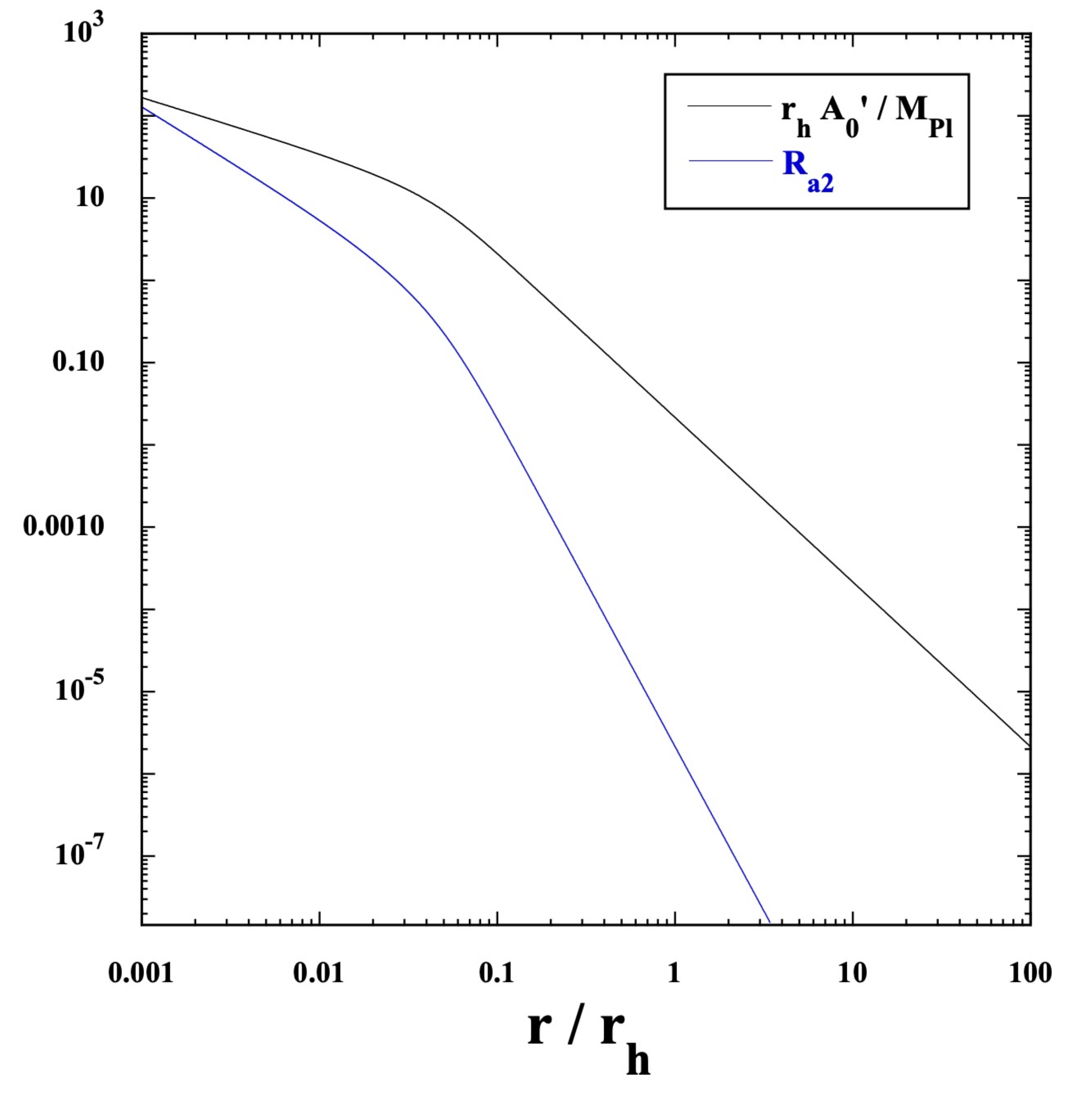}
\end{center}
\caption{\label{fig3} 
(Left panel) Metric component 
$h~(=f)$ as a function of $r/r_h$ 
for the power-law NED with 
$a_2 = 4.63 \times 10^{-3} r_h^2/\Mpl^2$ 
and $q_E = 2.15 \times 10^{-2} \Mpl r_h$, together 
with the boundary condition 
$h(r_i) = -1.0 \times 10^3$ 
at $r_i = 9.7 \times 10^{-4} r_h$. 
For comparison, we also show $h(r)$ for the RN BH 
($a_2=0$ and $q_E=0.155\,\Mpl r_h$) 
with the boundary condition 
$h(r_i) = 1.0 \times 10^2$ 
at $r_i = 7.0 \times 10^{-3} r_h$.  
(Right panel) Radial dependence of 
$r_h A_0'/\Mpl$ and $R_{a_2}$ 
for $a_2 = 4.63 \times 10^{-3} r_h^2/\Mpl^2$ 
and $q_E = 2.15 \times 10^{-2} \Mpl r_h$, 
with the same boundary condition as in the left panel.
}
\end{figure}
%%%%%%%%%%%%%%%%%%%%%%%%%%%%%%%%

To quantify the effect of the nonlinear 
Lagrangian $a_2 F^2$, we define 
\be
R_{a_2}=a_2 A_0'^2\,.
\ee
In the regime where $|R_{a_2}| > 1$, the contribution 
from $a_2 F^2$ dominates over the Maxwell term $F$. 

In Fig.~\ref{fig3}, we present a numerical example of $h~(=f)$ (left) and of $A_0'$ and $R_{a_2}$ (right) as functions of $r/r_h$, 
with $r_h$ denoting the radius of the single horizon.
This corresponds to the case with $a_2>0$ and $m>0$ in the expansion of Eq.~(\ref{hfr=0}).
As seen from the solid line in the left panel of Fig.~\ref{fig3}, the metric function undergoes a transition from the solution $h(r) \simeq 1 - 2M/r$ at large distances to $h(r) \simeq -2m/r$ at small distances. Unlike the RN BH, where $h(r)$ rises again toward $h(r \to 0) = +\infty$, the metric function for $a_2 \neq 0$ decreases toward $h(r \to 0) = -\infty$.
In the right panel of Fig.~\ref{fig3}, we see that $R_{a_2}$ exceeds 1 for $r < r_{a_2} = 2.7 \times 10^{-2} r_h$. In this regime, the radial dependence of the electric field changes from $A_0'(r) \propto r^{-2}$ to $A_0'(r) \propto r^{-2/3}$. This nontrivial behavior of $A_0'(r)$ for $r < r_{a_2}$, in turn, modifies the metric functions compared to those of the RN BH.

The above result corresponds to the case $p=2$, but 
one can also consider a general integer power $p$. 
At small distances, the leading behavior of the electric field 
scales as $A_0'(r) \propto r^{-2/(2p-1)}$. The metric 
functions near $r=0$ have the dependence $h(r)=f(r)
=-2m/r+c_p r^{-2/(2p-1)}+\cdots$, 
where $c_p$ is a constant.
Therefore, for $p \geq 2$, 
the leading-order contributions 
to $h(r)$ and $f(r)$ are $-2m/r$.
Defining 
$R_{a_p}=2^{1-p}p a_p A_0'^{2(p-1)}$, 
the regime in which the small-distance solutions are realized 
is characterized by the condition $|R_{a_p}| \gg 1$.  
In contrast, in the regime $|R_{a_p}| \ll 1$, the leading-order contributions of the metric functions reduce to the RN ones, with $A_0'(r) = q_E/r^2$, up to small corrections induced by the coupling $a_p$.  
The small- and large-distance solutions are smoothly connected 
around $|R_{a_p}| = 1$, as illustrated in Fig.~\ref{fig3}.
For increasing $p$, the variation of $A_0'(r)$ in the 
small-distance regime becomes less significant, 
approaching $A_0'(r) \propto \text{constant}$ 
in the limit $p \to \infty$.

\subsection{Born-Infeld theory 
with $\beta=0$}

We proceed to Born-Infeld theory without the HVT coupling. 
The NED sector is given by the Lagrangian 
\be
\cL(F)=\frac{2}{b}\left(1-\sqrt{1-b F}\right)\,,\label{modelBI}
\ee
where $b$ denotes a constant parameter. 
We will focus on the case $b>0$, in which 
$F~(>0)$ is bounded from above.

The electric field is simply given by 
\be
A_0'(r)=\frac{q_E}{\sqrt{r^4+r_b^4}}\,,
\label{A0BI}
\ee
where $r_b^4 \equiv b q_E^2/2$, and we have assumed 
$q_E>0$ and $A_0'>0$ without loss of generality.
In the limit 
$r \to 0$, the electric field 
approaches a constant value $q_E/r_b^2$.
Since we are now 
considering the case $\beta=0$, we have 
$f'/f=h'/h$ and hence $f(r)=h(r)$ upon imposing the boundary conditions 
$f(\infty) = h(\infty) = 1$. 
Introducing the mass function $\mu(r)$ 
in the form 
\be
h(r)=1-\frac{2\mu (r)}{r}\,,
\ee
it satisfies the differential equation 
\be
\mu'(r)=\frac{1}{b \Mpl^2} 
\left( \sqrt{r^4+r_b^4}-r^2 \right)\,.
\label{mud}
\ee
For small distances $r \ll r_b$, 
integrating Eq.~(\ref{mud}) yields
\be
\mu(r)=m+\frac{q_E}{\Mpl^2 \sqrt{2b}}r
-\frac{r^3}{3\Mpl^2 b}+{\cal O}(r^5)\,,
\ee
where $m$ is an integration constant. 
This translates to the metric functions 
\be
h(r)=f(r)=-\frac{2m}{r}+1
-\frac{q_E}{\Mpl^2} \sqrt{\frac{2}{b}}
+\frac{2r^2}{3b \Mpl^2}+{\cal O}(r^4)\,.
\label{BIS}
\ee
Unlike the RN BH, the leading-order terms 
of $h(r)$ and $f(r)$ are modified to 
$-2m/r$, as a result of the regularization 
of $A_0'(r)$ near $r=0$. 

At large-distances $r \gg r_b$, Eq.~(\ref{mud}) 
is integrated to give
\be
\mu(r)=M-\frac{q_E^2}{4\Mpl^2 r}
+\frac{b q_E^4}{160 \Mpl^2 r^5}
+{\cal O}(r^{-9})\,,
\ee
where $M$ is another integration constant.
This leads to 
\be
h(r)=f(r)=1-\frac{2M}{r}
+\frac{q_E^2}{2\Mpl^2 r^2}
-\frac{bq_E^4}{80 \Mpl^2 r^6}
+{\cal O} (r^{-10})\,,
\label{BIL}
\ee
which shows that the coupling $b$ works as 
a correction to the RN metric. The solutions 
(\ref{BIS}) and (\ref{BIL}) connect smoothly around $r = r_b$. 
The mass function $\mu(r)$ increases from its value $m$ at $r=0$ to the ADM mass $M$ at spatial infinity.

\subsection{Power-law or 
Born-Infeld NED theories with 
$\beta \neq 0$}
\label{NHVTsec}

We now turn to NED theories with the HVT coupling, examining in turn the power-law and 
Born-Infeld cases.

We first consider the power-law NED with ${\cal L}(F)=F+a_p F^p$, without imposing any restriction on the range of $p$
(in contrast to Sec.~\ref{powersec}, 
where $p$ was restricted to $p \geq 2$).
For $p<1$, the theory does not admit a proper Maxwell limit, ${\cal L}(F) \to F$ as $F \to 0$, which can lead to significant 
deviations from the RN solution at large distances. 
Here, we are primarily interested in the modifications 
induced by the nonlinear term $a_p F^p$ near $r=0$ 
(as in the case of nonsingular BHs in NED). 
Therefore, we do not exclude the possibility 
of $p<1$.

From Eqs.~\eqref{Fre} and \eqref{back4}, 
the equation of motion for $A_0'$ 
is given by 
\be
2^{1-p} a_p p \left( \frac{f}{h} \right)^{1-p}
r^2 A_0'^{2p-1}-q_E \sqrt{\frac{f}{h}}
+A_0' \left[ r^2 +8 \beta (1-h) \right]=0\,.
\label{Apeq}
\ee
For $a_p = 0$, we have already shown that the leading-order solutions for $h(r)$ and $f(r)$ near $r = 0$ scale as $r^{-1}$.
The same property holds for $\beta=0$ with $p \geq 2$.
Since the metric components are expected to exhibit a similar behavior for $a_p \neq 0$ and $\beta \neq 0$, we look for 
leading-order solutions of the form
\be
h(r)=\frac{h_1}{r}\,,\qquad 
f(r)=\frac{f_1}{r}\,,
\label{hfana}
\ee
in the vicinity of $r=0$. Substituting 
Eq.~(\ref{hfana}) into Eq.~(\ref{Apeq}), 
we find that the solution 
to $A_0'(r)$ depends on the power $p$. 
When $p > -1/2$, the last two terms 
in Eq.~(\ref{Apeq}) dominate, and the short-distance 
solution takes the form  
\be
A_0'(r)=-\frac{q_E}{8 \beta h_1} 
\sqrt{\frac{f_1}{h_1}}\,r\,,
\label{A0mi0}
\ee
which is analogous to the behavior in Eq.~(\ref{A0Ma}). 
This corresponds to the regime where the HVT coupling 
dominates over the power-law NED coupling $a_p F^p$.  

If $p=-1/2$, all three terms in Eq.~\eqref{Apeq} contribute. 
In this case, the electric field near the origin $r=0$ 
has the following behavior:
\be
A_0'(r)=C_{-1/2}r\,,
\label{A0112special}
\ee
where the coefficient $C_{-1/2}$ can be obtained by 
solving the leading-order contribution of Eq.~\eqref{Apeq}.

When $p < -1/2$, the first and third terms in Eq.~(\ref{Apeq}) dominate, and hence the short-distance solution is given by 
\be
A_0'(r)=\left(\frac{2^{1-p} a_p p}{8 \beta h_1} 
\right)^{1/[2(1-p)]}\sqrt{\frac{f_1}{h_1}} \, r^{3/[2(1-p)]}\,.
\label{A0mi}
\ee
The exponent of the $r$-dependent term in Eq.~(\ref{A0mi}) 
lies in the range $0 < 3/[2(1-p)] < 1$.  
Thus, the electric field remains regular in the limit $r \to 0$.  
For $p \geq -1/2$, we have $A_0'(r) \propto r$, so that the electric field is more strongly regularized 
by the HVT coupling than in the case $p < -1/2$.
Substituting Eq.~\eqref{A0mi} into the gravitational field 
Eqs.~(\ref{hele}) and (\ref{fhele}), the leading-order metric functions of the form (\ref{hfana}) are found to be consistent with them. The same property also 
holds for the electric-field solutions given 
in Eqs.~(\ref{A0mi0}) and \eqref{A0112special}.

The solutions derived above are valid in 
the small-distance regime where the HVT coupling 
and/or the NED power-law coupling contribute 
to the electric-field profile.
As we will see in Sec.~\ref{NEDHVTsec}, these solutions suffer from Laplacian instabilities of even-parity perturbations.
Therefore, we will not study how they are connected to the solutions in the large-distance regime.

In Born-Infeld theory, characterized by 
the Lagrangian ${\cal L}(F) =2/b \left(1 
- \sqrt{1 - b F}\right)$, we discuss the 
short-distance solutions in the presence 
of the HVT coupling.
In this case, the electric field obeys 
\be
q_E \sqrt{\frac{f}{h}}
+8 \beta (h-1) A_0'
=\frac{\sqrt{2}r^2 A_0'}
{\sqrt{2-bh A_0'^2/f}}\,.
\label{A0BIs}
\ee
Under the assumption that the metric functions are given by 
the forms (\ref{hfana}), the leading-order term of $A_0'$ is determined by setting the right-hand side of 
Eq.~(\ref{A0BIs}) to 0, yielding  
\be
A_0'(r)=-\frac{q_E}{8 \beta h_1} 
\sqrt{\frac{f_1}{h_1}}\,r\,,
\label{A0miD}
\ee
which is of the same form as Eq.~(\ref{A0mi0}) and 
Eq.~\eqref{A0Ma}. The Born-Infeld coupling $b$ does not 
affect the leading-order solution for $A_0'(r)$.
One can also verify that, with the electric field given by Eq.~(\ref{A0miD}), the leading-order metric components $h(r)=h_1/r$ and $f(r)=f_1/r$ are consistent with Eqs.~(\ref{hele}) and (\ref{fhele}).
For the same reason as in the case of the power-law NED, we will not discuss the behavior of the 
large-distance solutions.

We note that the solutions found in Eqs.~\eqref{A0mi0} and \eqref{A0miD}, corresponding to the power-law NED with $p>-1/2$ 
and the Born-Infeld NED, respectively, exhibit the same behavior 
as those in the Einstein-Maxwell-HVT theory, 
namely, $A_0'(r) \propto r$ near $r=0$. 
In fact, for any NED theory with a proper Maxwell limit, 
i.e., $\mathcal{L}(F) \rightarrow F$ as $F \rightarrow 0$, 
one can see from Eq.~\eqref{Fele} that there always exists 
a branch of solutions in which $h(r) \propto 1/r$, 
$f(r) \propto 1/r$, and $A_0'(r) \propto r$ around $r=0$.
This highlights the significant role of the HVT coupling $\beta$ 
in dominating over the NED contributions near the BH center.
It also implies that if the solutions in Maxwell-HVT 
theory exhibit instabilities near $r=0$, as we will 
demonstrate in Sec.~\ref{subsec:maxwellhvtinstablilty}, 
then solutions in general NED theories with $\beta \ne 0$, which behave similarly to those in  Maxwell-HVT theory near $r=0$, would likely suffer from the same 
instability, provided the NED theory has a proper Maxwell limit (see Sec.~\ref{NEDHVTsec}).

However, nontrivial solutions can exist if the assumption 
of a proper Maxwell limit for the NED theory is relaxed. 
An example is provided by the solutions in Eq.~\eqref{A0mi}, 
which correspond to the power-law NED with $p < -1/2$. 
Another possibility for obtaining nontrivial solutions 
is to consider branches in which $F$ does not approach 
0 at the origin, as we will demonstrate 
in the next subsection.

\subsection{Reconstructed NED theories with $\beta \neq 0$ }\label{subsec:recon_h}

The previous methods focused on solutions derived from 
physically motivated NED Lagrangians, ${\cal L}(F)$.
Alternatively, one can adopt a different approach by 
reconstructing the form of ${\cal L}(F)$ from a given metric function $h(r)$.
The idea is that certain forms of ${\cal L}(F)$ may have implications for the stability of BH solutions.
Let us consider the case in which one of the metric 
functions, $h(r)$, takes the form 
\begin{equation}
h(r)=1-\frac{2M}r+\frac{b_1}{r^2}\,,
\end{equation}
for arbitrary $r$, where $M$ and $b_1$ are constants.\footnote{
We reconstruct the theory corresponding to a given function $h(r)$. Although this theory may in principle admit infinitely many other solutions, they are generally inaccessible in analytic form. 
Furthermore, quantities such as $M$, which physically represent the ADM mass, cannot be derived from first principles within the theory, 
since their origin depends on the full set of solutions, which remains unknown. However, this method is interesting because it can generate 
solutions without Laplacian instabilities, as we will see 
later in Sec.~\ref{perresec}.} 
This form has been chosen because the term $b_1/r^2$ 
dominates over $-2M/r$ as $r \to 0$. 
If there is a NED Lagrangian compatible with such a behavior, it would have a different phenomenology from the other theories 
discussed above, i.e., $h(r) \propto 1/r$ for $r \to 0$.

We define a function $\tilde{f}$ as $\tilde{f} \equiv f/h$, 
with $\lim_{r \to \infty} \tilde{f}=1$.
We can solve Eqs.~\eqref{fhele} and \eqref{hele} for $F$ and $\mathcal{L}_{,F}$ to obtain
\begin{align}
F &= \frac{\Mpl^{2}{\tilde{f}'} r}{16 \beta  \tilde{f}}\,,\label{eq:F_recon}
\\  
{\mathcal L}_{,F} &= 
-\frac{4 \beta  
\bigl[r \Mpl^{2} \left( 2M r 
-b_{1} \right) {\tilde{f}'}-2 \left(r^{4} {\mathcal L}+b_{1} \Mpl^{2}\right)\tilde{f} \bigr]}{\Mpl^{2}{\tilde{f}'} r^{5}}\,,\label{eq:L_F_recon}
\end{align}
where we have assumed 
$\tilde{f} '\neq 0$. 
Substituting these expressions into Eq.~\eqref{Fele} 
yields an equation that determines $\mathcal{L}$, as follows:
\begin{equation}
\mathcal{L}=
\frac{\bigl(-4 \Mpl \tilde{f}' \beta  M \,r^{2}+2 \Mpl \tilde{f}' b_{1} \beta  r \pm\sqrt{2}\, q_{E} r^{2} 
\sqrt{\tilde{f}' \tilde{f} \beta  r}
-4 \Mpl \tilde{f} b_{1} \beta \bigr) 
\Mpl}{4 \beta \tilde{f} r^{4}}\,,
\label{eq:L_branch}
\end{equation}
where we have 
further assumed that 
$\tilde{f}'\tilde{f} \beta>0$. 
The plus/minus sign generally corresponds 
to two branches of solutions/theories. 
At this point, employing the previous 
three relations together with the integrability condition $\mathcal{L}'(r)=
\mathcal{L}_{,F}F'(r)$, 
we obtain a first-order differential equation 
for $\tilde{f}$, which can be written as
\begin{equation}
2\left[r \left( 3 M r -2 b_{1} \right) \tilde{f}'
+4 b_{1}\tilde{f} \right] \Mpl 
\sqrt{ \tilde{f}' \tilde{f} \beta r}
\mp \sqrt{2} \tilde{f}' \tilde{f} q_{E} r^{3}=0\,.
\end{equation}
Then, the profile for $\tilde{f}$ 
is determined as follows: 
\begin{equation}
\tilde{f}(r)=c_{1} \exp\! \left[ \int^r 
\frac{ q_{E}^{2} r^{4}
+16 \Mpl^{2} b_{1} \beta (2b_1-3Mr)
\mp q_{E} r^{2} 
\sqrt{q_{E}^{2} r^{4}
+32 \Mpl^{2} b_{1} \beta (2b_1-3Mr)}}
{4 \Mpl^2 \left( 3 M r-2 b_{1}\right)^{2} 
\beta r}
\,{\rm d}r \right],
\label{eq:f2_branch}
\end{equation}
where the constant $c_1$ needs to be determined by the condition $\lim_{r\to\infty} \tilde{f}(r)=1$. 
The minus/plus signs in Eq.~(\ref{eq:f2_branch}) 
give rise to two further branches, with the same minus/plus choices 
also serving as solutions for the two branches of $\mathcal{L}$. 
Thus, in total, four branches may arise: two associated 
with the choice of $\tilde{f}$ and two with the choice of $\mathcal{L}$. 
At lowest order, around the origin, we find that 
$\tilde{f}' \tilde{f} \beta >0$ leads to $2c_1^2\beta r^3>0$, 
and hence $\beta>0$ (for all branches). Since 
$\tilde{f}=c_1r^2+\mathcal{O}(r^3)$ near $r=0$, 
the metric function $f$ has the following dependence:
\begin{equation}
f(r)=b_{1} c_{1} +M c_{1}  r 
+\frac{c_{1} [ \mp
\sqrt{\beta}\, |b_{1}| q_{E}+\Mpl \beta  
( 3M^{2}+4 b_{1} )]}
{4\Mpl b_{1} 
\beta} r^{2}+\mathcal{O}\! \left(r^{3}\right).
\end{equation}
In what follows, we denote the cases with the minus and plus signs in Eq.~\eqref{eq:f2_branch} as branch A and branch B, respectively. 
For branch A, we also refer to the cases with the plus and minus signs in Eq.~(\ref{eq:L_branch}) as branch A1 and branch A2, respectively.
For branch A, the expansions of $F$ and 
${\cal L}$ near $r=0$ are given by
\begin{align}
F&=\frac{\Mpl^{2}}{8 \beta}
+ \frac{3 \Mpl^{2} M}{16 b_{1} \beta} r 
+\frac{9 \Mpl^{2} \beta  \,M^{2}-q_{E} \Mpl |b_{1}| 
\sqrt{\beta}}{32 b_{1}^{2} \beta^{2}} r^{2}+\mathcal{O}(r^{3})\,,\\
\mathcal{L}&=-\frac{\Mpl^{2} M}{2} r^{-3}
-\frac{\Mpl [3 \beta M^2 \Mpl-(2b_1\epsilon-|b_1|)q_E 
\sqrt{\beta}]}{4 b_1 \beta}+{\cal O}(r^{-1})\,,
\end{align}
where $\epsilon=+1$ $(\epsilon=-1)$ represents the branch A1 (A2).
This implies that $F$ approaches the finite value $\Mpl^2/(8\beta)$ 
as $r \to 0$. In the vicinity of the BH center, 
the NED Lagrangian behaves at leading order as
${\cal L}(F) \propto [F-\Mpl^2/(8\beta)]^{-3}$. 
Close to $r=0$, the electric field behaves 
as $A_0'(r) \propto r$, and is thus regular.
Using the properties 
$\mathcal{L}'(r)=
\mathcal{L}_{,F}F'(r)$ and 
$\mathcal{L}_{,F}'(r)
=\mathcal{L}_{,FF} F'(r)$,
we can further estimate $\mathcal{L}_{,F}$ and 
$\mathcal{L}_{,FF}$ near $r=0$ for branch A as
\begin{align}
\mathcal{L}_{,F}&=
\frac{8 \beta  b_{1}}{r^{4}}-\frac{16 \beta  M}{r^{3}}
+\epsilon\frac{2 \sqrt{\beta}\, q_{E}}{\Mpl r^{2}}
-\epsilon\frac{3 \sqrt{\beta}\, q_{E} M}{2 \Mpl b_{1} r}+\mathcal{O}(r^{0})\,,\\
\mathcal{L}_{,FF}&=-\frac{512 \beta^{2} b_{1}^{2}}{3 \Mpl^{2} M \,r^{5}}
+\frac{768 b_1 \beta}{\Mpl^{3} M^{2} r^{4}}
\left(\beta\,M^{2} \Mpl -\frac{2 \sqrt{\beta}\, q_{E} {| b_{1}|}}{27}\right)+\mathcal{O}(r^{-3})\,.
\end{align}
Any other relations can be obtained by taking derivatives with respect to $r$ of the previous quantities, such as $\mathcal{L}_{,FF}'(r)=
\mathcal{L}_{,FFF}\,F'(r)$ and $A_0'^2=2f_2\,F$. We will use them to compute quantities relevant to the linear stability of BHs in Sec.~\ref{perresec}. 
At lowest order in $r$, we have $A_0'^2=\Mpl^2 c_1 r^2/(4\beta)>0$ for all branches, so that  $c_1>0$ for consistency. Finally, although we have reported the results for branch A, at leading order, the results are 
the same for all the branches. 
Thus, it is possible to reconstruct the NED Lagrangian ${\cal L}(F)$ 
in such a way that $h(r)$ behaves as $h(r) \simeq b_1/r^2$ near $r=0$, 
with $f(r) \simeq {\rm constant}$ and $A_0'(r) \propto r$. 
Unlike the RN solution, $f(r)$ and $A_0'(r)$ remain finite at the BH center. However, this solution does not belong to the class of nonsingular BHs, since $h(r)$ diverges at $r=0$.

%%%%%%%%%%%%%%%%%%%%%%%%%%%%%%%%%%%%%%
\section{Black hole perturbations}
\label{BHpersec}
%%%%%%%%%%%%%%%%%%%%%%%%%%%%%%%%%%%%%%

The linear stability of electric BHs can be analyzed by considering metric perturbations $h_{\mu \nu}$ on the SSS background defined by the line element (\ref{metric_bg}) \cite{Regge:1957td,Zerilli:1970se,Zerilli:1970wzz}.
Each component of $h_{\mu \nu}$ can be expanded in 
spherical harmonics $Y_{lm}(\theta, \varphi)$ 
with coefficients that depend on $t$ and $r$.
Due to the spherical symmetry of the background, 
it suffices to consider the mode $m=0$, as the 
nonaxisymmetric modes ($m \neq 0$) can be obtained 
by an appropriate rotation.
In the following, we denote the spherical harmonics 
with $m=0$ by $Y_l(\theta)$. We also adopt a gauge 
in which the metric components $h_{t\theta}$, 
$h_{\varphi\varphi}$, and $h_{\theta\varphi}$ 
vanish \cite{DeFelice:2011ka,Kobayashi:2012kh,Kobayashi:2014wsa}. 
This choice completely fixes the residual gauge 
DOFs under infinitesimal coordinate 
transformations $x^\mu \to x^\mu + \xi^\mu$.
The nonvanishing metric components are then given by 
\ba
\hspace{-0.7cm}
& &
h_{tt}=f(r) H_0(t,r) Y_{l}(\theta)\,,
\qquad
h_{tr}=h_{rt}=H_1(t,r) Y_{l}(\theta)\,,
\qquad
h_{t \varphi}=h_{\varphi t}
=-Q(t,r) (\sin \theta) Y_{l, \theta} (\theta), 
\nonumber \\
\hspace{-0.7cm}
& &
h_{rr}=h^{-1}(r) H_2(t,r) Y_{l}(\theta)\,,
\qquad
h_{r \theta}=h_{\theta r}
=h_1 (t,r)Y_{l, \theta}(\theta)\,,\qquad
h_{r \varphi}=h_{\varphi r}=-W(t,r) 
(\sin \theta) Y_{l,\theta} (\theta)\,,
\label{hcom}
\ea
where $H_0$, $H_1$, $Q$, $H_2$, $h_1$, and $W$ 
are functions of $t$ and $r$, and we use the 
notation $Y_{l,\theta} \equiv \rd Y_l / \rd \theta$. 
Note that the summation over multipoles $l$ is omitted for each $h_{\mu\nu}$.

For the covector field $A_\mu$, the $U(1)$ gauge symmetry of the underlying theory allows us to choose
$\delta A_\theta=0$ \cite{Kase:2023kvq,DeFelice:2023rra,Chen:2024hkm}, 
where $\delta A_\theta$ denotes the $\theta$-component of 
the perturbation in $A_\mu$ \cite{Kase:2023kvq,DeFelice:2023rra,Chen:2024hkm}.
With this gauge choice, the perturbation in $A_\mu$ has the 
following nonvanishing components:
\be
\delta A_t=\delta A_0 (t,r) Y_{l}(\theta),\qquad 
\delta A_r=\delta A_1 (t,r) Y_{l}(\theta),\qquad
\delta A_{\varphi}=-\delta A(t,r) (\sin \theta) 
Y_{l,\theta}(\theta)\,.
\label{perma}
\ee
In Eqs.~(\ref{hcom}) and (\ref{perma}), the three fields 
$Q$, $W$, and $\delta A$ belong to the odd-parity 
sector, whereas the six fields 
$H_0$, $H_1$, $H_2$, $h_1$, $\delta A_0$, and $\delta A_1$ 
correspond to even-parity perturbations.

We expand the action (\ref{action}) up to second order in perturbations 
for $q_M=0$, imposing the condition (\ref{fhcon}), i.e., $f/h>0$. 
This analysis encompasses the behavior of perturbations 
in both the timelike region ($f>0$, $h>0$) and the 
spacelike region ($f<0$, $h<0$). 
After integrating the quadratic action over $\theta$ and $\varphi$, 
we obtain an action involving nine fields and their derivatives 
with respect to $t$ and $r$. 
Performing integration by parts and 
discarding irrelevant boundary terms, the second-order action 
can be written as ${\cal S}^{(2)}
=\int \rd t \rd r\,{\cal L}$. 
The Lagrangian ${\cal L}$ 
consists of two contributions,  
\be
{\cal L}={\cal L}_1+{\cal L}_2\,,
\label{Ltotal}
\ee
where
\ba
{\cal L}_1 &=& 
L \biggl[ p_1 \left( \dot{W}-Q'+\frac{2Q}{r} \right)^2 
+ \left( p_2 \delta A  + p_3 \delta A'  \right) 
\left( \dot{W}-Q'+\frac{2Q}{r} \right)
+ p_4\dot{\delta A}^2 + p_5 \delta A'^2 + p_6 \delta A^2 
+ p_7 W^2  \nonumber \\
& &\quad+ p_8 Q^2 + p_9 Q \delta A  \biggr]\,,
\label{L1}\\
{\cal L}_2 &=& 
a_0 H_0^2 + H_0 \left[ a_1 H_2' + L a_2 h_1' + (a_3+L a_4) H_2 
+ L a_5 h_1 \right]+ L b_0 H_1^2 
+ H_1 ( b_1 \dot{H}_2 + L b_2 \dot{h}_1 
+ L b_3 \delta A_1 ) \nonumber \\
& &
+ c_0 H_2^2 
+ L H_2 (c_1 h_1+ c_2 \delta A_0 ) + L d_0 \dot{h}_1^2 + L d_1 h_1^2 
+ L d_2 \dot{h}_1 \delta A_1 
+ L h_1 d_3 \delta A_0   \nonumber \\
& &
+ s_1 (\delta A_0'-\dot{\delta A}_1)^2 
+\left( s_2 H_0+s_3 H_2+L s_4 h_1 \right)
(\delta A_0'-\dot{\delta A}_1)
+ L s_5 \delta{A}_0^2 
+ L s_6 \delta A_1^2\,,
\label{L2}
\ea
with a dot being the derivative with 
respect to $t$, and 
\be
L \equiv l(l+1)\,.
\ee
The explicit expressions of $p_1$, etc., are presented in Appendix~A. 
The Lagrangians ${\cal L}_1$ and ${\cal L}_2$ describe perturbations 
in the odd- and even-parity sectors, respectively.

As studied in Refs.~\cite{Kase:2023kvq,DeFelice:2023rra,Chen:2024hkm},
there are four dynamical DOFs arising 
from the gravitational and vector-field sectors. 
They are given by 
\ba
\hspace{-0.5cm}
& &
\chi_1 \equiv 
\dot{W}-Q'+\frac{2Q}{r} 
+\frac{1}{2p_1} \left( p_2 \delta A
+p_3 \delta A' \right)\,,\label{chi1}\\
& &
\delta A\,,\\
& &
\chi_2 \equiv 
H_2-\frac{L}{r}h_1\,,\\
& &
V \equiv \delta A_0'-\dot{\delta A}_1+\frac{1}{2s_1} 
\left( s_2 H_0+s_3 H_2+L s_4 h_1 \right)\,.
\label{V}
\ea
Here, $\chi_1$ and $\delta A$ correspond to the gravitational and 
vector-field perturbations in the odd-parity sector, respectively. 
On the other hand, $\chi_2$ and $V$ represent the gravitational and 
vector-field perturbations in the even-parity sector, respectively.
Thus, the system of four dynamical perturbations 
is decomposed into two sectors:
\be
\vec{{\cal X}}_{\rm A}^t=\left( \chi_1, \delta A \right)\,,\qquad
\vec{{\cal X}}_{\rm B}^t=\left( \chi_2, V \right)\,,
\label{chiAB}
\ee
which correspond to the odd-parity and even-parity sectors, 
respectively. In the following, we investigate the linear 
stability of electric BHs for multipole modes $l \geq 2$.

\subsection{Stability conditions in the time-like region}
\subsubsection{Odd-parity perturbations}

Let us first derive the linear stability conditions of odd-parity 
perturbations $\vec{{\cal X}}_{\rm A}^t=\left( \chi_1, \delta A \right)$ 
in the time-like region ($f>0$ and $h>0$). 
For this purpose, it is convenient to deal with the field $\chi_1$ 
as a Lagrange multiplier and consider the following Lagrangian 
\ba
{\cal L}_{\rm A}={\cal L}_1
-L p_1 \left[ \dot{W}-Q'
+\frac{2Q}{r} 
+\frac{1}{2p_1} \left( p_2 \delta A
+p_3 \delta A' \right)
-\chi_1 \right]^2\,.
\label{Lodd}
\ea
The field equations of motion for $W$ and $Q$ are obtained 
by varying ${\cal L}_{\rm A}$ with respect to these perturbed fields. 
Provided that $p_7 \neq 0$ and $p_8 \neq 0$, 
these equations can be solved for $W$ and $Q$. 
This allows us to eliminate the terms $\dot{W}$, $Q'$, and $Q$ 
from ${\cal L}_{\rm A}$. 
After the integration by parts, the second-order Lagrangian is 
expressed in the form 
\be
{\cal L}_{\rm A}= L \left( \dot{\vec{\mathcal{X}}}^{t}_{\rm A}
{\bm K}_{\rm A}\dot{\vec{\mathcal{X}}}_{\rm A} 
+\vec{\mathcal{X}}_{\rm A}'^{t}
{\bm G}_{\rm A}\vec{\mathcal{X}}'_{\rm A}
+\vec{\mathcal{X}}_{\rm A}^{t}{\bm M}_{\rm A}
\vec{\mathcal{X}}_{\rm A}
+\vec{\mathcal{X}}_{\rm A}'^{t}{\bm S}_{\rm A}\vec{\mathcal{X}}_{\rm A} 
\right)\,, 
\label{Lodd2}
\ee
where ${\bm K}_{\rm A}$, ${\bm G}_{\rm A}$, and 
${\bm M}_{\rm A}$ are $2 \times 2$ symmetric matrices, 
${\bm S}_{\rm A}$ is a $2 \times 2$ 
antisymmetric matrix 
with the nonvanishing components 
$(S_{\rm A})_{12}=-(S_{\rm A})_{21} \neq 0$.

The kinetic matrix ${\bm K}_{\rm A}$ has the nonvanishing components $(K_{\rm A})_{11}=-p_1^2/p_7$ and $(K_{\rm A})_{22}=p_4$, 
where 
\be
p_1=\frac{\Mpl^2}{4} \sqrt{\frac{h}{f}}\,,\qquad
p_7=-\frac{\Mpl^2 f}{4r^2} 
\sqrt{\frac{h}{f}} \left( 1
-\frac{4\beta A_0'^2 h}{\Mpl^2 f} 
\right)(L-2)\,,\qquad 
p_4=\frac{1}{2h}\sqrt{\frac{h}{f}} 
\left( {\cal L}_{,F}-\frac{4\beta h'}{r} \right)\,.
\label{p185}
\ee
The ghost-free conditions correspond to 
$(K_{\rm A})_{11}>0$ and $(K_{\rm A})_{22}>0$, 
which translate to $p_7<0$ and $p_4>0$.
Since we are considering the time-like region with 
the multiples $l \geq 2$, the ghosts are absent if 
\ba
& &
{\cal G}_1 \equiv 
1-\frac{4\beta A_0'^2h}
{\Mpl^2 f}>0\,,\label{NG1a}\\
& & 
{\cal G}_2 \equiv 
{\cal L}_{,F}-\frac{4\beta h'}{r} >0\,.
\label{NG1b}
\ea
In NED with $\beta=0$, the inequality 
(\ref{NG1b}) translates to 
${\cal L}_{,F}>0$ \cite{DeFelice:2024seu}. 
In theories with ${\cal L}(F)=F$,  the two conditions 
(\ref{NG1a}) and (\ref{NG1b}) coincide with those derived in Ref.~\cite{Chen:2024hkm}.

The nonvanishing matrix components 
of ${\bm G}_{\rm A}$ are given by 
$(G_{\rm A})_{11}=-p_1^2/p_8$ and 
$(G_{\rm A})_{22}=-(p_3^2-4p_1 p_5)/(4p_1)$, 
where 
\be
p_8=-\frac{p_7}{fh}\,,\qquad 
\frac{p_3^2-4p_1 p_5}{4p_1}
\frac{1}{fh}=p_4\,.
\label{p95}
\ee
The propagation speeds of $\chi_1$ and $\delta A$ 
can be obtained by assuming the solutions to the perturbation equations for 
$\vec{\mathcal{X}}_{\rm A}^t$ of the form 
$\vec{\mathcal{X}}_{\rm A}^t=\vec{\mathcal{X}}_0^{t} e^{-i (\omega t-kr)}$, 
where $\vec{\mathcal{X}}_0^{t}$ is a constant vector. 
Taking the large $\omega$ and $k$ limits, we obtain the dispersion 
relations $\omega^2=
-[(G_{\rm A})_{11}/(K_{\rm A})_{11}]k^2$ and 
$\omega^2=-[(G_{\rm A})_{22}/(K_{\rm A})_{22}]k^2$. 
The squared propagation speeds $c_r$ along the radial direction, which are defined
in terms of the proper time $\tau=\int \sqrt{f}\,\rd t$ and the rescaled radial 
coordinate $\tilde{r}=\int \rd r/\sqrt{h}$ for $f>0$ and $h>0$, 
are given by 
$c_{r1}^2=-(G_{\rm A})_{11}/[fh (K_{\rm A})_{11}]$ and 
$c_{r2}^2=-(G_{\rm A})_{22}/[fh (K_{\rm A})_{22}]$. 
These reduce to
\ba
c_{r1}^2 &=& 
-\frac{1}{fh}\frac{p_7}{p_8}=1\,,\\
c_{r2}^2 &=& 
=\frac{1}{fh}\frac{p_3^2-4p_1 p_5}{4p_1 p_4}=1\,,
\label{crodd}
\ea
where we used Eq.~(\ref{p95}). 
Thus, the radial propagation speeds 
of $\chi_1$ and $\delta A$ are luminal. 
The diagonal matrix components of ${\bm M}$ are 
\ba
(M_{\rm A})_{11} &=&
-p_1-\frac{p_1 [r p_8'(r p_1'+2p_1)
+p_8(6p_1-r^2p_1'')]}
{r^2 p_8^2}\,,\\
(M_{\rm A})_{22}  &=& 
p_6-\frac{p_{9}^2}{4p_8}
-\frac{p_1' p_2 p_3+p_1(p_2^2-p_2' p_3-p_2 p_3')}{4p_1^2}\,.
\ea
In the eikonal limit ($l \gg 1$), 
these components have the dependence
$(M_{\rm A})_{11} \propto L^0$ and 
$(M_{\rm A})_{22} \propto L$. 
The off-diagonal components $(M_{\rm A})_{12}$ 
and $(S_{\rm A})_{12}$ are nonvanishing, 
with the large $l$ behavior 
$(M_{\rm A})_{12} \propto L^{0}$ and 
$(S_{\rm A})_{12}' \propto L^{0}$.
To derive the angular propagation speeds, we take the limits of large $\omega^2$ and $L \gg 1$ in the perturbation equations for 
$\vec{\mathcal{X}}_{\rm A}^t$.
The existence of nonzero solutions to $\vec{\mathcal{X}}_0^t$ requires that 
\be
\left[ \omega^2 (K_{\rm A})_{11}
+(M_{\rm A})_{11} \right]
\left[ \omega^2 (K_{\rm A})_{22}
+(M_{\rm A})_{22} \right]
-[(M_{\rm A})_{12}]^2
+\frac{1}{4}\left[{(S_{\rm A})_{12}}' 
\right]^2
=0\,.
\label{disan}
\ee
Since we are interested in solutions with  
$\omega^2={\cal O}(1) L \gg 1$, Eq.~(\ref{disan}) 
gives the two approximate dispersion relations
$\omega^2=-(M_{\rm A})_{11}/(K_{\rm A})_{11}$ and $\omega^2=-(M_{\rm A})_{22}/(K_{\rm A})_{22}$.
The angular propagation speed in proper time is defined by 
$c_{\Omega}=r \rd \theta/\rd \tau=\hat{c}_{\Omega}/\sqrt{f}$, 
where $\hat{c}_{\Omega}=r\rd \theta/\rd t$ obeys $\omega^2=\hat{c}_{\Omega}^2 l^2/r^2$. 
The two squared angular propagation speeds 
are then given by 
\ba
c_{\Omega 1}^2 &=& 
-\frac{r^2}{fL} \frac{(M_{\rm A})_{11}}
{(K_{\rm A})_{11}} \biggr|_{L \to \infty}
={\cal G}_1 \,,\label{cO1}\\
c_{\Omega 2}^2 &=& 
-\frac{r^2}{fL} \frac{(M_{\rm A})_{22}}
{(K_{\rm A})_{22}} 
\biggr|_{L \to \infty} 
=\frac{{\cal G}_3}{{\cal G}_1 {\cal G}_2}\,,
\label{cO2}
\ea
where 
\ba
{\cal G}_3 &\equiv& 
{\cal L}_{,F} 
+2\beta ( f'^2h -2f'' fh- f'f h')/f^2-4\beta h {\cal L}_{,F} A_0'^2/(\Mpl^2 f)+ 8\beta^2 
[ f (h'A_0' + 2h A_0'')^2 \nonumber \\
& &- f'h'h A_0'^2
+ 2h^2 A_0' (f''A_0' - 2f'A_0'')]/(\Mpl^2 f^2)\,.
\ea
Under the ghost-free condition 
(\ref{NG1a}), the right-hand 
side of Eq.~(\ref{cO1}) is positive. 
To avoid the angular Laplacian instability, we require $c_{\Omega 2}^2>0$, 
which, upon using the no-ghost conditions (\ref{NG1a}) and (\ref{NG1b}), reduces to 
\be
{\cal G}_3 > 0 \,.
\label{sta1}
\ee
We also note that, for $\beta=0$, one has 
${\cal G}_1=1$ and 
${\cal G}_2={\cal G}_3={\cal L}_{,F}$, 
so that $c_{\Omega 1}^2=c_{\Omega 2}^2=1$.

\subsubsection{Even-parity perturbations}

Let us now discuss the stability of electric BHs against even-parity perturbations 
$\vec{{\cal X}}_{\rm B}^t=\left( \chi_2, V \right)$ 
in the time-like region. 
We treat the field $V$ as a Lagrange multiplier 
and consider the following Lagrangian:
\ba
{\cal L}_{\rm B} &=& 
{\cal L}_2
-s_1 \left[ \delta A_0'-\dot{\delta A}_1+\frac{1}{2s_1} 
\left( s_2 H_0+s_3 H_2+L s_4 h_1 \right)-V \right]^2\,.
\label{Leven}
\ea
We vary ${\cal L}_{\rm B}$ with respect to 
$\delta A_0$, $\delta A_1$, and $H_1$. 
Provided that $s_5 \neq 0$, $s_6 \neq 0$, 
and $b_0 \neq 0$, the corresponding perturbation equations can be solved for these fields. 
We then eliminate $\delta A_0$, $\delta A_1$, $H_1$, 
and their derivatives from ${\cal L}_{\rm B}$. 
Variation with respect to $H_0$ yields a 
constraint equation for $h_1$, which is 
subsequently used to eliminate $h_1$ and 
$H_2$ ($= \chi_2 + L h_1 / r$) from the action.
The resulting Lagrangian can thus be expressed in terms of the two dynamical perturbations 
$\chi_2$, $V$, and their $t$ and $r$ derivatives. 
After the integration by parts, the second-order action takes the form
\be
{\cal L}_{\rm B}= 
\dot{\vec{\mathcal{X}}}_{\rm B}^{t}
{\bm K}_{\rm B} \dot{\vec{\mathcal{X}}}_{\rm B}
+\vec{\mathcal{X}}_{\rm B}'^{t}
{\bm G}_{\rm B}\vec{\mathcal{X}}'_{\rm B}
+\vec{\mathcal{X}}_{\rm B}^{t}
{\bm M}_{\rm B}\vec{\mathcal{X}}_{\rm B}
+\vec{\mathcal{X}}_{\rm B}'^{t}
{\bm S}_{\rm B}\vec{\mathcal{X}}_{\rm B}\,,
\label{Leven2}
\ee
where ${\bm K}_{\rm B}, {\bm G}_{\rm B}, {\bm M}_{\rm B}$ are $2 \times 2$ symmetric matrices, and 
${\bm S}_{\rm B}$ is a $2 \times 2$ antisymmetric matrix with the nonvanishing components 
$(S_{\rm B})_{12}=-(S_{\rm B})_{21} 
\neq 0$. We note that the off-diagonal components 
of ${\bm K}_{\rm B}$ and ${\bm G}_{\rm B}$ 
are nonvanishing for even-parity perturbations.

In the time-like region, the absence of 
ghosts requires that ${\rm det}\,{\bm K}_{\rm B}=
(K_{\rm B})_{11}(K_{\rm B})_{22}
-(K_{\rm B})_{12}^2>0$ and $(K_{\rm B})_{22}>0$. 
By considering the leading-order contribution to ${\rm det}\,{\bm K}_{\rm B}$ in the large $l$ 
limit, the ghost-free conditions can be 
expressed as
\ba
{\rm det}\,{\bm K}_{\rm B}
&=&\frac{\Mpl^2 [r^2(f {\cal L}_{,F}+h A_0'^2 
{\cal L}_{,FF})-8 \beta f (h-1)]^2 
r^2 h^2}{2f^4 L^3} \frac{{\cal G}_1}
{{\cal G}_2}>0
\,,\label{nogo3}\\
(K_{\rm B})_{22} &=& 
\frac{1}{2 f^3 {\cal G}_2 L} 
\sqrt{\frac{h}{f}}
\left[r^2 (f{\cal L}_{,F}+h A_0'^2 
{\cal L}_{,FF})-8 \beta f(h-1) 
\right]^2>0\,.
\label{nogo4}
\ea
These inequalities are satisfied if
\be
{\cal G}_1>0\,,\qquad 
{\cal G}_2>0\,.
\ee
Thus, the ghost-free conditions for even-parity 
perturbations are identical to those for 
odd-parity perturbations.

The radial propagation speeds $c_r$ in the time-like 
region can be found by solving 
\be
{\rm det}\,\left( fh c_r^2 {\bm K}_{\rm B}
+{\bm G}_{\rm B} 
\right)=0\,.
\ee
Taking the limit $l \gg 1$, we obtain the following 
two solutions for $c_r^2$ :
\be
c_{r3}^2=1\,,\qquad 
c_{r4}^2=1\,.
\label{cr34}
\ee
Thus, both $\chi_2$ and $V$ have luminal 
propagation speeds.

For the propagation along the angular direction, 
the matrix components in ${\bm K}_{\rm B}$, 
${\bm M}_{\rm B}$, and ${\bm S}_{\rm B}$ 
contribute to the dispersion relation. 
In the large $l$ limit, $(S_{\rm B})_{12}$ is proportional to 
$L^{-1}$, while the components in ${\bm K}_{\rm B}$ and 
${\bm M}_{\rm B}$ have leading-order contributions 
proportional to $L^{-1}$ and $L^0$, respectively.
In this eikonal limit, the angular propagation speeds 
$c_{\Omega}$ can be obtained by solving 
\be
{\rm det} \left( fL c_{\Omega}^2 {\bm K}_{\rm B}
+r^2 {\bm M}_{\rm B} 
\right)=0\,.
\label{detan}
\ee
Eq.~(\ref{detan}) has two solutions for $c_{\Omega}^2$.
One of them corresponds to 
the perturbation $\chi_2$, 
\be
c_{\Omega 3}^2={\rm Eq.~}(\ref{cO3Ap})\,, 
\ee
whose explicit form is given in Appendix~B.
In the limit $\beta \to 0$, we find that $c_{\Omega 3}^2 \to 1$.
The other solution, which corresponds to 
the squared propagation speed of $V$, 
is given by 
\ba
c_{\Omega 4}^2 &=& \frac{r (r {\cal L}_{,F}
-4\beta h')}{r^2({\cal L}_{,F}
+2F{\cal L}_{,FF})-8 \beta (h-1)}\,, 
\label{cO4}
\ea
where $F=h A_0'^2/(2f)$.  
In the limit $\beta \to 0$, 
Eq.~(\ref{cO4}) reduces to the value
$c_{\Omega 4}^2={\cal L}_{,F}/
({\cal L}_{,F}+2F{\cal L}_{,FF})$ in NED.\footnote{The four squared angular propagation speeds $c_{\Omega i}^2$ all differing from one another when $\beta\ne0$ imply the violation of the eikonal correspondence 
\cite{Cardoso:2008bp} between eikonal quasinormal modes and bound photon orbits around the BH. Such a violation could happen in the presence of nonminimal couplings between gravity and matter fields \cite{Chen:2018vuw,Chen:2019dip,Chen:2021cts}, and could have interesting observational implications \cite{Chen:2022nlw}.}
To avoid the Laplacian instability along the angular direction, we require that 
\be
c_{\Omega 3}^2>0\,,\qquad 
c_{\Omega 4}^2>0\,.
\ee

From the above discussions, the linear stability of BHs in the even-parity sector is ensured 
under the conditions ${\cal G}_1>0$, 
${\cal G}_2>0$, $c_{\Omega 3}^2>0$, and 
$c_{\Omega 4}^2>0$. 

\subsection{Stability conditions in the 
space-like region}

We also consider the linear stability of electric BHs 
in the space-like region ($f<0$ and $h<0$). 
In the odd-parity sector, the ghost-free conditions 
are determined by the positivity of the matrix 
${\bm G}_{\rm A}$. 
Using the relations in Eq.~(\ref{p95}), we find 
$(G_{\rm A})_{11}>0$ and $(G_{\rm A})_{22}>0$ 
for $p_7>0$ and $p_4<0$. 
From the expressions of $p_7$ and 
$p_4$ given in Eq.~(\ref{p185}), the no-ghost 
conditions are satisfied under the 
two inequalities ${\cal G}_1>0$ 
and ${\cal G}_2>0$.

To derive the propagation speeds of 
odd-parity perturbations 
$\chi_1$ and $\delta A$, we assume solutions of the 
perturbation equations in the form 
$\vec{\mathcal{X}}_{\rm A}^t = 
\vec{\mathcal{X}}_0^t \, 
e^{-i (\omega r - k t)}$. 
The squared radial propagation speeds, 
measured with respect to the proper time 
$\tau = \int {\rm d}r/\sqrt{-h}$ and the 
rescaled radial coordinate 
$\tilde{r} = \int \sqrt{-f}\,{\rm d}t$, are given by 
$c_{r1}^2=-fh (K_{\rm A})_{11}/(G_{\rm A})_{11}=1$ and 
$c_{r2}^2=-fh (K_{\rm A})_{22}/(G_{\rm A})_{22}=1$, 
so that both are luminal. 
The squared angular propagation speeds, measured 
using the proper time, are expressed as  
$c_{\Omega 1}^2=h r^2 (M_{\rm A})_{11}/
[L(G_{\rm A})_{11}]$ 
and 
$c_{\Omega 2}^2=h r^2 (M_{\rm A})_{22}/
[L(G_{\rm A})_{22}]$. 
In the limit $l \to \infty$, these reduce to the same values 
as those given in Eqs.~(\ref{cO1}) and (\ref{cO2}). 
Hence, the Laplacian instability is absent 
if ${\cal G}_1 > 0$ and ${\cal G}_3 > 0$.

For even-parity perturbations $\chi_2$ and $V$, the ghost-free conditions correspond to 
${\rm det}\,{\bm G}_{\rm B}=
(G_{\rm B})_{11}(G_{\rm B})_{22}
-(G_{\rm B})_{12}^2
>0$ and $(G_{\rm B})_{22}>0$. 
For $f<0$ and $h<0$, these conditions are 
satisfied if ${\cal G}_1>0$ and ${\cal G}_2>0$. 
The radial propagation speeds $c_r$ 
can be derived by solving
${\rm det} ( c_r^2 {\bm G}_{\rm B}
+fh {\bm K}_{\rm B} )=0$. Taking the limit 
$l \gg 1$, we obtain the same two luminal 
values of $c_r^2$ as those given in Eq.~(\ref{cr34}). 
The angular propagation speeds 
$c_{\Omega}$ can be found by solving  
${\rm det} ( L c_{\Omega}^2 {\bm G}_{\rm B}
-h r^2 {\bm M}_{\rm B} )=0$ with the limit 
$l \gg 1$. This equation leads to the 
same values of $c_{\Omega 3}^2$ and 
$c_{\Omega 4}^2$ as those given in 
Eqs.~(\ref{cO3Ap}) and (\ref{cO4}). 

In summary, the stability of electric BHs in the space-like region is ensured if
\be
{\cal G}_1>0\,,\qquad 
{\cal G}_2>0\,,\qquad
{\cal G}_3>0\,,\qquad 
c_{\Omega 3}^2>0\,,\qquad 
c_{\Omega 4}^2>0\,.
\ee
These conditions are the same as those derived 
in the time-like region.

%%%%%%%%%%%%%%%%%%%%%%%%%%%%%%%%%%%%%%%%%%
\section{Instability of nonsingular electric BHs}
\label{elesinstasec}
%%%%%%%%%%%%%%%%%%%%%%%%%%%%%%%%%%%%%%%%%%

In this section, we study the linear stability of nonsingular electric BHs with regular centers. 
The consistent solution for $A_0'(r)$ 
that has a continuous limit to NED 
as $\beta \to 0$ corresponds to the minus branch of Eq.~(\ref{A0so}), i.e., 
\be
A_{0-}'(r)=-\frac{q_E}
{4\beta (2h-2-rh')} \sqrt{\frac{f}{h}}
\left( 1-\sqrt{1-\xi} \right)\,.
\label{A0ele}
\ee
We also recall that the Lagrangian 
${\cal L}(F)$ is expressed in the form (\ref{cL}), 
which contains $A_0'$, $f$, $h$, 
and their $r$ derivatives. 
Near $r=0$, the metric components 
$f$ and $h$ of regular BHs can be expanded as Eq.~(\ref{fhexpansion}). 

In NED with $\beta=0$, the Laplacian instability arises from the negativity of the squared angular 
propagation speed $c_{\Omega 4}^2$, 
given by Eq.~(\ref{cO4}).
For $\beta \neq 0$, we examine whether 
a similar property holds.  
In Eq.~(\ref{cO4}), the $F$ derivatives of ${\cal L}$ can be computed as 
${\cal L}_{,F}={\cal L}'(r)/F'(r)$ and 
${\cal L}_{,FF}
=[{\cal L}''(r)F'(r)
-{\cal L}'(r)F''(r)]/F'(r)^3$, 
where $F=h A_0'^2/(2f)$.  
Using Eqs.~(\ref{A0ele}) and (\ref{cL}), together with the expansion 
(\ref{fhexpansion}) of 
metric functions around $r=0$, 
we find that $c_{\Omega 4}^2$ 
is expanded as 
\ba
c_{\Omega 4}^2=
\begin{cases}
-\dfrac{3}{2}-\dfrac{5h_4}{4h_3}r
+{\cal O}(r^2) & ({\rm for}~h_3 \neq 0)\,,\\
-2-\dfrac{9h_5}{10 h_4}r
+{\cal O}(r^2) & ({\rm for}~h_3=0,~h_4 \neq 0)\,,
\\
-\dfrac{5}{2}-\dfrac{7h_6}{9 h_5}r
+{\cal O}(r^2) & ({\rm for}~h_3=0,~h_4=0,~h_5 \neq 0)\,,
\end{cases}
\label{cOr=0}
\ea
whose leading-order terms are always negative. 
The leading-order contribution to 
$c_{\Omega 4}^2$ arises from the next-order term 
$h_2 r^2$ in the expansion of $h(r)$.  
If this next-order term is $h_n r^n$ with $n \geq 3$, 
then the leading-order contribution to $c_{\Omega 4}^2$ is $-n/2$, so that $c_{\Omega 4}^2 \leq -3/2$.  
This behavior is the same as that found for regular electric BHs with $\beta=0$ \cite{DeFelice:2024seu,DeFelice:2024ops}. 
Thus, the nonvanishing HVT coupling $\beta$ does not help to circumvent the angular instability 
associated with the perturbation $V$.

As discussed in Ref.~\cite{DeFelice:2024seu}, 
the instability of $V$ occurs on a short time scale 
of order $t_{\rm ins} \approx r/(\sqrt{-c_{\Omega 4}^2}\, l)$, 
where $r$ is roughly the size of the inner horizon.
Since $V$ is coupled to the gravitational 
perturbation $\chi_2$, the latter is also 
subject to exponential growth.
This implies that the regular metric of the form 
\eqref{fhexpansion} cannot be sustained in a 
steady state, so that the nonsingular electric 
BH is ruled out by the angular instability. 
Since regular BHs with magnetic charges cannot exist at the background level, we have excluded the presence of all stable nonsingular BHs in theories described by the action (\ref{action}). 

The fact that nonsingular BHs are linearly 
unstable indicates that they are unlikely to be the 
endpoint of gravitational collapse. 
To examine whether such linearly unstable BHs
indeed fail to form through gravitational collapse, 
a full numerical analysis that incorporates nonlinear 
effects will be required. 
Performing such nonlinear simulations is beyond 
the scope of this paper.

%%%%%%%%%%%%%%%%%%%%%%%%%%%%%%%%%%%%%%%%
\section{Stability of singular electric BHs}
\label{elesstasec}
%%%%%%%%%%%%%%%%%%%%%%%%%%%%%%%%%%%%%%%%

We now analyze the linear stability of the singular electric BH solutions, considering in turn the five classes of theories presented in Sec.~\ref{NSBHsec0}.

\subsection{Maxwell-HVT theory}
\label{subsec:maxwellhvtinstablilty}

In Maxwell-HVT theory, the metric 
components near $r=0$ are given 
by Eqs.~(\ref{hr=0}) and (\ref{fr=0}). 
Very close to the BH center, the singularity at $r=0$ prevents the direct applicability of linear perturbation theory. However, at distances near $r=0$ where the curvature scalars remain finite, we can still estimate the quantities relevant for the linear stability of BHs.
In this regime, the quantities associated with the ghost-free conditions can be expanded as
\be
{\cal G}_1=1-\frac{q_E^2}
{64\beta m^2 \Mpl^2}r^2
+{\cal O}(r^4),\qquad 
{\cal G}_2=-\frac{8 \beta m}{r^3}
+\frac{q_E^2}{16 m \Mpl^2 r}+{\cal O}(r^0),
\qquad
{\cal G}_3=\frac{16 \beta m}{r^3}
-\frac{ 5q_E^2}{8 m \Mpl^2 r}
+{\cal O}(r^0)\,.
\label{G123}
\ee
One can also estimate the squared angular propagation speeds as
\ba
c_{\Omega 1}^2 &=& 
1-\frac{q_E^2}{64\beta m^2 
\Mpl^2}r^2+{\cal O}(r^4)\,,
\qquad c_{\Omega 2}^2=-2
+\frac{q_E^2}{32\beta m^2 
\Mpl^2}r^2+{\cal O}(r^3)\,,\nonumber\\
c_{\Omega 3}^2 &=& 
1-\frac{q_E^2}{64\beta m^2 
\Mpl^2}r^2+{\cal O}(r^4)\,,
\qquad c_{\Omega 4}^2=-\frac{1}{2}
+\frac{q_E^2}{128\beta m^2 
\Mpl^2}r^2+{\cal O}(r^3)\,.
\label{MHeleomega}
\ea
Since the leading-order contributions to 
$c_{\Omega 2}^2$ and $c_{\Omega 4}^2$ are negative, the vector-field perturbations in both the odd- and even-parity sectors are subject to Laplacian instabilities around the BH center. 
In Sec.~\ref{MHVTback}, it was shown that consistent background BH solutions exist for $\beta m > 0$.  
In this case, the leading-order term of ${\cal G}_2$ in Eq.~(\ref{G123}) is negative, so that the no-ghost conditions for 
vector-field perturbations 
are violated.
Indeed, for arbitrary signs of $\beta$ and $m$, either ${\cal G}_2$ or 
${\cal G}_3$ is necessarily negative, 
so that at least one of the stability conditions is always violated.

%%%%%%%%%%%%%%%%%%%%%%%%%%%%%%%%
\begin{figure}[ht]
\begin{center}
\includegraphics[height=3.0in,width=3.4in]{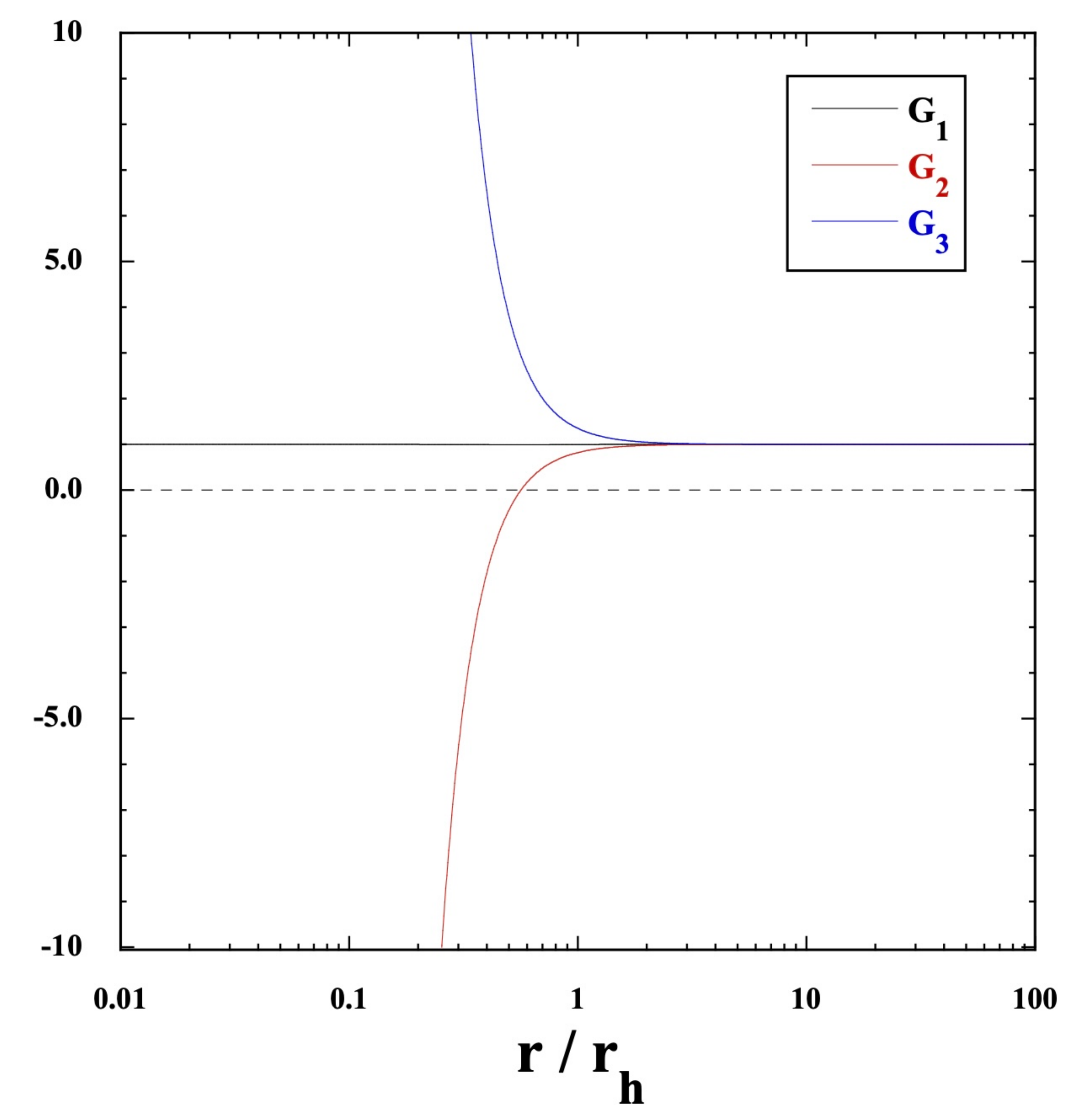}
\includegraphics[height=3.0in,width=3.4in]{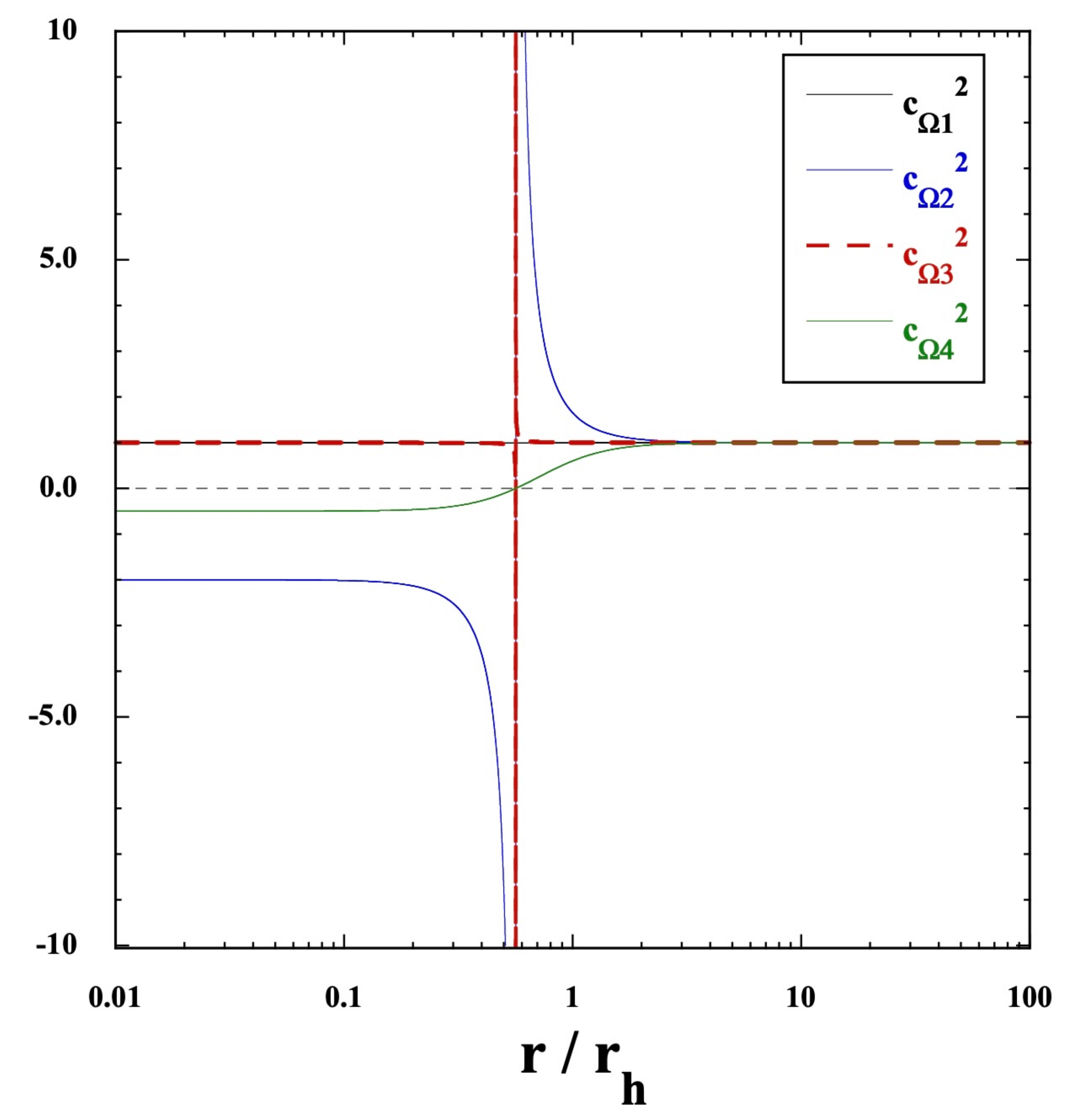}
\end{center}
\caption{\label{fig4} 
We plot ${\cal G}_1$, ${\cal G}_2$, and ${\cal G}_3$ (left panel), together with 
$c_{\Omega 1}^2$, $c_{\Omega 2}^2$, $c_{\Omega 3}^2$, and $c_{\Omega 4}^2$ (right panel), 
as functions of $r/r_h$, using the same model parameters and boundary conditions as in Fig.~\ref{fig1}.
We observe that ${\cal G}_2$ 
changes sign 
at $r = 0.56\, r_h$, where $c_{\Omega 2}^2$, 
$c_{\Omega 3}^2$, and $c_{\Omega 4}^2$ 
simultaneously flip their signs.
}
\end{figure}
%%%%%%%%%%%%%%%%%%%%%%%%%%%%%%%%

%%%%%%%%%%%%%%%%%%%%%%%%%%%%%%%%
\begin{figure}[ht]
\begin{center}
\includegraphics[height=3.0in,width=3.4in]{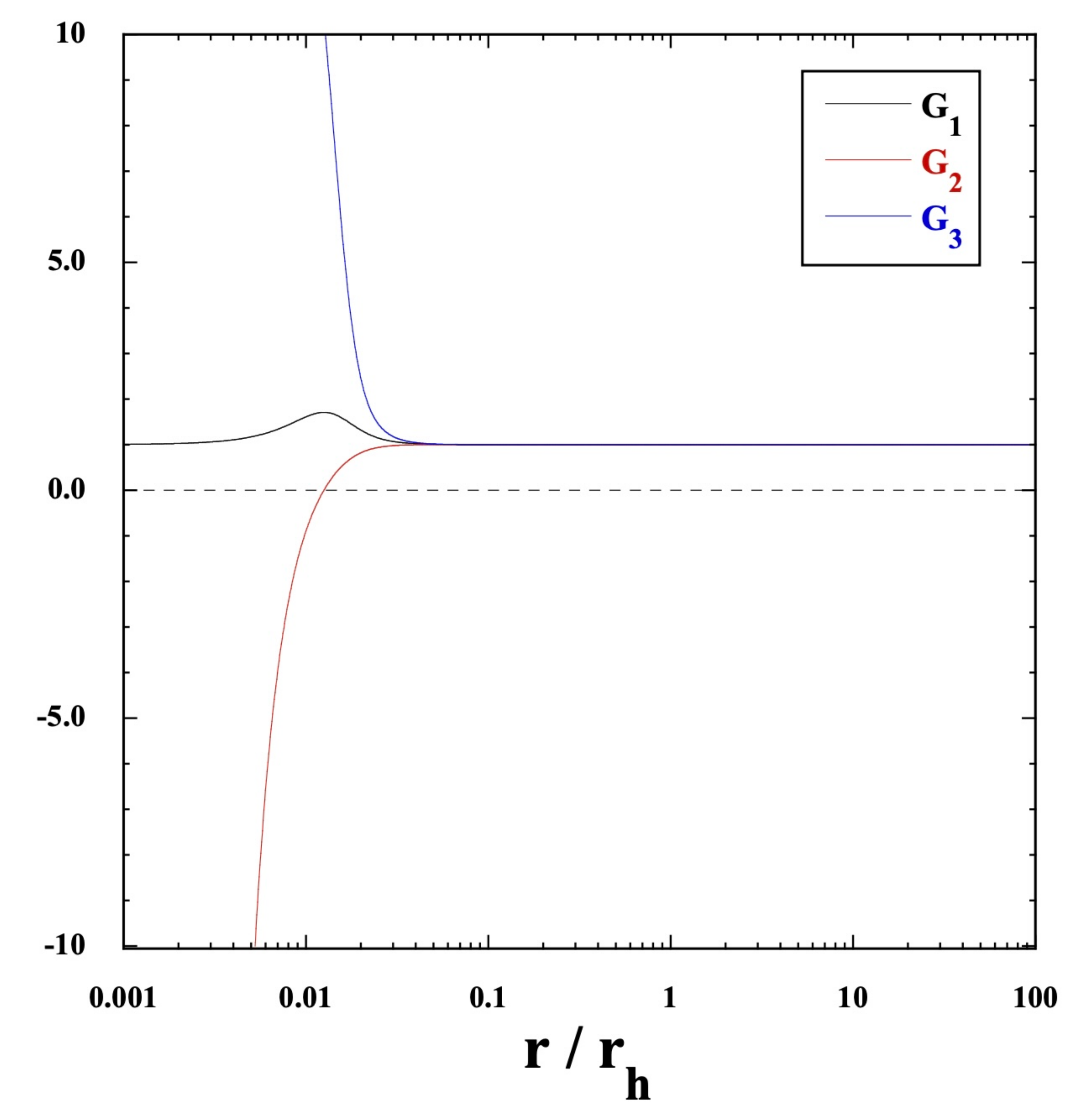}
\includegraphics[height=3.0in,width=3.4in]{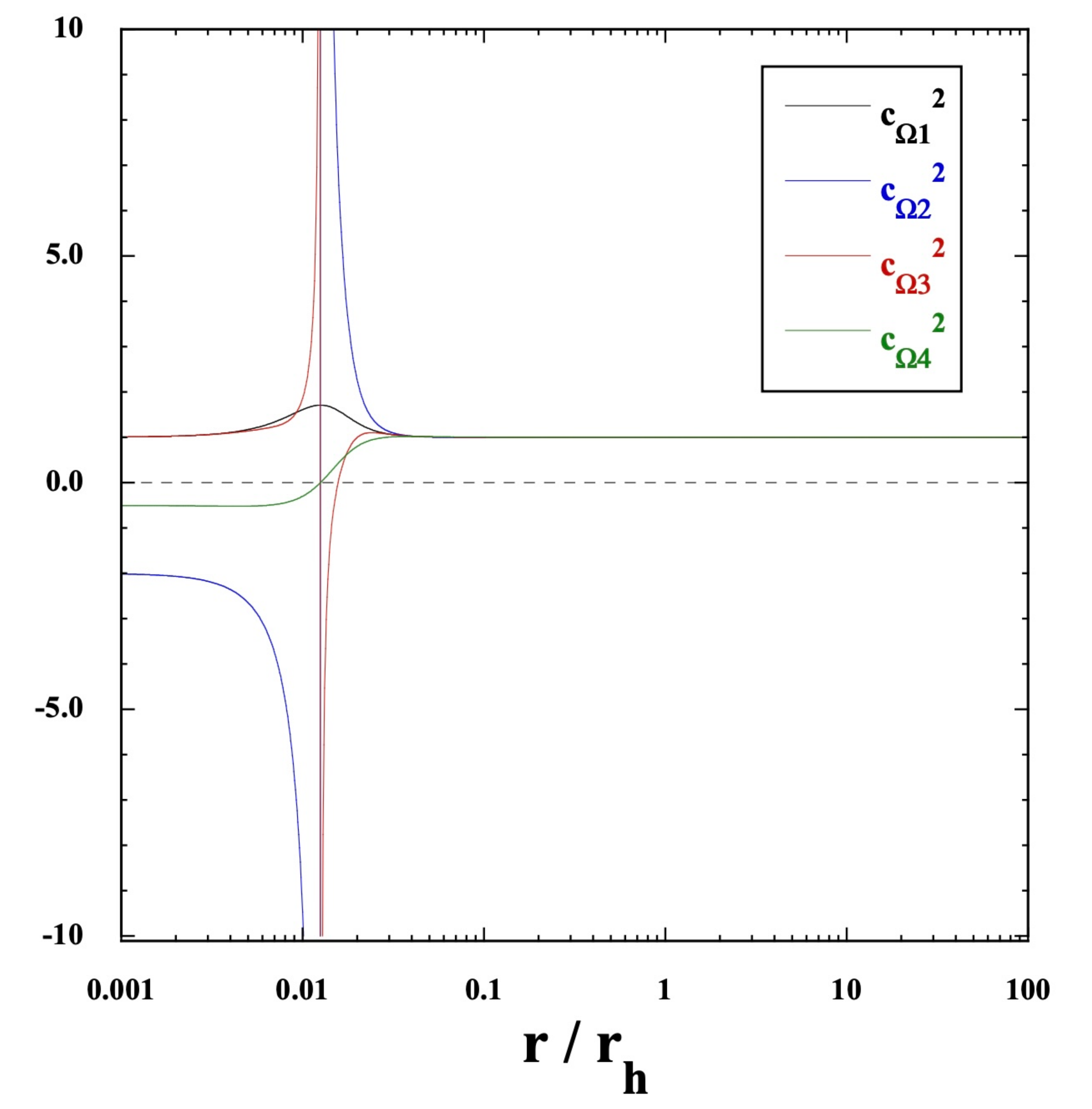}
\end{center}
\caption{\label{fig5}
Plots of ${\cal G}_1$, ${\cal G}_2$, and ${\cal G}_3$ (left panel), together with 
$c_{\Omega 1}^2$, $c_{\Omega 2}^2$, $c_{\Omega 3}^2$, and $c_{\Omega 4}^2$ (right panel), 
as functions of $r/r_h$, obtained with the same model parameters and boundary conditions as in Fig.~\ref{fig2}.
In this case, ${\cal G}_2$ crosses zero at $r = 1.25 \times 10^{-2} r_h$, where 
$c_{\Omega 2}^2$, $c_{\Omega 3}^2$, and $c_{\Omega 4}^2$ change sign.
}
\end{figure}
%%%%%%%%%%%%%%%%%%%%%%%%%%%%%%%%

To identify the regions where Laplacian and ghost instabilities arise, 
we analyze the behavior of the 
quantity ${\cal G}_2 = 1 - 4\beta h'/r$. 
Substituting the relation 
$h(r) \simeq -2m/r$ 
(valid for $r \ll r_h$) 
into this expression, 
we obtain ${\cal G}_2 \simeq 
1-8\beta m/r^3$. 
Thus, ${\cal G}_2$ becomes negative 
for $r<r_g$, where 
\be
r_g=\left( 8\beta m \right)^{1/3}\,.
\ee
Since $c_{\Omega 2}^2 = {\cal G}_3 
/({\cal G}_1 {\cal G}_2)$, 
crossing ${\cal G}_2=0$ at the distance 
$r=r_g$ leads to the divergence of $c_{\Omega 2}^2$ for nonvanishing 
values of ${\cal G}_1$ and ${\cal G}_3$. 
The fourth squared angular propagation speed is given by 
\be
c_{\Omega 4}^2 = \frac{r^2 {\cal G}_2}
{r^2 - 8 \beta (h-1)}\,,
\label{cO4D}
\ee
which also vanishes at $r=r_g$. 
As we estimated in 
Eq.~(\ref{rc}), the term 
$-8 \beta (h-1)$ in the denominator 
of Eq.~(\ref{cO4D}) becomes of 
the same order as $r^2$ at 
$r_c = (16 \beta m )^{1/3}$, 
which is about 1.26 times larger 
than $r_g$. 
In Sec.~\ref{MHVTback}, we focused on the cases in which the denominator of Eq.~(\ref{cO4D}) does not vanish with $r^2 - 8 \beta (h-1)>0$, 
like those plotted 
in Figs.~\ref{fig1} and \ref{fig2}. 
Therefore, both $c_{\Omega 2}^2$ and $c_{\Omega 4}^2$ are negative for $r < r_g$, with their signs changing 
at $r = r_g$.

To confirm the above analytic estimates, we numerically compute ${\cal G}_1$, ${\cal G}_2$, and ${\cal G}_3$, as well as $c_{\Omega 1}^2$, $c_{\Omega 2}^2$, $c_{\Omega 3}^2$, 
and $c_{\Omega 4}^2$, and plot them as 
functions of $r/r_h$ in Figs.~\ref{fig4} 
and \ref{fig5}, using the same model parameters and boundary conditions as in Figs.~\ref{fig1} and \ref{fig2}, respectively.
In Fig.~\ref{fig4}, corresponding to the case $\beta>0$ and $m>0$, we observe that ${\cal G}_2 < 0$ for 
$r < r_g = 0.56\, r_h$, 
indicating that a ghost appears already 
at a distance close to the horizon.
The distance at which the HVT coupling alters the behavior of the background solution is 
$r_c = 0.71\, r_h$, which is approximately 
$1.26$ times larger than $r_g$ 
as expected. 
In Fig.~\ref{fig5}, corresponding to the case $\beta < 0$ and $m < 0$, ${\cal G}_2$ is also negative for $r < r_g=1.25 \times 10^{-2} r_h$.

In the right panel of Fig.~\ref{fig4}, we find that $c_{\Omega 1}^2$ is positive both inside 
and outside the horizon. 
On the other hand, 
$c_{\Omega 2}^2$ and $c_{\Omega 4}^2$ are negative for 
$r < r_g = 0.56\, r_h$, 
with their signs changing 
at $r = r_g$.
After crossing ${\cal G}_2=0$, they quickly approach their asymptotic values, 
$c_{\Omega 2}^2 \to -2$ and $c_{\Omega 4}^2 \to -1/2$, 
for $r$ of order $0.1\, r_h$.
In other words, the ghost and Laplacian 
instabilities are present not only in the vicinity of $r=0$ 
but also in the region close to the horizon, 
$0.1\, r_h \lesssim r < 0.56\, r_h$.
In this regime, the linear perturbation theory remains valid, and therefore the BH solution shown in Fig.~\ref{fig1} is ruled out due to the presence of instabilities inside the horizon.
We note that the denominator of $c_{\Omega 3}^2$ 
is proportional to ${\cal G}_2$ for ${\cal L}(F) = F$, 
so that $c_{\Omega 3}^2$ changes sign at $r = r_g$.

In the right panel of Fig.~\ref{fig5}, we observe 
that both $c_{\Omega 2}^2$ and $c_{\Omega 4}^2$ are negative 
for $r<r_g=1.25 \times 10^{-2}r_h$, with the divergence of $c_{\Omega 3}^2$ at $r=r_g$. 
Again, the presence of ghost and 
Laplacian instabilities in the region 
$r<r_g$ excludes the BH 
solution shown in Fig.~\ref{fig2} 
as a stable configuration. 
The difference from the $\beta > 0$ case in Fig.~\ref{fig4} is that instabilities appear at a smaller value of $r_g$, of order $10^{-2} r_h$. 
This is due to the fact that the HVT coupling in Fig.~\ref{fig5} is $\beta = -{\cal O}(10^{-7}) r_h^2$, whose magnitude is much smaller than $\beta = {\cal O}(10^{-2})r_h^2$ chosen in Fig.~\ref{fig4}.

To avoid ghost and Laplacian instabilities, 
we must consider the case where $r_g$ is 
smaller than an EFT scale $r_{\rm EFT}$, 
below which linear perturbation theory 
breaks down. Alternatively, one can introduce a 
cutoff length scale, below which the theory is modified to achieve an ultraviolet completion.
The condition $r_g<r_{\rm EFT}$ translates to 
\be
|\beta| < \frac{r_{\rm EFT}^3}{8|m|}\,.
\label{betacon}
\ee
In Fig.~\ref{fig4}, the HVT coupling is chosen as $\beta = 4.59 \times 10^{-2} r_h^2$, yielding $2m = 0.96 r_h$.
In Fig.~\ref{fig5}, we have $\beta = -3.062 \times 10^{-7}r_h^2$ and $|2m| = 1.59 r_h$. 
In both cases, $|2m|$ is of order 
$r_h$. As long as $|2m|$ is comparable to $r_h$, 
the inequality (\ref{betacon}) translates to
\be
|\beta| \lesssim 
\frac{r_{\rm EFT}^3}{r_h}\,.
\label{betacon2}
\ee
The linear perturbation theory is expected to be valid down to the EFT scale $r_{\rm EFT}$.
Except very close to the BH center, the 
perturbation analysis can be trusted; 
thus, $r_{\rm EFT}$ can be much smaller than $r_h$, i.e., $r_{\rm EFT} \ll r_h$.
For example, if $r_{\rm EFT} = 10^{-2} r_h$, it follows that $|\beta| \lesssim 10^{-6} r_h^2$.
Under such a stringent upper bound on $|\beta|$, the term $-8 \beta (h-1)$ in Eqs.~(\ref{hele2}) and (\ref{fhele2}) is suppressed relative to $r^2$ 
outside the horizon, so that the background solution is nearly indistinguishable from the $\beta = 0$ case. 
The same is expected to hold for the quasinormal modes of BHs, making it difficult to observe signatures of the HVT coupling.

\subsection{Power-law NED theory 
with $\beta=0$}
\label{powerstasec}

We now turn to the stability analysis of BHs in power-law NED theories, 
described by the Lagrangian (\ref{powerNED}) with 
$p \geq 2$. Using the background 
Eq.~(\ref{back4}) with $f=h$, we find 
that the quantities associated with 
the linear stability of BHs reduce to 
\be
{\cal G}_1=1\,,\qquad 
{\cal G}_2={\cal G}_3=
1+ 2^{1-p} p\,a_p A_0'^{2(p-1)}\,,
\ee
and 
\be
c_{\Omega 1}^2=c_{\Omega 2}^2
=c_{\Omega 3}^2=1\,,\qquad
c_{\Omega 4}^2=
\frac{1+2^{1-p}p\,a_p 
A_0'^{2(p-1)}}{1+2^{1-p}(2p-1)p\,
a_p A_0'^{2(p-1)}}\,.
\ee
In the small-distance regime where the condition 
$|a_p F^p| \gg F$ is satisfied, the electric 
field behaves as $A_0'(r) \propto r^{-2/(2p-1)}$. 
In this region, the contribution 
of the term $R_{a_p}=2^{1-p} p\,a_p A_0'^{2(p-1)}$ dominates over 1 for $p \geq 2$, 
so that ${\cal G}_2={\cal G}_3 
\simeq 2^{1-p} p\,a_p A_0'^{2(p-1)}$. 
Provided that 
\be
a_p >0\,,
\ee
the conditions ${\cal G}_2
={\cal G}_3>0$ are always satisfied. 
In the regime $R_{a_p} \gg 1$, 
the leading-order term of 
$c_{\Omega 4}^2$ is given by 
\be
c_{\Omega 4}^2 \simeq 
\frac{1}{2p-1}\,,
\ee
which is positive for $p \geq 2$.
In contrast, in the large-distance regime characterized by $R_{a_p} \ll 1$, 
the asymptotic value of $c_{\Omega 4}^2$ approaches 1. Numerically, we computed $c_{\Omega 4}^2$ for several values of 
$p$ with $p \geq 2$, and confirmed that it increases smoothly from 
$1/(2p-1)$ in the small-distance region 
to 1 in the large-distance region.
Therefore, as long as $a_p > 0$ and $p \geq 2$, 
neither ghost nor Laplacian instabilities appear in timelike or spacelike regions. 

\subsection{Born-Infeld theory}

In Born-Infeld theories with the HVT coupling, 
the solution for $A_0'(r)$ is given by Eq.~\eqref{A0BI}. 
Using this background solution, the quantities 
relevant for the linear stability of BHs 
reduce to
\be
{\cal G}_1=1\,,\qquad 
{\cal G}_2={\cal G}_3=
\frac{\sqrt{r^4+r_b^4}}{r^2}\,,
\ee
and
\be
c_{\Omega 1}^2=
c_{\Omega 2}^2=
c_{\Omega 3}^2=1\,, \qquad c_{\Omega 4}^2=\frac{r^4}{r^4+r_b^4}\,.
\ee
For $r > 0$, all of these quantities are positive, and therefore ghost and Laplacian instabilities are absent. 
However, $c_{\Omega 4}^2 \to 0$ as $r \to 0$, indicating the emergence of a strong coupling problem. To avoid this issue near the 
singular point $r=0$, we need to assume 
that the linear perturbation theory starts 
to lose validity below an EFT scale $r_{\rm EFT}$.
In other words, we require the condition 
$r_b<r_{\rm EFT}$, i.e., 
\be
b<\frac{2r_{\rm EFT}^4}{q_E^2}\,.
\ee
Since $r_{\rm EFT}$ can be much smaller than the horizon radius $r_h$, the coupling $b$ is constrained by $b \ll 2 r_h^4 / q_E^2$. 
Under such a bound, the metric functions outside the horizon, as well as the electric field, are almost indistinguishable from those of the RN BH.

\subsection{Power-law or 
Born-Infeld NED theories with 
$\beta \neq 0$}
\label{NEDHVTsec}

In power-law NED theories with the HVT coupling, the leading-order metric functions $h(r)$ and $f(r)$ near $r=0$ exhibit the dependence given in Eq.~(\ref{hfana}), with $A_0'(r)$ 
taking the forms of 
Eqs.~(\ref{A0mi0}), \eqref{A0112special}, and (\ref{A0mi}) for the cases $p>-1/2$, $p=-1/2$, and $p < -1/2$, respectively. 
In this small-distance regime, we examine the behavior of the fourth 
squared angular propagation speed:  
\be
c_{\Omega 4}^2=
-\frac{r (a_p\,p\,r F^{p-1}+r
-4\beta h')}
{(1-2p)a_p\,p\,r^2 F^{p-1}
+8 \beta (h-1)-r^2}\,.
\label{cO4r=0}
\ee
Near $r=0$, the electric-field strength 
behaves as $F=h A_0'^2/(2f) 
\simeq h_1 A_0'^2/(2f_1) \propto 
A_0'^2$, where we have used the expansions 
$h \simeq h_1/r$ and $f \simeq f_1/r$.
When $p>-1/2$, the dominant contributions to 
Eq.~(\ref{cO4r=0}) come from the terms involving the coupling $\beta$. At leading order, we have
\be
c_{\Omega 4}^2 = -\frac{1}{2}\,.
\ee
For $p=-1/2$, the leading-order term of $c_{\Omega 4}^2$ in Eq.~\eqref{cO4r=0} is independent of the coefficient $C_{-1/2}$, defined in Eq.~\eqref{A0112special}. 
As a result, we have
\be
c_{\Omega 4}^2 = -\frac{1}{2}\,,
\ee
as well in this case. 
For $p < -1/2$, the coupling $a_p$ also contributes to $c_{\Omega 4}^2$ alongside $\beta$. 
In this case, the leading-order term of 
Eq.~(\ref{cO4r=0}) reads
\be
c_{\Omega 4}^2 = -\frac{3}{4(1-p)}\,,
\ee
which lies in the range 
$-1/2 < c_{\Omega 4}^2 < 0$.
Since $c_{\Omega 4}^2$ is negative 
for all $p$, the BH solution exhibits 
a Laplacian instability near $r=0$. 
To avoid this problem, the couplings $\beta$ and $a_p$ must be chosen sufficiently small so that the radius $r_g$, below which the instability occurs, lies within the EFT scale $r_{\rm EFT}$ that ensures the validity of linear perturbation theory. While we have considered only the angular propagation of $V$, the fact that $c_{\Omega 4}^2$ always becomes negative near $r=0$ is sufficient to exclude BHs with $r_g$ larger than $r_{\rm EFT}$. 
We also note that this instability arises as a consequence of the regularization of $A_0'(r)$ in the vicinity of $r=0$.

In Born-Infeld theory with the HVT coupling, the metric functions near $r=0$ are given by Eq.~(\ref{hfana}), with $A_0'(r)$ of the form (\ref{A0miD}). In this theory, the fourth squared angular propagation speed takes 
the form 
\be
c_{\Omega 4}^2=-\frac{r(1-bF) 
(r-4\beta \sqrt{1-bF}\,h')}
{8 \beta (1-bF)^{3/2} (h-1)-r^2}\,.
\label{cO4aBI}
\ee
Near $r=0$, the electric-field strength behaves as 
$F \simeq h_1 A_0'^2/(2f_1) \propto r^2$. This shows that the $b$-dependent terms do not contribute to Eq.~(\ref{cO4aBI}) in the small-distance region. 
Therefore, the leading-order term of $c_{\Omega 4}^2$ around $r=0$ reduces to that of Maxwell-HVT theory, namely
\be
c_{\Omega 4}^2=-\frac{1}{2}\,,
\ee
showing the presence of Laplacian instability. 
This instability manifests even in the region close to the horizon $r_h$, unless $\beta$ is much smaller than $r_h^2$.

The above results show that the presence of the HVT coupling induces Laplacian instabilities in both the power-law NED and Born-Infeld theories. 
To avoid this problem, one must either set $\beta=0$ or choose $\beta$ sufficiently small such that $r_g$ lies within the EFT scale $r_{\rm EFT}$. 
In addition, as noted at the end of Sec.~\ref{NHVTsec}, for any NED with 
a proper Maxwell limit, i.e., 
$\mathcal{L}(F) \to F$ as $F \to 0$, 
there is always a branch of solutions for 
which $h(r) \approx 1/r$, $f(r) \approx 1/r$, 
and $A_0'(r)\propto r$ near the origin. For this branch of solutions, the HVT effect dominates over NED contributions, 
and the behavior of solutions, including 
at the perturbation level, should be similar to that in 
Maxwell-HVT theory. 
Our results in this section, particularly the squared propagation speed $c_{\Omega 4}^2$ 
for power-law NED theories with 
$p \geq -1/2$ and for 
Born-Infeld theory, 
support this expectation. 
Nontrivial solutions can be obtained either by considering the NED without a proper Maxwell limit, 
such as power-law NED theories 
with $p < -1/2$, 
or by taking branches of solutions where $F$ does not approach 0 at the origin 
(see Sec.~\ref{subsec:recon_h}). In the next subsection, we present the stability analysis of the solutions 
derived in Sec.~\ref{subsec:recon_h} 
and show that they indeed exhibit nontrivial behavior at the perturbation level.

\subsection{Reconstructed NED 
theories with $\beta \neq 0$}
\label{perresec}

In Sec.~\ref{subsec:recon_h}, we reconstructed background BH solutions with the metric function 
$h(r) = 1 - 2M/r + b_1/r^2$ in NED theories with the HVT coupling. There are four branches of solutions, 
depending on the signs of Eqs.~\eqref{eq:L_branch} 
and \eqref{eq:f2_branch}, 
but their leading-order behaviors around 
the origin are the same in all cases.
At leading order, the quantities relevant to 
the linear stability of BHs near $r=0$ 
are given by 
\ba
\mathcal{G}_1 &=&-\frac{3M}{2b_1}\,r\,,
\qquad
\mathcal{G}_2=\frac{16b_1\beta}{r^4}\,,\qquad 
\mathcal{G}_3=-\frac{12M\beta}{r^3}\,,\\
c_{\Omega1}^2 &=&-\frac{3M}{2b_1}\,r\,,\qquad
c_{\Omega2}^2=\frac{1}{2}\,,\qquad
c_{\Omega3}^2=\frac{3}{2}\,,\qquad
c_{\Omega4}^2=-\frac{3M}{8b_1}\,r\,.
\ea
Since the metric component $h(r)=1 - 2M/r + b_1/r^2$ 
is valid for arbitrary $r$, the constant $M$ corresponds to the ADM mass, and thus $M > 0$.
To satisfy the no-ghost condition ${\cal G}_1>0$, 
we require that $b_1<0$. 
In Sec.~\ref{subsec:recon_h}, we showed that 
consistent background BH solutions exist only 
for $\beta>0$. With $b_1 < 0$ and $\beta > 0$, 
we have ${\cal G}_2 < 0$, 
so the other ghost-free condition 
is violated. Under the same inequalities, 
all squared angular propagation speeds are 
positive, and hence Laplacian instabilities 
are absent. However, a strong coupling issue 
arises because $c_{\Omega 1}^2 \to 0$ 
and $c_{\Omega 4}^2 \to 0$ as $r \to 0$.
On the other hand, linear perturbation theory 
can break down as an EFT in the limit $r \to 0$ 
due to the presence of curvature singularities. 

The difference from the models discussed in 
Secs.~\ref{subsec:maxwellhvtinstablilty} and \ref{NEDHVTsec} is that the HVT coupling does 
not necessarily induce Laplacian instabilities. 
This implies that, at the classical level, 
the model in this section is less harmful 
than the other cases. Although a ghost arises
from ${\cal G}_2 < 0$, it may not be 
problematic if: 
i) the perturbations remain classically stable, and 
ii) a consistent quantization prescription, such as the fakeon approach \cite{Anselmi:2018kgz}, is adopted. However, a detailed discussion on the quantization of these modes is beyond the scope 
of this paper.

In any case, the stability of solutions in the vicinity of the origin is necessary, though not sufficient, to guarantee stability throughout. 
For instance, by focusing on branch A, we can numerically solve the differential equation: 
\begin{equation}
\tilde{f}'=\tilde{f}\, 
\frac{q_{E}^{2} r^{4}
+16 \Mpl^{2} b_{1} \beta (2b_1-3Mr)
-  q_{E} r^{2} 
\sqrt{q_{E}^{2} r^{4}
+32 \Mpl^{2} b_{1} \beta (2b_1-3Mr)}}
{4 \Mpl^2 \left( 3 M r-2 b_{1}\right)^{2} 
\beta r}\,,
\label{eq:f2prime_recon}
\end{equation}
to find any other quantities relevant to 
the linear stability of BHs. 
Since we have imposed the condition 
$\beta \tilde{f}\tilde{f}'>0$, together with 
$\lim_{r\to\infty} \tilde{f}=1$, we will consider the case where $\tilde{f}$ is monotonically increasing, 
i.e., $\tilde{f}'>0$ and $\tilde{f}>0$.

It should be noted that 
Eq.~\eqref{eq:f2prime_recon} determines the value of 
$\tilde{f}'/\tilde{f}$ as a function of $r$.
From Eq.~\eqref{eq:L_branch}, once this ratio is fixed, $\mathcal{L}$ becomes a known function of $r$ (up to the choice of the plus or minus sign, which corresponds to branch A1 and branch A2, respectively).
Along the same lines, we can see that $F$ 
and $\mathcal{L}_{,F}$ in 
Eqs.~\eqref{eq:F_recon} and \eqref{eq:L_F_recon} can be expressed explicitly 
as functions of $r$; that is, they no longer depend on the exact form of $\tilde{f}$, but only on the ratio $\tilde{f}'/\tilde{f}$. 
This allows us, for instance, to construct analytically the angular squared 
propagation speeds as functions of $r$.\footnote{For this purpose, it is convenient to use the 
relations ${A_0'}^2 = 2 \tilde{f} F$ and 
$A_0''/A_0'=( \tilde{f}'/\tilde{f} 
+F'/F )/2$.}
For instance, we explicitly consider the case 
of $c_{\Omega1}^2$, which can be written as
\begin{equation}
c_{\Omega1}^2=\frac{r 
\bigl[ q_E r 
\sqrt{32 \beta  b_1 \Mpl^2 \left(2b_1-3 M r\right)+q_E^2 r^4}+24 \beta  M \Mpl^2
\left(3 M r-2 b_1\right)-q_E^2 r^3\bigr]}
{8 \beta
\Mpl^2 \left(2 b_1-3 M r\right){}^2}\,.
\end{equation}
Analogous expressions can be obtained for the other three propagation speeds, although these are more involved.

Assuming $q_E>0$ (and $\beta>0$, $b_1<0$) and expanding this function around $r \to \infty$ 
for branch A2, the squared angular 
propagation speeds are expressed as
\ba
c_{\Omega1}^2 &=&
1-\frac{16 b_1^2\beta \Mpl^2}{q_E^2 r^4}+\mathcal{O}(r^{-7})\,,\qquad
c_{\Omega2}^2 = 1+\frac{48 b_1 \beta M \Mpl^2}{q_E^2 r^3}+\mathcal{O}(r^{-4}) \,,\nonumber \\
c_{\Omega3}^2 &=& 
1-\frac{16 b_1^2\beta \Mpl^2}{q_E^2 r^4}+\mathcal{O}(r^{-6})\,,\qquad
c_{\Omega4}^2 =1+\frac{36 b_1\beta M \Mpl^2}{q_E^2 r^3}+\mathcal{O}(r^{-4})\,.
\ea
This result does not hold for all branches. 
In fact, the two branches corresponding to the plus sign in Eq.~\eqref{eq:f2_branch}, which are further distinguished by the plus/minus sign in Eq.~\eqref{eq:L_branch} 
(i.e., branches B1 and B2), are unstable at infinity, since $c_{\Omega4}^2 \to -1/2$. 
Therefore, they should be discarded due to 
Laplacian instabilities 
(at least for this choice of parameter signs).
It can also be shown that $c_{\Omega2}^2$ becomes negative near the horizon for branch A1, and, as a result, this branch solution is also unstable.

\begin{figure}[ht]
\centering
{\includegraphics[width=0.49\linewidth]{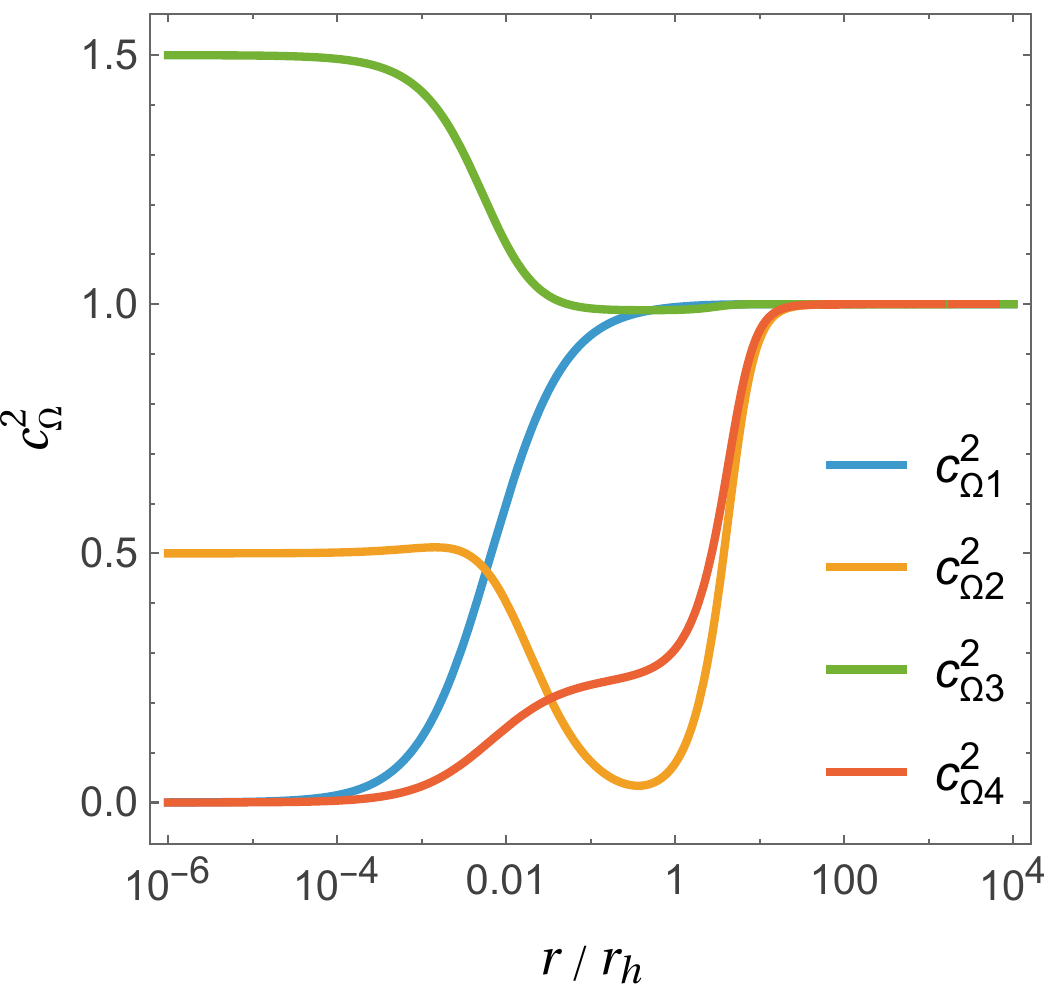}\hfill \includegraphics[width=0.49\linewidth]{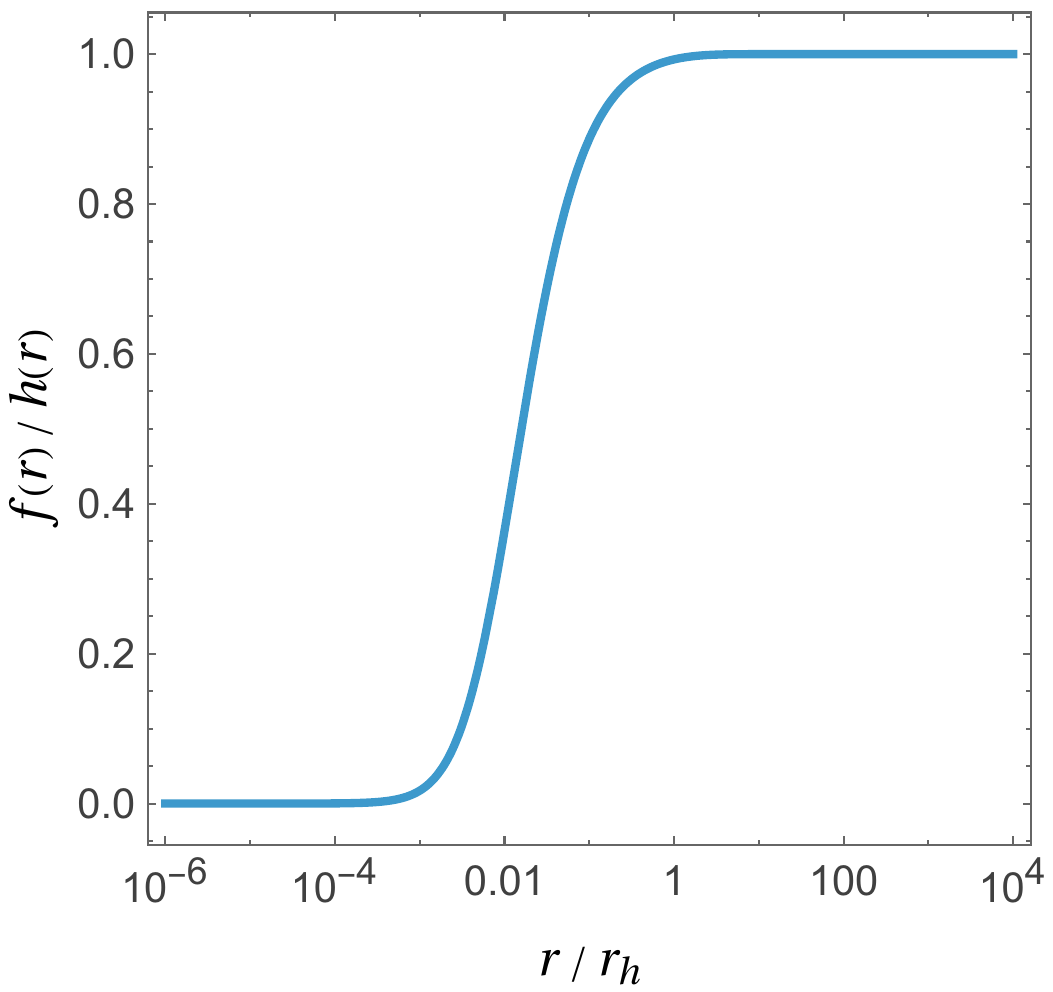}}
\caption{(Left) Plot of the four squared 
angular propagation speeds for 
the reconstructed solution 
corresponding to the choice 
$h(r) = 1 - 2M/r + b_1/r^2$.
The plot is generated for branch A2 using the parameters $b_1 = -10^{-2} M r_h$, $\beta = 2 r_h^2$, $q_E = M \Mpl / 9$, and $r_h / M = 2.01$. The integration is started well inside the horizon 
(at $r = 10^{-6} r_h$), providing the initial conditions for $\tilde{f}$. 
Since multiplying $\tilde{f}$ by any non-zero constant also yields a solution of the differential Eq.~\eqref{eq:f2prime_recon}, 
we finally redefine $\tilde{f}$ as 
$\tilde{f} \to \tilde{f}/\tilde{f}(r = \infty)$, 
so that $\tilde{f}$ satisfies the desired boundary conditions. The condition $h(r = r_h) = 0$ determines $r_h$ as a function of $M$ and $b_1$. 
(Right) Plot of $\tilde{f} = f / h$, showing that the solution is positive, monotonically increasing (as well as continuous and differentiable everywhere), and, in particular, that $f$ and $h$ vanish at the same point.
}
\label{fig:cOmega4_recon}
\end{figure}

For branch A2, namely the branch for which 
$\tilde{f}'$ is defined via Eq.~\eqref{eq:f2prime_recon} and 
$\mathcal{L}$ corresponds to the minus sign in Eq.~\eqref{eq:L_branch}, 
we observe in Fig.~\ref{fig:cOmega4_recon} 
that all squared angular propagation speeds are positive at the distance $r>0$. 
We also find that ${\cal G}_1$ remains positive and ${\cal G}_2$ stays negative, as is the case around the origin. Furthermore, we numerically confirm that the dimensionless quantity 
${A_0'}^2 r^4 / (\Mpl^2 r_h^2)$, expressed 
in terms of $\tilde{f} F$, remains positive everywhere and approaches a constant as $r \to \infty$. 
This is a necessary condition for the solution to remain classically stable. Whether this also constitutes a sufficient condition for classical stability---namely, whether no other type of classical instability is present (for instance, a tachyonic instability of the perturbations)---is beyond the scope of this work.

%%%%%%%%%%%%%%%%%%%%%%%%%%%%%%%%%%%
%%%%%%%%%%%%%%%%%%%%%%%%%%%%%%%%

%%%%%%%%%%%%%%%%%%%%%%%%%%%%
\section{Conclusions}
\label{consec}
%%%%%%%%%%%%%%%%%%%%%%%%%%%%

In this paper, we have studied the existence and stability of BH solutions on the SSS background in theories described by the action (\ref{action}). 
In NED without the HVT coupling, it is known that nonsingular BHs with electric or magnetic charge are prone to Laplacian instabilities 
near the center \cite{DeFelice:2024seu,DeFelice:2024ops}. The primary motivation of this work is to clarify whether the presence of the HVT coupling in NED allows for the existence of linearly stable, nonsingular BHs.
By including both electric and magnetic charges, we found that the existence of regular BHs compatible with the metric ansatz (\ref{fhexpansion}) requires the magnetic charge $q_M$ to vanish. 
Therefore, we restricted our analysis to purely electric BHs and investigated their linear stability for both nonsingular and singular cases.

In Sec.~\ref{pureelesec}, we showed that purely electric BHs with regular centers can exist at the background level. There is a branch where the electric field remains finite near $r=0$, with $A_{0-}'(r) \propto r^3$. The NED Lagrangian ${\cal L}(F)$ can then be reconstructed to yield the regular metric functions expanded as in Eq.~(\ref{fhexpansion}) around the origin. Depending on whether $h_3$ vanishes, the resulting functional form of ${\cal L}(F)$ near $r=0$ differs.

In Sec.~\ref{NSBHsec0}, we studied the background profiles of singular BHs 
for five classes of 
theories with an Einstein-Hilbert 
term: 
(A) Maxwell-HVT theory, 
(B) Power-law NED theory with $\beta=0$, 
(C) Born-Infeld theory with $\beta=0$, 
(D) Power-law or 
Born-Infeld NED theories with 
$\beta \neq 0$, and 
(E) Reconstructed NED theories with $\beta \neq 0$. 
In theories (A)--(D), the metric functions behave as $h(r) \propto r^{-1}$ and $f(r) \propto r^{-1}$ near $r=0$, exhibiting properties that 
are different from those of the 
RN solution. 
This behavior is related to the regularization of the electric field near the origin, as summarized in 
Table \ref{table}. In theory (E), we reconstructed $f(r)$ and $A_0'(r)$ by assuming the other metric function 
to take the form 
$h(r) = 1 - 2m/r + b_1/r^2$. 
In this case, we showed that $f(r)$ 
approaches a constant as $r \to 0$, 
with the regular electric field 
$A_0'(r)$ proportional to $r$.

In Sec.~\ref{BHpersec}, we derived the second-order action for perturbations on the SSS background for electric BHs in NED with the HVT coupling. 
There are four dynamical perturbations originating from the gravitational and vector-field sectors, which can be classified into two odd-parity and two even-parity modes. 
We showed that neither ghosts nor Laplacian instabilities appear in odd- and even-parity perturbations under the conditions 
${\cal G}_1>0$, ${\cal G}_2>0$, 
${\cal G}_3>0$, 
$c_{\Omega 3}^2>0$, and 
$c_{\Omega 4}^2>0$.
These stability criteria are valid in both timelike and spacelike regions.

In Sec.~\ref{elesinstasec}, we applied the linear stability conditions to nonsingular BHs and found that $c_{\Omega 4}^2$, which corresponds to the squared propagation speed of the even-parity vector perturbation $V$, is always negative near the regular center. Since the even-parity perturbation $\chi_2$ is subject to Laplacian instabilities through its coupling with $V$, the regular metric cannot be maintained in a steady state. Thus, as in NED, linearly stable nonsingular BHs cannot exist, even with the inclusion of the HVT coupling.

In Sec.~\ref{elesstasec}, we studied 
the linear stability of singular BHs 
for the five classes of theories discussed in Sec.~\ref{NSBHsec0}. 
In Maxwell-HVT theory (A), we found that $c_{\Omega 2}^2$ and $c_{\Omega 4}^2$ are negative near $r=0$, leading to Laplacian instabilities unless the transition distance $r_g$ is smaller than the EFT scale $r_{\rm EFT}$, which marks the limit of validity for linear perturbation theory. 
This imposes a bound on the HVT coupling, $|\beta| \lesssim 
r_{\rm EFT}^3 / r_h$, where $r_h$ is the outer horizon radius.
Since $r_{\rm EFT}$ is expected to be 
much smaller than $r_h$, the coupling 
$|\beta|$ is constrained to be 
$|\beta| \ll r_h^2$. 
This fact was not recognized in Ref.~\cite{Chen:2024hkm}, as the stability analysis there was restricted to the region outside the outer horizon.

In power-law NED with $\beta=0$, i.e., 
theory (B) with ${\cal L}(F) = F + a_p F^p$, we showed that neither ghost nor Laplacian instabilities occur for $a_p > 0$ and $p > 2$ in both timelike and spacelike regions. 
In Born-Infeld theory with $\beta=0$, i.e., theory (C) with 
${\cal L}(F) = (2/b) 
(1 - \sqrt{1 - bF})$, all linear stability conditions are satisfied for $r>0$. However, a strong coupling can arise due to $c_{\Omega 4}^2$ vanishing as $r \to 0$. 
To avoid this problem, the coupling $b$ is constrained to be $b<2r_{\rm EFT}^4/q_E^2 \ll 
2r_h^4/q_E^2$. 
Under such a bound, the metric outside the outer horizon is almost indistinguishable from that of the RN BH. 
In theories (D), i.e., power-law or Born-Infeld NED with $\beta \neq 0$, the HVT coupling dominates over the NED terms near $r=0$, leading to Laplacian instabilities due to negative values of $c_{\Omega 4}^2$. 
Thus, unless the HVT coupling is sufficiently small, as in theory (A), i.e., $|\beta| \lesssim r_{\rm EFT}^3 / r_h$, the BH solutions in theories (D) become unstable. 

In theories (E), Laplacian instabilities near $r=0$ are absent for $M>0$ and $b_1<0$. 
As shown in Fig.~\ref{fig:cOmega4_recon}, there exists a branch of solutions with $\beta>0$ where all squared angular propagation speeds remain positive 
for all $r>0$.
This property is different from 
those in theories (A) and (D).
However, $c_{\Omega 1}^2$ and $c_{\Omega 4}^2$ approach 0 
as $r \to 0$, giving rise to a strong coupling issue around the BH center. Since ${\cal G}_2$ is negative, a ghost 
is also present. The Laplacian instabilities can be avoided 
for classical perturbations, 
but the presence of ghosts 
can be problematic at the quantum level.
In Table \ref{table}, we summarize the 
linear stability of BHs in 
theories (A)--(E).

%%%%%%%%%%%%%%%%%%%%%%%%%
\begin{table}[h]
\centering
\begin{tabular}{|c|c|c|c|c|c|c|}
\hline
 & $h(r)$ near $r=0$ & 
 $A_0'(r)$ near $r=0$ & Ghosts & Laplacian instabilities & Strong coupling & 
 Parameter constraints \\ 
\hline
(A) & $\propto r^{-1}$ & 
$\propto r$ & Yes & Yes & No 
&  $|\beta|<r_{\rm EFT}^3/r_h$ \\ 
\hline
(B) & $\propto r^{-1}$ & 
$\propto r^{-2/(2p-1)}$ & No 
& No & No & $a_p>0$
\\ 
\hline
(C) & $\propto r^{-1}$ & 
$\propto r^0$ & No & No 
& Yes & 
$b<2r_{\rm EFT}^4/q_E^2$ \\
\hline
(D) & $\propto r^{-1}$ & 
$\propto r$ & Yes & Yes 
& No & $|\beta|<r_{\rm EFT}^3/r_h$
\\
\hline
(E) & $\propto r^{-2}$ & 
$\propto r$ & Yes & No 
& Yes & $\beta>0$, $M>0$, 
$b_1<0$ \\
\hline
\end{tabular}
\caption{We summarize the behavior of $h(r)$ and $A_0'(r)$ near $r=0$, as well as the presence of ghosts, Laplacian instabilities, and strong-coupling problems across five classes of theories. 
In the last column, we also present the bounds on the couplings imposed by theoretical consistency.
The theories considered are:
(A) Maxwell-HVT theory, 
(B) Power-law NED theory with $\beta=0$ and $p \geq 2$, 
(C) Born-Infeld theory with $\beta=0$, 
(D) Power-law and
Born-Infeld NED theories with 
$\beta \neq 0$ ($p \geq -1/2$ in the power-law case), and 
(E) Reconstructed NED theories with $\beta \neq 0$ and 
$h(r)=1-2M/r+b_1/r^2$.
}
\label{table}
\end{table}
%%%%%%%%%%%%%%%%%%%%%%%%%%%%%%%

We have thus shown that the 
HVT coupling generally 
induces ghost or Laplacian instabilities near the center 
of electrically charged BHs. 
To avoid this problem, 
we need to choose the coupling 
$\beta$ sufficiently small. 
For $\beta=0$, the power-law 
NED Lagrangian 
${\cal L}(F) = F + a_p F^p$, 
with $a_p>0$ and $p>2$, can 
realize linearly stable singular electric BHs 
without a strong coupling. 
Our results indicate that an alternative ultraviolet completion of vector-tensor theories, other than the HVT Lagrangian, is required to stabilize electric BHs in the high-curvature regime. 
It would be of interest to 
study whether a similar property holds for magnetically 
charged singular BHs present in 
NED theories with $\beta \neq 0$, 
which we leave for a future work.

%%%%%%%%%%%%%%%%%%%%%%%%%%%
\section*{Acknowledgements}
%%%%%%%%%%%%%%%%%%%%%%%%%%%%

CYC is supported by the Special Postdoctoral Researcher (SPDR) Program at RIKEN and RIKEN Incentive Research Grant (Shoreikadai) 2025. 
ST thanks JSPS KAKENHI Grant No.~22K03642 and Waseda University Special Research Projects (Nos.~2025C-488 and 2025R-028) 
for support. 
TS is supported by Waseda University Special Research Projects (Nos.~2025C-493) and the 15th Early Bird Program of Waseda University. 

%%%%%%%%%%%%%%%%%%%%%%%%%%%%%%%%%%%%%%%%%%%%%%%%%
\section*{Appendix~A:~Coefficients in the second-order action}
\label{AppA}
%%%%%%%%%%%%%%%%%%%%%%%%%%%%%%%%%%%%%%%%%%%%%%%%%

The coefficients appearing in Eqs.~(\ref{L1}) and (\ref{L2}) are given by
\ba
& &
p_1=-\frac{a_1}{2rf}\,,\qquad 
p_2=\sqrt{\frac{h}{f}} \frac{(8\beta h-r^2{\cal L}_{,F})A_0'}{r^2}\,,\qquad 
p_3=-\sqrt{\frac{h}{f}} \frac{4 \beta h A_0'}{r}\,,\qquad 
p_4=s_5\,,\qquad p_5=s_6\,,
\nonumber \\
& &
p_6=-\sqrt{\frac{h}{f}} \frac{L}{2r^6 fh}
\left[ f^2 r^4 {\cal L}_{,F} -2 \beta r^4 
f( 2 f'' h + f'h')+2\beta r^4 f'^2 h\right]
\,,\nonumber \\
& &
p_7=-\frac{\Mpl^2 f}{4r^2} 
\sqrt{\frac{h}{f}} \left( 1
-\frac{4\beta A_0'^2 h}{\Mpl^2 f} 
\right)(L-2)\,,
\qquad 
p_8=-\frac{p_7}{fh}\,, 
\nonumber \\
& &
p_9=\frac{1}{2r^2 fh} \sqrt{\frac{h}{f}}
[4 \beta fh' (L - 6h ) A_0' 
-r^2 f' h {\cal L}_{,F} A_0' 
-4 \beta h (L - 2 h) (f' A_0' - 2 fA_0'') 
\nonumber \\
& &\qquad \qquad \qquad \quad~
+ rf \{ 2rhF' A_0'{\cal L}_{,FF} 
+ {\cal L}_{,F} ( r h' A_0' 
+ 2 rh A_0'' + 4 h A_0') \}] 
\,,
\nonumber \\
& &
a_0=-A_0'^2\sqrt{\frac{h}{f}} \left[ 
\beta (h-1)-\frac{1}{8}r^2 {\cal L}_{,F}
-\frac{r^2 h {\cal L}_{,FF}A_0'^2}{8f} \right]\,,\qquad 
a_1=-\frac{r f}{2}\sqrt{\frac{h}{f}}\Mpl^2\,,
\qquad 
a_2=-\frac{a_1}{r}\,,\nonumber \\
& &
a_3=a_1'+\frac{1}{2}s_3 A_0'\,,\qquad
a_4=\frac{a_1}{2rh}\,,\qquad 
a_5=\frac{1}{4r^4 h} \sqrt{\frac{h}{f}} 
\left[ \Mpl^2 r^3 f (rh'+2h) -8 \beta r^3 h^2 A_0'^2
\right]\,,\nonumber \\
& &
b_0=-\frac{a_1}{2rf}\,,\qquad 
b_1=-\frac{2}{f}a_1\,,\qquad 
b_2=\frac{a_1}{rf}\,,\qquad 
b_3= \frac{4\beta h A_0'}{r}\sqrt{\frac{h}{f}}\,,\nonumber \\
& &
c_0=-\frac{1}{2} a_3
-2 \beta h A_0'^2 \sqrt{\frac{h}{f}}\,,\qquad
c_1=-\frac{1}{4r^4}\sqrt{\frac{h}{f}} 
\left[ \Mpl^2 r^3 (rf'+2f)
-24\beta r^3 h A_0'^2\right]\,,\qquad
c_2=-\frac{b_3}{h}\,,\nonumber \\
& &
d_0=-\frac{a_1}{2rf}\,,\qquad
d_1=\frac{1}{2r^5}\sqrt{\frac{h}{f}}
\left[ \Mpl^2 r^3 h(rf' + f)
-12\beta r^3 h^2 A_0'^2-r^5 f {\cal L}
+r^5{\cal L}_{,F} h A_0'^2 \right]\,,\qquad 
d_2=b_3\,,\nonumber \\
& &
d_3=p_2\,,\qquad 
s_1=\frac{1}{2}\sqrt{\frac{h}{f}}
\left[ 8\beta (1-h)
+r^2 {\cal L}_{,F}+\frac{r^2 h 
A_0'^2{\cal L}_{,FF}} {f}\right]\,,\qquad 
s_2=A_0' s_1\,,\nonumber \\
& &
s_3=\frac{1}{2f}\sqrt{\frac{h}{f}} 
\left[ f \{ 8 \beta (3h-1) 
-r^2 {\cal L}_{,F} \} A_0' 
-r^2 h A_0'^3{\cal L}_{,FF}
\right]\,,\qquad
s_4=-2b_3\,,
\nonumber \\
& &
s_5=\frac{1}{2rh}\sqrt{\frac{h}{f}} 
\left( r{\cal L}_{,F}-4\beta h' \right)\,,\qquad
s_6=-\frac{1}{2r}\sqrt{\frac{h}{f}} 
\left( rf {\cal L}_{,F}-4\beta f'h \right)\,.
\ea
%

%%%%%%%%%%%%%%%%%%%%%%%%%%%%%%%%%%%%%%%%%%%%%%%%%
\section*{Appendix~B:~One of the 
angular propagation speeds for electric BHs}
\label{AppB}
%%%%%%%%%%%%%%%%%%%%%%%%%%%%%%%%%%%%%%%%%%%%%%%%%

The even-parity perturbation $\chi_2$ has the following squared propagation 
speed in the angular direction:
\ba
c_{\Omega 3}^2 &=& 
1-A_0' (8192 \beta^5 f^2 (h-1)^2 h^3A_0'^5 
-4 \beta \Mpl^2 r^4 f ({\cal L}_{,F} A_0' 
(4 f^3 {\cal L}_{,F}(r^2{\cal L} + \Mpl^2 h) 
+ f^2 h (4 {\cal L}_{,FF}(r^2{\cal L} + \Mpl^2)
\nonumber \\
& &
-3r^2 {\cal L}_{,F}^2) A_0'^2 
- 2 r^2 fh^2 {\cal L}_{,F}{\cal L}_{,FF} A_0'^4
+r^2 h^3 {\cal L}_{,FF}^2  A_0'^6) 
+ r ({\cal L}_{,FF} h A_0'^2+ {\cal L}_{,F}f) 
(f^2 {\cal L}_{,F} (3(h-1) \Mpl^2 + r^2 {\cal L}) 
\nonumber \\
& &
+ fh ({\cal L}_{,FF}(1-h)\Mpl^2 
+ ({\cal L}{\cal L}_{,FF} -{\cal L}_{,F}^2) r^2)
A_0'^2-{\cal L}_{,F}{\cal L}_{,FF} h^2 r^2 A_0'^4)A_0'') 
+f^2 {\cal L}_{,F} \Mpl^4 r^6 ({\cal L}_{,FF} h A_0'^2 
+{\cal L}_{,F} f) \nonumber \\
& & \times
({\cal L}_{,FF} h r A_0'^2 A_0'' 
+ f {\cal L}_{,F}(r A_0'' + 2A_0')) 
+ 512 \beta^4 f (h-1 ) hA_0'^2 ({\cal L}_{,FF}
(1 - 3 h) h^2 r^2 A_0'^5-fh {\cal L}_{,F} (1 + 5h)
r^2 A_0'^3 \nonumber \\
& & 
+ 2 f^2 ({\cal L}(1 + h) r^2A_0' 
+ \Mpl^2 (A_0' - 3(h-2) h A_0' 
- (1 + h)^2 r A_0''))) 
- 64 \beta^3 ({\cal L}_{,FF}^2 (1 - h) h^4 r^4 A_0'^9 
\nonumber \\
& &
+ 2 f h^3 r^2 A_0'^6 ({\cal L}_{,F} {\cal L}_{,FF}
(1 - 3 h) r^2 A_0' -4 {\cal L}_{,FFF} h^2 \Mpl^2 
(A_0'-r A_0'' )) 
- 4 {\cal L}_{,F}f^3  h r^2 A_0'^2 ({\cal L}(h-2) r^2A_0'
\nonumber \\
& &
+ \Mpl^2 ((3(h-2)h-1)A_0'+ 2(1 + h) rA_0'')) 
+ 4 f^4 ({\cal L}^2 (h-1) r^4 A_0' 
+ {\cal L}(1 + h)
\Mpl^2 r^2 ((1 + h)rA_0''-2A_0') \nonumber \\
& &
+ (h-1) \Mpl^4 
((1 + h)(1 + 3h)A_0' -(1 + h (2 + 5 h)) r A_0'')) 
+ f^2 h^2 r^2A_0'^4 (-{\cal L}_{,F}^2 (3 + h) r^2 A_0' 
\nonumber \\
& &
+ 4{\cal L}_{,FF} (h (2 (2 - 5 h) \Mpl^2 + {\cal L}r^2)A_0'
+ (7 h^2 - 1) \Mpl^2 r A_0''))) 
+ 16 \beta^2 r^2 (-{\cal L}_{,F} {\cal L}_{,FF}^2 
h^4 r^4 A_0'^9 \nonumber \\
& &
+ {\cal L}_{,FF}^2 f h^3 r^2 
A_0'^6 ({\cal L} r^2 A_0' - (h-1) \Mpl^2 (rA_0'' + A_0')) 
+ 2 f^4{\cal L}_{,F} ({\cal L}^2 r^4A_0' 
- 2{\cal L} \Mpl^2 r^2 (A_0'- 4h A_0' + 
(1 + h) r A_0'') \nonumber \\
& &
+ \Mpl^4 (((8 - 9h)h-3) A_0' 
+ 4 h(2 h-1) r A_0'')) + f^2 h^2 r A_0'^3 
({\cal L}_{,F}^3 r^3 A_0'^2 
+ 8 {\cal L}_{,FFF} h^2 \Mpl^4 A_0''(r A_0'' - A_0') 
\nonumber \\
& &
+ 2{\cal L}_{,F}{\cal L}_{,FF} r A_0' (-{\cal L}r^2A_0' 
+ \Mpl^2 ((3 + h) r A_0''-(1 + 3 h)A_0' ))) 
+ f^3 h A_0' ({\cal L}_{,F}^2 r^2 A_0'(-3{\cal L}r^2A_0' 
+ \Mpl^2 ((5 - 13 h)A_0' \nonumber \\
& &
+ (5 + 3 h) r A_0'')) 
+ 2{\cal L}_{,FF}({\cal L}^2 r^4 A_0'^2 
+ 2{\cal L}\Mpl^2 r^2 A_0' (A_0' 
+ 2h A_0' - (1 + h)r A_0'') 
+ \Mpl^4 ((1 + (4 - 9 h) h) A_0'^2 \nonumber \\
& &
- 2 (3 h^2 + 1) 
rA_0'A_0'' + 12 h^2 r^2 A_0''^2)))))
/(4 f^2 \Mpl^2 ( \Mpl^2f-4 \beta h A_0'^2) 
(4 \beta f (h-1) \Mpl^2 + f ({\cal L}_{,F} \Mpl^2 
- 4 {\cal L} \beta) r^2 \nonumber \\
& & 
+ 4 \beta h (4 \beta (1 - h) 
+ {\cal L}_{,F} r^2) A_0'^2) 
(8 \beta f (h-1) -r^2({\cal L}_{,FF} h A_0'^2 + {\cal L}_{,F}f)))
\,.
\label{cO3Ap}
\ea

\bibliographystyle{mybibstyle}
\bibliography{bib}

\end{document}